\documentclass[11pt,letter]{article}
\pdfoutput=1 

\usepackage{jheppub}
\usepackage{amssymb,amsmath,amsthm,graphicx}
\usepackage{graphics, color}
\usepackage{subfigure}
\usepackage{latexsym}
\usepackage{bm}
\usepackage[hang,flushmargin]{footmisc}
\usepackage{epsfig}
\usepackage{textcomp}
\hypersetup{
 pdfauthor={Gary T. Horowitz, Maciej Kolanowski, Grant N. Remmen, Jorge  E. Santos},
 citecolor=black,linkcolor=black,urlcolor=black}
\usepackage[utf8]{inputenc}
\allowdisplaybreaks[1]

\begin{document}
\title{\centering{Sudden breakdown of effective field theory near \\ cool Kerr-Newman black holes}}
\author[1]{Gary~T.~Horowitz,}
\author[1]{Maciej~Kolanowski,}
\author[2]{Grant~N.~Remmen,}
\author[3]{Jorge~E.~Santos}

\affiliation[1]{Department of Physics, University of California, Santa Barbara, CA 93106, U.S.A.}
\affiliation[2]{Center for Cosmology and Particle Physics, Department of Physics, New York University, \\ New York, NY 10003, U.S.A.}
\affiliation[3]{DAMTP, Centre for Mathematical Sciences, University of Cambridge, Wilberforce Road, \\ Cambridge CB3 0WA, UK}
\emailAdd{horowitz@ucsb.edu}
\emailAdd{mkolanowski@ucsb.edu}
\emailAdd{grant.remmen@nyu.edu}
\emailAdd{jss55@cam.ac.uk}

\newcommand{\blue}{\color{blue}}
\newcommand{\red}{\color{red}}

\abstract{
It was recently shown that (near-)extremal Kerr black holes are sensitive probes of small higher-derivative corrections to general relativity. In particular, these corrections produce diverging tidal forces on the horizon in the extremal limit. We show that adding a black hole charge makes this effect qualitatively stronger. Higher-derivative corrections to the Kerr-Newman solution produce tidal forces that scale inversely in the black hole temperature. We find that, unlike the Kerr case, for realistic values of the black hole charge large tidal forces can arise before quantum corrections due to the Schwarzian mode become important, so that the near-horizon behavior of the black hole is dictated by higher-derivative terms in the effective theory.
}

\maketitle

\newpage
\section{Introduction}

 General relativity often arises as the leading term in a low-energy effective field theory (EFT), in which the Einstein-Hilbert term receives higher-derivative corrections associated with new physics. These higher-derivative terms are suppressed by powers of a mass scale associated with degrees of freedom that have been integrated out, which sets the limit of validity of the EFT. As a result, such terms usually produce only small corrections to the low-energy physics. It was recently shown that maximally rotating black holes are an exception: small higher-derivative corrections to Einstein's equation can result in tidal force singularities on the horizon of large extremal Kerr black holes~\cite{Horowitz:2023xyl}. These singularities are unusual in that all scalar curvature invariants remain finite. Nonextremal black holes remain smooth, but the tidal forces blow up like a power of $1/T$, where $T$ is the black hole temperature.  This power is very small, so $T$ must be exponentially small before large tidal forces appear.

We show that the
inclusion of a nonzero black hole charge $Q$ results in a much stronger singularity at the horizon.
In particular, the divergence of tidal forces as $T \to 0$ is much more rapid. Moreover, this result does not depend too much on the matter content of the ultraviolet theory, as long as it generates higher-derivative terms containing the ${\rm U}(1)$ gauge field, resulting in an Einstein-Maxwell EFT.
For example, these terms---the so-called Euler-Heisenberg Lagrangian---are generated if the ultraviolet contains charged particles (in the case of our universe, the leading contribution to the Einstein-Maxwell EFT comes from the electron loops), though a massive dilaton coupled to the Maxwell kinetic term can also generate the effects of interest at tree level. 

As we review in Secs.~\ref{sec:settingscene} and \ref{sec:scaling_EFT}, the leading EFT corrections to Einstein-Maxwell theory in four spacetime dimensions can be reduced to three four-derivative terms~\cite{Kats:2006xp,Cheung:2014ega},
\begin{equation}
\int_{\mathcal{M}}{\rm d}^4 x\sqrt{-g}\Bigg(c_6\,R^{abcd}\,F_{ab}\,F_{cd} +c_7\,F_{ab}\,F^{ab}\,F_{cd}\,F^{cd}+c_8\,F_{ab}\,F^{bc}\,F_{cd}\,F^{da}\Bigg).
\label{eq:xhd}
\end{equation}
These terms produce singularities on the  horizon for the following reason. Near an extremal horizon the metric may be approximated as
\begin{equation}
    g \approx g_{NH} + \rho^\gamma h,
\end{equation}
where $g_{NH}$ is the near-horizon metric, $\rho$ is an affine distance from the horizon, $h$ is a smooth tensor, and the scaling dimension $\gamma$ is determined from the equations of motion. For Einstein-Maxwell solutions, the horizons are smooth and the leading corrections have integer $\gamma$, starting with $\gamma = 1$.  The EFT corrections to the action perturb this solution. While $g_{NH}$ is smooth, $\gamma$ can be shifted away from its integer value. For Kerr, the $\gamma =1$ mode is not shifted, but $\gamma \ge 2$ is. We will show that for Kerr-Newman, $\gamma = 1$ is shifted by EFT corrections, and this makes a significant difference in the strength of the singularity.

In ingoing null coordinates where $\ell =\partial/\partial \rho$ is tangent to affinely parameterized null geodesics, and setting $m = \partial/\partial \phi$ for $\phi$ the coordinate of axisymmetry,
the Weyl tensor near the extremal horizon is
\begin{equation}
    C_{abcd} \ell^a m^b\ell^c m^d  \sim \gamma (\gamma -1) \rho^{\gamma-2}.
\end{equation}
Analogously, at finite temperature the tidal forces at the horizon are
\begin{equation}
    C_{abcd} \ell^a m^b\ell^c m^d |_{H} \sim \gamma (\gamma -1) T^{\gamma-2}.
\end{equation}
For Kerr, we showed in Ref.~\cite{Horowitz:2023xyl} that Riemann-cubed corrections, with a sign determined by the spin of the lightest massive particles integrated out at one loop (and dominated in the standard model by the neutrinos), decrease $\gamma$ slightly below 2, so there is a singularity. For Kerr-Newman, since $\gamma = 1$ is shifted, for any sign of $\delta \gamma$ the singularity is stronger. To first order in the coefficients $c_k$,
    \begin{equation}
  C_{abcd} \ell^a m^b\ell^c m^d |_{H} \sim \delta\gamma / T.\label{eq:NP}
\end{equation}

Since the leading EFT corrections cause a large change to the horizon, one might expect that higher-order terms will also cause large changes, so including just the four-derivative terms is not reliable. As we discuss in Sec.~\ref{sec:Discussion}, this is not the case. The breakdown of EFT is more subtle.

\subsection{Rough estimates}\label{sec:rough}

We now estimate at what $T$  the EFT-generated tidal forces will exceed the ambient curvature. 
A physically motivated ultraviolet completion of the higher-dimension operator terms containing $F_{ab}$ is a charged, massive particle integrated out at one loop, that is, a gravitational extension of the Euler-Heisenberg Lagrangian~\cite{Cheung:2014ega,Drummond:1979pp,Dunne,Bastianelli:2008cu,Bastianelli:2012bz,EulerHeisenberg}.
In that case, $c_{7,8}$ will go like $(q/m)^4$ for the particle integrated out, while $c_6$ will go like $(q/m)^2$, all multiplied by loop factors of order $10^{-4}$.
The standard model particle with the largest charge-to-mass ratio that we can integrate out is an electron. Since in geometric units ${q_e}/{m_e} \approx 2 \times 10^{21}$, we may expect that $c_7$ and $c_8$ will yield larger first-order contributions than $c_6$.

Astrophysical black holes can carry a small charge. One often hears that a charged black hole will preferentially accrete matter of the opposite charge, and so realistic black holes are neutral. But this only means that their charge $Q$ is negligible in comparison to their mass, $M \gg Q$, and thus one may safely use the Kerr solution to describe their exterior. However, black holes are often in environments with external magnetic fields, and Wald~\cite{Wald:1974np} noticed that a rotating black hole embedded in an external magnetic field produces a nonzero electric field nearby (just as a rotating surface would do in classical electrodynamics). As a result, the black hole attracts charges of one sign (depending on the orientation of the spin with respect to the magnetic field). This process continues until the accumulated electric charge is enough to balance the electric field produced by rotation, yielding $Q = 2BJ$. So $Q/M$ grows with the size of the black hole (if $B$ is constant).  For the supermassive black hole in the center of our galaxy, it has been estimated that $Q/M \approx 10^{-12}$~\cite{zajavcek2019electric}.
   Black holes can temporarily have much larger charge in certain dynamical situations. Since a pulsar has very strong magnetic fields, a black hole that is about to merge with a pulsar can have~\cite{Levin:2018mzg}
    \begin{equation}
        Q/M \approx 10^{-7}.\label{eq:bestcase}
    \end{equation}

We can estimate the size of the shift in $\gamma$ generated by the last two terms in Eq.~\eqref{eq:xhd} as follows. Metric perturbations are sourced by $ c_{7,8} F^4 $, so since $\gamma$ is dimensionless,
\begin{equation}
    \delta \gamma \sim G  c_{7,8}  F^4 (GM)^2 \sim  c_{7,8} Z^4 / G^3 M^2,
\end{equation}
where $Z=Q/M$ 
and we have restored powers of Newton's constant. Using  $c_{7,8} \approx 10^{-4} (q_e/m_e)^4 \approx  10^{81} G^2$, we get
\begin{equation}
    \delta \gamma\sim  10^{81} Z^4 / (G M^2) \sim 
    10^{81} Z^4 / S,
\end{equation}
where $S$ is the black hole entropy.

Since the curvature scales like $\delta \gamma /T$ we have 
\begin{equation}\label{eq:estimate2}
    \delta C_{\rho \phi \rho \phi}|_{H} \sim 10^{81} \frac{Z^4}{  G M S T}.
\end{equation}
The factor $GM$ in the denominator arises since $\phi$ is dimensionless, so the curvature component $C_{\rho \phi \rho \phi}|_{H}$ is dimensionless. Since it is order one in the background, the new tidal forces will be important when $T < T_{EFT}$, with
\begin{equation}\label{eq:TEFT}
    T_{EFT} \sim \frac{10^{81} Z^4}{G M S}  .
\end{equation}

It was shown in Refs.~\cite{Rakic:2023vhv,Kapec:2023ruw} that at very low $T$, there is an important quantum correction to Kerr (and Kerr-Newman \cite{Rakic:2023vhv,Maulik:2024dwq}) that is not captured by the EFT. It comes from a graviton mode in the near-horizon throat that becomes very light in the extremal limit. Including this mode modifies the low-energy density of states and removes the puzzling ground-state degeneracy suggested by the nonzero entropy of the extremal solution.\footnote{This is an extension of earlier work~\cite{Iliesiu:2020qvm}, which established a similar result for (nonsupersymmetric) extremal charged black holes.} The effects of this mode become important at a temperature
\begin{equation}
    T_q = \frac{\pi}{G M S} .
\end{equation}
 We see that for any $Z$ above $10^{-20}$ the effect of the EFT loop corrections that we are discussing are important before this quantum gravity effect.

  It should be noted that unless $Z$ is of order one, $T_{EFT}$ is a very low temperature in the EFT of the standard model. A typical nonextremal black hole has $T \sim 1/GM$, and a solar mass black hole has $S \sim 10^{78}$. So for $Z$ given in Eq.~\eqref{eq:bestcase}, $T_{EFT}$ is exceedingly small, of order $10^{-25}/GM$. Said another way, to see these new tidal forces the black hole has to be much closer to extremality, i.e., $J\to M^2$, than any astrophysical black hole is expected to be.
We will return to the question of possible astrophysical relevance in Sec.~\ref{sec:numerics}, after we numerically construct EFT-corrected black holes farther from extremality.
  
Of course,  properties of charged black holes are of fundamental theoretical interest, independent of possible astrophysical applications.  In particular, the behavior of black holes under higher-derivative corrections is of great interest for the weak gravity conjecture~\cite{Arkani-Hamed:2006emk,Kats:2006xp,Cheung:2018cwt,Cheung:2019cwi,Arkani-Hamed:2021ajd,Hamada:2018dde,Aalsma:2022knj}. For this reason we will not restrict to small $Q$, but consider EFT corrections to the general extreme Kerr-Newman black hole.

In more exotic scenarios, it is of course possible that the leading contributions to $F^4$ terms do not come from integrating out the electron at one loop, but are larger and generated via other mechanisms.
Possibilities include a massive dilaton coupled to $F^2$, the Born-Infeld Lagrangian, and more, though the Wilson coefficients have been constrained by experiment to not be parametrically larger than the Euler-Heisenberg values~\cite{DellaValle:2014xoa,Fouche:2016qqj}.
However, the obstacle to seeing our result in the standard model can be alleviated by new physics; for example, a new ultralight dark sector with charged matter and a dark ${\rm U}(1)$ can dramatically enhance the values of the Wilson coefficients for the analogous operators for the dark photon, as long as the new gauge field remains unbroken~\cite{Preskill:1990ty} and the black hole is charged under it (e.g., ${\rm U}(1)_{B-L}$ scenarios~\cite{Cheung:2014vva,Heeck:2014zfa}).

 Fully exploring the parameter space for new physics leading to $F^4$ terms, as well as their astrophysical phenomenology, is a task we leave to future work.
  The upshot of the present paper will be to again demonstrate that black holes can be sensitive probes of new physics, in a manner even more dramatic than the pure Kerr case of Ref.~\cite{Horowitz:2023xyl}, and in a way that persists beyond the strict extremal limit.

\subsection{Summary of the results and plan of the paper}
In this paper we investigate  higher-curvature corrections to the extreme (and near extreme) Kerr-Newman spacetime. These corrections are of fourth order in the derivatives of the background metric and the Maxwell field. 

We start by reviewing the general EFT for the Einstein-Maxwell theory in  Sec.~\ref{sec:settingscene}. We also review  the Kerr-Newman solution and its near-horizon geometry (and gauge field). The next three sections are focused on this near-horizon region, and our results are all obtained analytically. In Sec.~\ref{sec:scaling_KN} we  solve for the axially symmetric, stationary perturbations of this near-horizon geometry (in Einstein-Maxwell theory) and calculate their scaling dimensions. Previously, these were known only for Kerr(-AdS) and Reissner-Nordstr\"om(-AdS) \cite{Horowitz:2022mly}. In Sec.~\ref{sec:EFT_NHG}, the EFT corrections to the near-horizon geometry of Kerr-Newman are found. Then in Sec.~\ref{sec:scaling_EFT}, the EFT corrections to the scaling dimensions are computed. We  explicitly verify  that these shifts are field redefinition invariant. In particular, it is shown that the mode with $\gamma =1$ has a shift in its scaling dimension, which implies the appearance of a singularity on the horizon. 

To confirm this result, in the next two sections we numerically compute the full asymptotically flat EFT-corrected solutions. In Sec.~\ref{sec:finite_temp} we first construct solutions at finite temperature. We measure the tidal forces at the horizon and verify than they blow up as $T \to 0$ in the manner predicted above. Next, in Sec.~\ref{sec:extremal} we directly compute the extremal EFT-corrected solution. We recover the change in the scaling exponents from the full solution and show that they agree with the near-horizon analysis. Most importantly, we show that the tidal forces indeed blow up as $\rho \to 0$ as expected. We also confirm that scalar curvature invariants remain finite at the horizon.

In Sec.~\ref{sec:numerics} we explore whether there are astrophysical applications of our results. Finally, in Sec.~\ref{sec:Discussion} we discuss some implications of our results and potential future directions.
\section{Setting the scene}\label{sec:settingscene}
In this section, we review background material that will be needed for our analysis.

\subsection{EFT with Maxwell fields}

We start by considering the leading parity-preserving higher-derivative corrections to the Einstein-Maxwell effective theory,
\begin{equation}
\begin{aligned}
S&\equiv\int_{\mathcal{M}}{\rm d}^4 x\sqrt{-g}\mathcal{L}
\\&=\int_{\mathcal{M}}{\rm d}^4 x\sqrt{-g}\Bigg(\frac{1}{2\kappa^2}R-\frac{1}{4}F_{ab}\,F^{ab}+c_1\,R^2+c_2\,R^{ab}\,R_{ab}+c_3\,R_{abcd}\,R^{abcd}
\\
&\qquad\qquad\qquad\qquad\qquad\qquad\qquad +c_4\,R\,F^{ab}\,F_{ab}+c_5\,R^{ab}\,F_a^{\phantom{a}c}\,F_{bc}+c_6\,R^{abcd}\,F_{ab}\,F_{cd}
\\& \qquad\qquad\qquad\qquad\qquad\qquad\qquad+c_7\,F_{ab}\,F^{ab}\,F_{cd}\,F^{cd}+c_8\,F_{ab}\,F^{bc}\,F_{cd}\,F^{da}\Bigg)\,,
\label{eq:hd}
\end{aligned}
\end{equation}
where $\kappa^2=8 \pi G$, $G$ is Newton's constant, and $F={\rm d}A$. 
If the Einstein-Maxwell EFT is generated by integrating out a fermion of charge $q$ and mass $m$, one has~\cite{Cheung:2014ega,Drummond:1979pp,Dunne,Bastianelli:2008cu,Bastianelli:2012bz,EulerHeisenberg} 
\begin{equation}
\begin{aligned}\label{eq:cvalue}
(c_7,c_8) &= \frac{q^4}{\pi^2 m^4}\times \left(-\frac{1}{576},\frac{7}{1440}\right) \\
(c_4,c_5,c_6) &= \frac{q^4}{\pi^2 m^2}\times\left(\frac{1}{576},-\frac{13}{1440},\frac{1}{1440}\right),
\end{aligned}
\end{equation}
while for a scalar, one finds
\begin{equation}
\begin{aligned}
(c_7,c_8) &= \frac{q^4}{\pi^2 m^4}\times\left(\frac{1}{4608},\frac{1}{5760}\right) \\
(c_4,c_5,c_6) &= \frac{q^2}{\pi^2 m^2}\times\left(-\frac{1}{1152},-\frac{1}{1440},-\frac{1}{2880}\right).
\end{aligned}
\end{equation}
Threshold values of $c_{1,2,3}$ are not calculable within quantum field theory (though their beta functions are~\cite{Deser:1974cz,Arkani-Hamed:2021ajd}); they are fixed by quantum gravity and are expected to be string- or Planck-suppressed.

Throughout this paper, we will always treat all the higher-corrections perturbatively, and in particular we will work only to first order in Wilson coefficients $c_i$. These are not fundamental quantities and may be changed by field redefinitions~\cite{Cheung:2018cwt,Cheung:2019cwi}.\footnote{This ambiguity exists because the effective action is derived in the context of scattering processes, and the S-matrix is invariant upon such redefinitions.} The most general field redefinition relevant at this order in derivatives is 
\begin{equation}
    g_{ab} \rightarrow g_{ab} + r_1 R_{ab} + r_2 R g_{ab} + r_3 \kappa^2 F_{ac} F_{b}^{\ \ c} + r_4 \kappa^2 g_{ab} F_{cd} F^{cd}\label{eq:redef}
\end{equation}
for some arbitrary coefficients $r_i$. 
For later use, it is convenient to introduce rescaled Wilsonian coefficients $d_i$ as 
\begin{align}\label{eq:ddef}
&d_{1,2,3}=\kappa^2\,c_{1,2,3}\nonumber
\\
&d_{4,5,6}=c_{4,5,6}
\\
&d_{7,8}=\kappa^{-2}\,c_{7,8}\,,\nonumber
\end{align}
so that all the $d_i$ have uniform mass dimension $-2$; that is, in a tree-level completion, $d_i \sim 1/\Lambda^2$, where $\Lambda$ is the scale of new physics. 
The only four combinations of the Wilson coefficients that are left invariant (to first order in $r_i, c_i$) are 
    \begin{equation}
    \begin{aligned}
        d_0 &\equiv d_2 +4 d_3 + d_5 + d_6 + 4 d_7 + 2 d_8 \\
        d_9 &\equiv d_2+4 d_3 + d_5 + 2d_6 + d_8\,,
        \end{aligned}\label{eq:goodbasis}
    \end{equation}
    along with $d_3$ and $d_6$. Any physical observable must be a function of $d_0$, $d_3$, $d_6$, and $d_9$ only. Moreover, we know that the Gauss-Bonnet term,
\begin{equation}
\int_{\mathcal{M}}{\rm d}^4x \sqrt{-g}\left(R^2-4R_{ab}R^{ab}+R_{abcd}R^{abcd}\right),
\end{equation}
is topological in four spacetime dimensions and thus cannot affect the equations of motion. The net result is that one can take $d_6$, $d_7$, and $d_8$ as a basis for the EFT corrections (as in Eq. \eqref{eq:xhd}), but for now we include all terms to explicitly check that our results are field redefinition invariant.

The equations of motion that follow from Eq.~\eqref{eq:hd} read~\cite{Kats:2006xp}
\begin{equation}
\begin{aligned}
\label{eqs:EOM}
\nabla^aF_{ab}&=c_4 J^{c_4}_{b}+c_5 J^{c_5}_{b}+c_6 J^{c_6}_{b}+c_7\,J^{c_7}_{b}+c_8 J^{c_8}_{b} \\ \\
R_{ab}-\frac{1}{2}Rg_{ab}-\kappa^2\left(F_{a}^{\phantom{a}c}F_{bc}{-}\frac{1}{4}g_{ab}F_{cd}F^{cd}\right) &=\kappa^2\big(c_1 T^{c_1}_{ab}+c_2 T^{c_2}_{ab}+c_3 T^{c_3}_{ab}+c_4 T^{c_4}_{ab}
\\
&\qquad +c_5 T^{c_5}_{ab}+c_6 T^{c_6}_{ab}+c_7 T^{c_7}_{ab}+c_8 T^{c_8}_{ab}\big)\,,
\end{aligned}
\end{equation}
where
\begin{equation}
\begin{aligned}
J^{c_4}_a&=4\left(R\nabla^bF_{ba}-F_{ab}\nabla^b R\right)\\
J^{c_5}_a&=-2\left(R_{a}^{\phantom{a}c}\nabla_{b}F_c^{\phantom{c}b}+R^{cb}\nabla_b F_{ac}+F_{a}^{\phantom{a}c}\nabla_bR_{c}^{\phantom{c}b}+F^{cb}\nabla_{b}R_{ac}\right)\\
J^{c_6}_a&=-4 R_{adbc}\nabla^dF^{bc}-4 F^{bc}\nabla^d R_{adbc}\\
J^{c_7}_a&=8 \nabla^e\left(F_{ea} F^{cd}F_{cd}\right)\\
J^{c_8}_a&=-8 \nabla^b\left(F_{a}^{\phantom{a}p}F_{cp} F^{c}_{\phantom{c}b}\right)
\end{aligned}
\end{equation}
and
\begin{equation}
\begin{aligned}
T^{c_1}_{ab}& =4\nabla_a \nabla_b R-4 g_{ab}\Box R-4 R_{ab} R+g_{ab}R^2
\\
T^{c_2}_{ab}& =4\nabla_c \nabla_{(a} R_{b)}^{\phantom{b}c}-2\Box R_{ab}-2 g_{ab}\nabla_d\nabla_cR^{cd}-4 R_a^{\phantom{a}c}R_{bc}+g_{ab}R_{cd}R^{cd}
\\
T^{c_3}_{ab}& =-\left(4R_{a}^{\phantom{a}cde}R_{bcde}-g_{ab}R_{cdef}R^{cdef}+8 \nabla_c\nabla_d R_{(a\phantom{c}b)}^{\phantom{(a}c\phantom{b)}d}\right)
\\
T^{c_4}_{ab}& =4F^{cd}\nabla_{(a}\nabla_{b)}F_{cd}+4\nabla_a F^{cd}\nabla_b F_{cd}-4 g_{ab}F^{cd}\Box F_{cd}
\\
&\qquad-4 g_{ab}\nabla_e F_{cd}\nabla^e F^{cd}-2 R_{ab}F_{cd}F^{cd}-4 F_{a}^{\phantom{a}c}F_{bc}R+g_{ab}F_{cd}F^{cd}R
\\
T^{c_5}_{ab}& =4F_{(a}^{\phantom{(a}c}R_{b)d}F_{c}^{\phantom{c}d}-2F_{a}^{\phantom{a}c}F_{b}^{\phantom{b}d}R_{cd}+g_{ab}F_c^{\phantom{c}e}F^{cd}R_{de}-2\nabla_{(a}F_{b)}^{\phantom{b)}c}\nabla_d F_{c}^{\phantom{c}d}
\\
&\qquad-2\nabla_d \nabla_{(a}F_{b)c}F^{cd}-2\nabla_d\nabla_{(a}F^{cd}F_{b)c}-2 \Box F_{(a}^{\phantom{(a}c}F_{b)c}
\\
&\qquad -g_{ab}F^{cd}\nabla_d \nabla^e F_{ce}-2\nabla^d F_{(a}^{\phantom{(a}c}\nabla_{b)}F_{cd}-2 \nabla^d F_{a}^{\phantom{a}c}\nabla_d F_{bc}
\\
&\qquad+g_{ab}\nabla_c F^{cd}\nabla_e F_{d}^{\phantom{d}e}-g_{ab}F^{cd}\nabla_e\nabla_d F_{c}^{\phantom{c}e}-g_{ab}\nabla_d F_{ce}\nabla^e F^{cd}
\\
T^{c_6}_{ab}& =-\Big(6F_{(a}^{\phantom{b}c}F^{de}R_{b)cde}-g_{ab}F^{cd}F^{ef}R_{cdef}-4F_{c(a}\nabla^{c}\nabla^{d}F_{b)d}
\\
&\qquad -4F_{d(a}\nabla^{d}\nabla^{c}F_{b)d}+4\nabla_{c}F_{a}^{\phantom{a}c}\nabla_dF_{b}^{\phantom{b}d}+4 \nabla_{c}F_{bd}\nabla^dF_{a}^{\phantom{a}c}\Big) \\
T^{c_7}_{ab}&=F^{pq}F_{pq}\left(g_{ab}F^{cd}F_{cd}-8 F_{a c}F_{b}^{\phantom{b}c}\right) \\
T^{c_8}_{ab}&=g_{ab}F_{c}^{\phantom{c}p}F_{dp}F^{cq}F^{d}_{\phantom{d}q}-8 F_{a}^{\phantom{a}p}F_{cp}F_{b}^{\phantom{b}q}F^c_{\phantom{c}q} \,.
\end{aligned}
\end{equation}
In the following sections, we are going to study a number of solutions to these equations of motion. As mentioned above, we are always going to work assuming that the right-hand sides of both expressions in Eq.~\eqref{eqs:EOM} are perturbatively small compared with the background solution, which we take to be a Kerr-Newman black hole.

\subsection{Kerr-Newman black holes}

The metric and Maxwell potential of a Kerr-Newman black hole are~\cite{Newman:1965my}
\begin{equation}
\begin{aligned}
\label{eqs:KNgA}
{\rm d}s^2_{\rm KN}=&-\frac{\Delta(r)}{\Sigma(r,\theta)}({\rm d}t-a\,\sin^2\theta\, {\rm d}\phi)^2+\Sigma(r,\theta)\left(\frac{{\rm d}r^2}{\Delta(r)}+{\rm d}\theta^2\right)
\\
&+\frac{\sin^2\theta}{\Sigma(r,\theta)}\left[a\,{\rm d}t-(r^2+a^2){\rm d}\phi\right]^2 \\
A_{\rm KN}  =&-\frac{\sqrt{2}\,Q\,r}{\kappa\,\Sigma(r,\theta)}({\rm d}t-a\,\sin^2\theta\, {\rm d}\phi)-\frac{\sqrt{2}\,P\,\cos \theta}{\kappa\,\Sigma(r,\theta)}\left[a\,{\rm d}t-(r^2+a^2){\rm d}\phi\right],
\end{aligned}
\end{equation}
with
\begin{equation}
\Sigma(r,\theta)=r^2+a^2\cos^2\theta\quad\text{and}\quad \Delta(r)=r^2+a^2-2\,M\,r+Q^2 +P^2.
\end{equation}
In the above, $\phi$ and $\theta$ can be regarded as standard azimuthal and polar angle coordinates on the $S^2$, respectively, with $\phi\sim\phi+2\pi$ and $\theta\in[0,\pi]$.

Once we are given an axisymmetric and stationary metric, we can compute the total electric charge, magnetic charge, angular momentum, and mass using standard Komar integrals,
\begin{equation}
\begin{aligned}
\label{eqs:komar}
Q_{\rm e}&=\lim_{r\to+\infty}\int_{S^2_r} \star F, &&\qquad & J&=\frac{1}{2\kappa^2}\lim_{r\to+\infty}\int_{S^2_r} \star {\rm d}m, \\
Q_{\rm m}&=\lim_{r\to+\infty}\int_{S^2_r} F, && &
E& =-\frac{1}{\kappa^2}\lim_{r\to+\infty}\int_{S^2_r} \star {\rm d}k\,,
\end{aligned}
\end{equation}
with $m=\partial/\partial \phi$ and $k=\partial/\partial t$ the axial and stationary Killing vector fields of the Kerr-Newman spacetime. Note that these expressions will remain valid even in the presence of higher-derivative corrections, as long as the novel EFT terms decay sufficiently rapidly near spatial infinity~\cite{Cheung:2018cwt}. All the higher-derivative terms appearing in Eq.~(\ref{eq:hd}) fall under this class. 

For the Kerr-Newman spacetime, we have
\begin{equation}
Q_{\rm e}=\frac{4\pi \sqrt{2}}{\kappa}Q\,,\quad Q_{\rm m}=\frac{4\pi \sqrt{2}}{\kappa}P\,,\quad J = a E\,,\quad\text{and}\quad E=\frac{8\pi M}{\kappa^2}\,.\label{eq:charges}
\end{equation}
In what follows, we will be interested in purely electrically charged solutions, so we set $P=0$. For brevity, we will assume throughout that $J>0$ and $Q>0$ without loss of generality.
For $M\geq \sqrt{a^2+Q^2}$, the Kerr-Newman spacetime describes a black hole. The upper bound, $M=\sqrt{a^2+Q^2}$, corresponds to an \emph{extremal} black hole and plays the leading role in this work. Extremal black holes have minimum energy for given values of the electric charge and angular momentum. The black hole event horizon corresponds to the null hypersurface\footnote{To see that $r=r_+$ is a null hypersurface, one needs to first change to regular coordinates at $r=r_+$. These are akin to the so-called Kerr coordinates,
\begin{align*}
&\mathrm{d}v_{\pm}=\mathrm{d}t\pm\frac{r^2+a^2}{\Delta(r)}\mathrm{d}r
\\
&\mathrm{d}\varphi_{\pm}=\mathrm{d}\phi\pm\frac{a}{\Delta(r)}\mathrm{d}r\,,
\end{align*}
with $\varphi_{\pm}\sim\varphi_{\pm}+2\pi$. The coordinates $(v_+,\varphi_+)$ show that $r=r_+$ is a future event horizon, while the coordinates $(v_-,\varphi_-)$ cover a different extension of the Kerr-Newman metric and reveal that, in that extension, $r=r_+$ is a white hole horizon.} $r=r_+$, with $r_{\pm}=M\pm\sqrt{M^2-a^2-Q^2}$. The null hypersurface $r=r_-$, on the other hand, is a \emph{Cauchy horizon}. When the black hole becomes extremal, $r_+=r_-$, at which point the Cauchy horizon and event horizon coalesce. 

As expected from Hawking's rigidity theorem, the horizon is a \emph{Killing horizon}, with the horizon generator being
\begin{equation}
K=k+\Omega\,m\,,
\end{equation}
where $\Omega = a/(a^2 + r_+^2)$ is the black hole's angular velocity. To the Killing horizon we can associate a Hawking temperature \cite{Hawking:1974rv},
\begin{equation}
T=\frac{1}{2\pi}\sqrt{-\frac{1}{4}\left.\frac{\nabla_a(K^cK_c)\nabla^a (K^d K_d)}{K^e K_e}\right|_{H}}=\frac{r_+^2-a^2-Q^2}{4\,\pi\,r_+\,(r_+^2+a^2)}\,,
\end{equation}
from which we note that extremal black holes have $T=0$. The two-surface of constant $t$ and $r=r_+$ is the so-called bifurcating Killing surface where $K$ vanishes identically, which we denote by $\mathcal{B}^+$.

Another quantity that we can associate with a Killing horizon is the so-called electric chemical potential $\mu$ defined as
\begin{equation}
\mu=-\frac{\kappa}{\sqrt{2}}\left(\left.K^aA_a\right|_{H}-\lim_{r\to+\infty}K^aA_a\right)=\frac{Q}{r_+^2+a^2}\,.
\end{equation}
So long as the horizon is Killing, the expressions above for the temperature and chemical potential are valid even in the presence of higher-derivative corrections~\cite{Wald:1993nt}.

The last quantity of interest for us is the \emph{Wald entropy} \cite{Wald:1993nt}, defined for \emph{stationary} solutions only and given by
\begin{equation}
S_{\rm W}= -2\pi \oint_{\mathcal{B}^+} {\rm d}^{2}x\, \sqrt{\sigma} \, \frac{\delta \mathcal{L}}{\delta R_{abcd}} \varepsilon_{ab} \varepsilon_{cd}\,,
\end{equation}
where $\varepsilon^{ab}$ is the binormal to the bifurcating Killing surface, with normalization $\varepsilon_{ab}\varepsilon^{ab} = -2$, $\sigma$ is the determinant of the two-dimensional metric on $\mathcal{B}^+$, and $\mathcal{L}$ is the Lagrangian of the theory under consideration, which for us is defined in Eq.~(\ref{eq:hd}).

It is a simple exercise to show that when no higher-derivative terms are present, we recover the usual Bekenstein-Hawking  entropy~\cite{Bekenstein:1973ur,Hawking:1975vcx},
\begin{equation}
\lim_{c_i\to0}S_{\rm W}=S_{\rm BH}=\frac{2\pi\,A}{\kappa^2},
\end{equation}
with $A$ the area of $\mathcal{B}^+$.   For the terms proportional to $c_7$ and $c_8$, we will only need $S_{\text{BH}}$, as the corresponding higher-derivative corrections do not depend on the Riemann tensor. However, for the {remaining $c_i$}, the Wald entropy will be different from the standard Bekenstein-Hawking entropy and will play a role later in the paper.

One can show that, with the definitions above, the electrically charged Kerr-Newman black hole satisfies a first law of black hole mechanics,
\begin{equation}
{\rm d}M=T\,{\rm d}S_{\rm W}+\mu\,{\rm d}Q+\Omega\,{\rm d}J\,.
\end{equation}
Indeed, we expect the above to be valid even in the presence of higher-derivative corrections; in fact, this is how the Wald entropy was initially derived and the reason why it only applies to stationary solutions~\cite{Wald:1993nt}.
\subsection{The near-horizon geometry of Kerr-Newman black holes} \label{sec:NHG-KN}
There is a limit of the extremal Kerr-Newman black hole that will be important in what follows. This is the so-called near-horizon limit and in the current example generalizes the Bardeen-Horowitz construction~\cite{Bardeen:1999px}. To take this limit, we set the spin parameter  $a$ to $\sqrt{M^2-Q^2}$, introduce $x=\cos \theta$, and take 
\begin{equation}
t=4M(2-Z^2)\frac{\tau}{\lambda}\,,\quad r=r_+\left(1+\frac{\lambda\,\rho}{4}\right)\,,\quad\text{and}\quad \phi=\varphi+\frac{a}{a^2+r_+^2}\,t\,,
\end{equation}
where we again used $Z\equiv Q/M$ for the charge-to-mass ratio in geometric units. The limit that we are interested in sends $\lambda\to0$ while keeping $(\tau,\rho,x,\varphi)$ fixed. The resulting metric and Maxwell potential read\footnote{This was first derived in Ref.~\cite{Hartman:2008pb}, including a nonzero cosmological constant.} 
\begin{equation}
\begin{aligned}
\label{eq:NHKN}
\mathrm{d}s^2_{\rm NH-KN}&=2M^2[{F^{(0)}_1(x)}]^2\biggl[-\rho^2 \mathrm{d}\tau^2  + \frac{\mathrm{d}\rho^2}{\rho^2} + \frac{\mathrm{d}x^2}{1 - x^2} 
\\&\qquad\qquad\qquad\qquad + [{F^{(0)}_2(x)}]^2(1 - x^2)\left(\mathrm{d}\varphi+\rho\,\omega_{\rm NH}^{(0)}\,\mathrm{d}\tau\right)^2\biggr]\\
A_{\rm NH-KN}&=\frac{\sqrt{2}\,M}{\kappa}\left[Q_{\rm NH}^{(0)}\,\rho\,{\rm d}\tau+(1-x^2)F^{(0)}_2(x)F^{(0)}_3(x)\left(\mathrm{d}\varphi+\rho\,\omega_{\rm NH}^{(0)}\,\mathrm{d}\tau\right)\right]\,,
\end{aligned}
\end{equation}
where
\begin{equation}
\begin{aligned}
F^{(0)}_1(x)&=\frac{\sqrt{1+(1-Z^2)x^2}}{\sqrt{2}}\,,\;\; F^{(0)}_2(x)=\frac{(2-Z^2)}{1+(1-Z^2)x^2}\,,\;\;\,F^{(0)}_3(x)=\frac{Z\sqrt{1-Z^2}}{2-Z^2}\,,\label{eq:NHEKvalues}
\\
\omega_{\rm NH}^{(0)}&=\frac{2\sqrt{1-Z^2}}{2-Z^2},\quad \text{and}\quad Q_{\rm NH}^{(0)}=\frac{Z^3}{2-Z^2}\,.
\end{aligned}
\end{equation}
We note that in going from the full extremal Kerr-Newman black hole to its near-horizon geometry, we also apply a gauge transformation to $A$ of the form
\begin{equation}
A+\mathrm{d}\chi\quad\text{with}\quad \chi = \frac{\sqrt{2}}{\kappa}\frac{Z}{2-Z^2}\,t
\end{equation}
\emph{before} taking the near-horizon limit $\lambda\to0$. Note that Eq.~\eqref{eq:NHKN}, with symmetry group ${\rm O}(2,1) \times {\rm U}(1)$, is more symmetric than the initial Kerr-Newman spacetime, as is typical of near-horizon extremal geometries~\cite{Bardeen:1999px,Kunduri:2013gce,Horowitz:2023xyl,Hadar:2020kry,Porfyriadis:2021psx,Charalambous:2022rre,Porfyriadis:2021zfb}. 

\section{Scaling dimensions for extremal Kerr-Newman black holes} \label{sec:scaling_KN}

Asymptotically flat extremal black holes have corrections to the near-horizon geometry that decay as the horizon is approached. In this section, we start with Einstein-Maxwell theory and determine how fast stationary, axisymmetric
 deformations can decay as we approach the Kerr-Newman near-horizon geometry. These are the deformations that we will EFT-correct in the following sections. 
 
 In order to study these deformations, we first consider a more general ansatz for the metric and gauge field of the form\footnote{We use a superscript $(0)$ on quantities in this section to reflect the fact that we are not yet including any higher derivative corrections.}
\begin{equation}
\begin{aligned}
{\rm d}s^2&=2\,M^2\,[{f^{(0)}_1(x,\rho)}]^2\left[-\rho^2 \mathrm{d}\tau^2+\frac{\mathrm{d}\rho^2}{\rho^2}f^{(0)}_7(x,\rho)+\frac{\left(\mathrm{d}x+f^{(0)}_8(x,\rho) \mathrm{d}\rho\right)^2}{f^{(0)}_6(x,\rho)\left(1-x^2\right)}\right.
\\
&\qquad\qquad\qquad\qquad\qquad \left.+{f^{(0)}_2(x,\rho)}^2\left(1-x^2\right)\left(\mathrm{d}\varphi+\rho\,f^{(0)}_4(x,\rho)\,\mathrm{d}\tau\right)^2 \vphantom{\frac{\left(\mathrm{d}x+f^{(0)}_8(x,\rho) \mathrm{d}\rho\right)^2}{f^{(0)}_6(x,\rho)\left(1-x^2\right)}} \right]
\end{aligned}
\end{equation}
and
\begin{equation}
\begin{aligned}
A&=\frac{\sqrt{2}\,M}{\kappa}\left[f^{(0)}_5(x,\rho)\,\rho\,{\rm d}\tau+(1-x^2)f^{(0)}_3(x,\rho)f^{(0)}_2(x,\rho)\left(\mathrm{d}\varphi+\rho\,f^{(0)}_4(x,\rho)\,\mathrm{d}\tau\right)\right]\,.
\end{aligned}
\end{equation}
Next, we note that in these expressions, we have not fixed diffeomorphisms, i.e., the above metric and gauge field remain invariant under arbitrary redefinitions of $\rho$ and $x$ that are not dependent on $\tau$ or $\varphi$. To fix these coordinate redundancies, we set $f^{(0)}_7(\rho,x)=1$ and $f^{(0)}_8(x,\rho)=0$, so that our (partially) gauge-fixed metric takes the simpler form
\begin{equation}
\begin{aligned}
{\rm d}s^2& =2\,M^2\,[{f^{(0)}_1(x,\rho)}]^2\bigg[-\rho^2 \mathrm{d}\tau^2+\frac{\mathrm{d}\rho^2}{\rho^2}+\frac{\mathrm{d}x^2}{f^{(0)}_6(x,\rho)\left(1-x^2\right)}
\\
&\qquad\qquad\qquad\qquad\qquad +{f^{(0)}_2(x,\rho)}^2\left(1-x^2\right)\left(\mathrm{d}\varphi+\rho\,f^{(0)}_4(x,\rho)\,\mathrm{d}\tau\right)^2\bigg].
\label{eq:fixed}
\end{aligned}
\end{equation}
To proceed, we further consider
\begin{equation}
f^{(0)}_i(x,\rho)=F^{(0)}_i(x)\left[1+\delta \hat{f}^{(0)}_i(x,\rho)\right]\,\quad\text{for}\quad i = 1,2,3,
\end{equation}
along with 
\begin{equation}
\begin{aligned}
f^{(0)}_4(x,\rho)&=\omega^{(0)}_{\rm NH}\left[1+\delta \hat{f}^{(0)}_4(x,\rho)\right]
\\
f^{(0)}_5(x,\rho)&=Q^{(0)}_{\rm NH}\left[1+\delta \hat{f}^{(0)}_5(x,\rho)\right]
\\
f^{(0)}_6(x,\rho)&=1+\delta \hat{f}^{(0)}_6(x,\rho),
\end{aligned}
\end{equation}
with the $F^{(0)}_i(x)$ given in Eq.~(\ref{eq:NHEKvalues}). The $\delta \hat{f}^{(0)}_i(x,\rho)$ are assumed to be arbitrarily small, reflecting the expectation that we envisage the near-horizon geometry of extremal Kerr-Newman to be robust with respect to the deformations under consideration. So the $\delta \hat{f}^{(0)}_i$ satisfy linearized equations on the background of the near-horizon geometry. 

So far, we have not made any use of the ${\rm O}(2,1)$ symmetry of the background solution. Indeed, we can further expand $\delta \hat{f}^{(0)}_i(x,\rho)$ into harmonics of ${\rm O}(2,1)$. These harmonics will be labeled by a real number $\gamma^{(0)}$, and it turns out that modes with $\gamma^{(0)}\neq1$ behave very differently than those with $\gamma^{(0)}=1$. For simplicity, we will start with the former case.

\subsection{Modes with $\gamma^{(0)}\neq1$}
For time-independent perturbations, the harmonics of ${\rm O}(2,1)$ are simply $\rho^{\gamma^{(0)}}$, and for this reason we take
\begin{equation}
\delta\hat{f}^{(0)}_i(x,\rho)=\rho^{\gamma^{(0)}}\delta f^{(0)}_i(x)\,,
\label{eq:ads2de}
\end{equation}
where we assume in this subsection that $\gamma^{(0)}\neq1$.

The perturbed equations governing the $\delta f^{(0)}_i$, $i=1,\ldots,6$, are too daunting to explicitly write in the main text. However, they are by construction linear in the $\delta f^{(0)}_i(x)$ and depend on $\gamma^{(0)}$. Schematically, these equations take the form
\begin{equation}
{}^{(0)}\Delta_{ij}\delta f_{j}^{(0)}=0\,,
\label{eq:newzeroth}
\end{equation}
where ${}^{(0)}\Delta_{ij}$ is a differential operator that depends on $x$ and $\gamma^{(0)}$. Indeed, $\gamma^{(0)}$ turns out to appear as an eigenvalue for the resulting equations in a generalized St\"urm-Liouville problem. By manipulating the perturbed Einstein-Maxwell equations we find, so long as $\gamma^{(0)}\neq1$, that
\begin{equation}
\delta f^{(0)}_{6}(x)=2\left[2 \delta f^{(0)}_{1}(x)+\delta f^{(0)}_{2}(x)\right]\,.
\label{eq:algebraic}
\end{equation}

With some effort, we can reduce the full system of equations to two first-order equations in $\delta f_1^{(0)}$ and $\delta f_2^{(0)}$, along with three second-order equations for $\delta f_3^{(0)}$, $\delta f_4^{(0)}$, and $\delta f_5^{(0)}$. Perhaps surprisingly, the resulting equations do indeed admit a general solution in terms of simple functions. However, to see this remarkable solution, one has to redefine the functions in a fairly nonobvious way, which we found via inspection.
The result will be to exchange the five remaining $\delta f_i^{(0)}$---in terms of which the equations of motion have total order equal to eight---for four functions $v_i$, each of which will be required to satisfy a second-order equation. 

Explicitly, the transformations relating the $\delta f_i^{(0)}$ to the $v_i$ 
(and the first derivatives of the $v_i$) are given as follows:
\begin{equation}
\begin{aligned}\label{eq:magic}
\delta f^{(0)}_1(x)&= \frac{v_1(x)}{2}+\frac{1}{1+(1-Z^2) x^2}\Bigg[\frac{Z^2(1-x^2)}{2+Z^2}v_4(x)-(2-Z^2)x(1-x^2)\frac{v_1^\prime(x)}{4}
\\
&\hspace{5mm} +(2+Z^2)(1-Z^2)(1-x^2)^2\frac{v_2^\prime(x)}{x} -Z^4(2+Z^2)(1-Z^2)x(1-x^2)v_3^\prime(x)\Bigg]\\
\delta f^{(0)}_2(x)&= -\frac{v_1(x)}{2}-\frac{2}{1+(1-Z^2) x^2}\Bigg[\frac{Z^2(1-x^2)}{2+Z^2}v_4(x)-(2-Z^2)x(1-x^2)\frac{v_1^\prime(x)}{4}
\\
&\hspace{5mm}+(2+Z^2)(1-Z^2)(1-x^2)^2\frac{v_2^\prime(x)}{x}-Z^4(2+Z^2)(1-Z^2)x(1-x^2)v_3^\prime(x)\Bigg]\\
\delta f^{(0)}_3(x)&=\frac{v_1(x)}{2}\,{+}\,v_4(x)\,{-}\,Z^2(2\,{+}\,Z^2)(1\,{-}\,x^2)\frac{v_2^\prime(x)}{x} \,{-}\,Z^2(2\,{+}\,Z^2)(2\,{-}\,3Z^2)x v_3^\prime(x)\\
\delta f^{(0)}_4(x)&=\frac{1}{\gamma^{(0)}+1}\Bigg[\frac{2(1+x^2)+\lambda^{(0)}(1-x^2)}{2}\frac{v_1(x)}{2}+\lambda^{(0)}(4-Z^4)v_2(x)
\\
&\hspace{20mm} +\lambda^{(0)}Z^4(2+Z^2)v_3(x)+\frac{Z^2}{2+Z^2}(1+x^2)v_4(x)
\\&\hspace{20mm}-x(1-x^2)\frac{v_1^\prime(x)}{2} -\frac{Z^2}{2+Z^2}x(1-x^2)v_4^\prime(x)\Bigg]\\
\delta f^{(0)}_5(x)&=\frac{1}{\gamma^{(0)}+1}\Bigg[\frac{2(1+x^2)+\lambda^{(0)}(1-x^2)}{2}\frac{v_1(x)}{2}-\lambda^{(0)}(2+Z^2)^2(1-Z^2)v_3(x)
\\
&\hspace{20mm} -\frac{2-3Z^2}{2+Z^2}(1+x^2)\frac{v_4(x)}{Z^2}-x(1-x^2)\frac{v_1^\prime(x)}{2}\\&\hspace{20mm}+\frac{2-3Z^2}{2+Z^2}x(1-x^2)\frac{v_4^\prime(x)}{Z^2}\Bigg]\\
\delta f^{(0)}_6(x)&=v_1(x)\,.
\end{aligned}
\end{equation}
Here, we defined $\lambda^{(0)}\equiv \gamma^{(0)}(\gamma^{(0)}+1)$. 
One finds four \emph{decoupled} second-order equations for $v_1$, $v_2$, $v_3$, and $v_4$, which take the following rather simple form:
\begin{equation}
\begin{aligned}
\label{eq:simplev}
\left[(1-x^2)^2v_1^\prime\right]^\prime+(\lambda^{(0)}-2)(1-x^2)v_1&=0\\
\left[\frac{(1-x^2)^2}{x^2}v_2^\prime\right]^\prime+\lambda^{(0)}\frac{1-x^2}{x^2}v_2&=0\\
\left[(1-x^2)v_3^\prime\right]^\prime+\lambda^{(0)}v_3&=0\\
\left[(1-x^2)^2v_4^\prime\right]^\prime+(\lambda^{(0)}-2)(1-x^2)v_4&=0\,.
\end{aligned}
\end{equation}
It is a tedious but straightforward exercise to show that deformations generated by $v_1$, $v_2$, $v_3$, and $v_4$ are actually independent from each other.
In fact, we have engineered the $v_i$ such that they decouple, giving us the four equations of motion in Eq.~\eqref{eq:simplev}.
Of course, each $v_i$ itself corresponds to a particular superposition of metric and electromagnetic perturbations; this is generically inevitable in a nonzero electromagnetic field, where the photon and graviton kinetic terms mix.
Furthermore, it is also possible to show that any smooth solution for $v_1$, $v_2$, $v_3$, and $v_4$ provides a good solution for the $\delta f^{(0)}_{i}$ and vice versa. Note that we could have expected  to have four independent modes, since we expect two independent degrees of freedom for the gravitational field and two degrees of freedom for the electromagnetic field.

One can solve for the $v$ functions directly, and we  label the modes as $\{\gamma^{(0)}_{1}{,}\gamma^{(0)}_{2}{,}\gamma^{(0)}_{3}{,}\gamma^{(0)}_{4}\}$. They are given compactly by
\begin{equation}
\begin{aligned}
\label{eqs:vs}
v_1(x)&=P_{\ell_1}^\prime(x),&&\quad \gamma^{(0)}_{1}=\ell_{1}\\
v_2(x)&=x(\ell_{2}+1)\ell_2 P_{\ell_{2}}(x)+(1+x^2\ell_{2})P_{\ell_{2}}^\prime(x),&&\quad \gamma^{(0)}_{2}=\ell_{2}+1\\
v_3(x)&=P_{\ell_{3}}(x),&&\quad \gamma^{(0)}_{3}=\ell_{3} \\
v_4(x)&=P_{\ell_{4}}^\prime(x),&&\quad \gamma^{(0)}_{4}=\ell_{4}\,,
\end{aligned}
\end{equation}
where $P_{\ell}(x)$ is the Legendre polynomial of order $\ell$. As we are considering a background supported by an electric field, the sense of polar vs.~axial perturbations labeling the parity transformation properties of the modes match for the gravity and gauge degrees of freedom~\cite{Regge:1957td,Zerilli:1970se,Zerilli:1974ai} (unlike the magnetic case, cf.~Ref.~\cite{Porfyriadis:2021zfb}); specifically, $v_{1,2,4}$ are axial (picking up a sign $(-1)^{\ell+1}$ under parity inversion), while $v_3$ is polar (picking up a sign $(-1)^\ell$ under parity inversion). 
We will now comment on the ranges of the several $\ell$. Recall that modes with $\gamma^{(0)}=1$ are excluded from this analysis, implying that $\ell_1=\ell_3=\ell_4=1$ and $\ell_2=0$ are excluded and will be addressed in the next section. Furthermore, modes with $\ell_1=\ell_3=\ell_4=0$ vanish identically. We are thus left with the range $\ell_{1,3,4}\geq2$ and $\ell_2\geq1$.

The fact that the exponents $\gamma_i^{(0)}$ are all integers means that stationary, axisymmetric perturbations of extreme Kerr-Newman remain smooth in Einstein-Maxwell theory. In principle, the scaling exponents could have depended on the dimensionless ratio $a/Q$, but we see that they do not.

Next we turn our attention to modes with $\gamma^{(0)}=1$, which turn out to be the most relevant modes to us, as explained in the introduction.

\subsection{Modes with $\gamma^{(0)}=1$}
For modes with $\gamma^{(0)}=1$, one could be tempted to take Eq.~(\ref{eq:ads2de}) and simply set $\gamma^{(0)}$ to $1$. The answer would be partially correct, but incomplete. In particular, for $\gamma^{(0)}=1$  there are other deformations that do not fit the power law decomposition~\eqref{eq:ads2de}, but will play an important role later on. These turn out to be proportional to $\rho \log \rho$. For this sector of deformations, we take a decomposition of the $\delta\hat f_i(x,\rho)$ functions with $\rho$ dependence organized into the form,
\begin{equation}
\begin{aligned}
\label{eq:expand}
\delta \hat{f}^{(0)}_1(x,\rho)&=\delta f^{(0)}_1(x)\rho+V^{(0)}\rho \log \rho\\
\delta \hat{f}^{(0)}_2(x,\rho)&=\left[\delta f^{(0)}_2(x)-\frac{V^{(0)}}{2}\right]\rho-V^{(0)}\rho \log \rho\\
\delta \hat{f}^{(0)}_3(x,\rho)&=\left[\delta f^{(0)}_3(x)+\frac{V^{(0)}}{Z^2}\right]\rho+V^{(0)}\rho \log \rho\\
\delta \hat{f}^{(0)}_4(x,\rho)&=\delta f^{(0)}_4(x)\rho+V^{(0)}\rho \log \rho\\
\delta \hat{f}^{(0)}_5(x,\rho)&=\left[\delta f^{(0)}_5(x)-\frac{(2-Z^2)(1-Z^2)(1+x^2)}{4Z^4}V^{(0)}\right]\rho+V^{(0)}\rho \log \rho\\
\delta \hat{f}^{(0)}_6(x,\rho)&=\left[\delta f^{(0)}_6(x)+V^{(0)}\right]\rho+2V^{(0)}\rho \log \rho\,,
\end{aligned}
\end{equation}
with $V^{(0)}$ constant. We will see that despite the presence of the low-differentiability term $\rho \log \rho$, the above mode generates \emph{no divergent} tidal forces. In fact, this mode is already present for extremal Kerr-Newman black holes, as we will see later, and arises since our coordinates are not smooth on the horizon. 

However, the subtleties with the $\gamma^{(0)}=1$ mode do not stop here. In particular, there is a \emph{residual gauge symmetry}, which we now comment upon. Under a coordinate transformation of the form
\begin{equation}
\begin{aligned}
x&\to x-(1-x^2)\rho\,\delta C^\prime(x)
\\
\rho &\to \rho+\rho^2 \delta C(x),
\end{aligned}
\end{equation}
where $\delta C(x)$ is an arbitrary function of $x$, the metric and gauge field deformations transform as 
\begin{equation}
\label{eq:gaugesym}
\begin{aligned}
\Delta \delta f^{(0)}_1(x)&=\delta C(x)-(1-x^2)\frac{{F_1^{(0)}}^\prime(x)}{F^{(0)}_1(x)}\delta C^\prime(x)\\
\Delta \delta f^{(0)}_2(x)&=-\delta C(x)+x \delta C^\prime(x)-(1-x^2) \frac{{F_2^{(0)}}^\prime(x)}{F^{(0)}_2(x)} \delta C^\prime(x)\\
\Delta \delta f^{(0)}_3(x)&=\delta C(x)+x \delta C'(x)-(1-x^2)\frac{{F_3^{(0)}}^\prime(x)}{F^{(0)}_3(x)} \delta C^\prime(x)\\
\Delta \delta f^{(0)}_4(x)&=\delta C(x)\\
\Delta \delta f^{(0)}_5(x)&=\delta C(x)\\
\Delta \delta f^{(0)}_6(x)&=2 \left[\delta C(x)-x \delta C^\prime(x)+(1-x^2)\delta C^{\prime\prime}(x)\right].
\end{aligned}
\end{equation}
We will use this freedom to set
\begin{equation}
\delta f^{(0)}_{6}(x)=2\left[2 \delta f^{(0)}_{1}(x)+\delta f^{(0)}_{2}(x)\right]\,,
\label{eq:algebraicell1}
\end{equation}
which specifies $\delta C$ up to a constant, which we will fix later on. This is a particularly nice choice of gauge because it allows us to discuss these modes on equal footing with those with $\gamma^{(0)}\neq1$, modulo the dependence on $V^{(0)}$. Naturally, this gauge choice is also consistent with regularity at the poles $x=\pm1$.

Just as in the case where $\gamma^{(0)}\neq1$, we can take the $\delta f^{(0)}_i$ as in Eq.~(\ref{eq:magic}) with $\lambda^{(0)}=2$ and $\gamma^{(0)}=1$. The resulting equations for the $v_i$ are exactly those given in Eq.~(\ref{eq:simplev}) with $\lambda^{(0)}=2$, and the corresponding solutions are provided in Eq.~(\ref{eqs:vs}). Modes with $\ell_2=0$ vanish identically, leaving us with $\ell_1=\ell_4=1$ as the only even parity  modes. Furthermore, as we show below, the $\ell_1=1$ mode is pure gauge and can be eliminated by using the constant part of the gauge transformation $\delta C$ appearing in Eq.~(\ref{eq:gaugesym}). We are thus left with a single $\gamma^{(0)}=1$ mode with two unknown coefficients: $V^{(0)}$ (the constant parameterizing the $\rho \log \rho$ term in Eq.~(\ref{eq:expand})) and the amplitude of the $\ell_4=1$ mode parameterized by $v_4(x)=A_4$, with $A_4$ being a constant. 

Both of these coefficients are nonvanishing for extremal Kerr-Newman black holes if we use the $(\rho,x)$ coordinates used in Eq.~(\ref{eq:fixed}). To see this, consider the Kerr-Newman line element and gauge field in Eq.~(\ref{eqs:KNgA}), under a coordinate transformation, 
\begin{equation}
\begin{aligned}
t&=4M(2-Z^2)\tau\\
\phi &= \varphi+\frac{\sqrt{1-Z^2}}{2-Z^2}\frac{t}{M}\\
r&=r_+\left[1+\frac{\rho}{4}+\rho^2 G_0+G_1 \rho^2 \log \rho+\mathcal{O}(\rho^3 \log^2\rho)\right] \label{eq:rp}\\
\cos \theta &= x+\mathcal{O}(\rho^2),
\end{aligned}
\end{equation}
with $G_0$ and $G_1$ being constants. Expanding to leading order in $\rho$ gives the near-horizon geometry~(\ref{eq:NHKN}), while expanding to subleading order in $\rho$ and setting
\begin{equation}
G_0= -\frac{1}{2} G_1 = \frac{1}{16(2-Z^2)}
\end{equation}
reveals that for an extremal Kerr-Newman black hole we have
\begin{equation}
\begin{aligned}
V^{(0)}&=-\frac{1}{2(2-Z^2)}\quad \text{and}\quad A_4=\frac{(1-Z^2)(2+Z^2)}{4Z^2(2-Z^2)}
\label{eq:money}
\end{aligned}
\end{equation}
in our chosen gauge. Note that $G_0$ is related to the constant part of $\delta C$ appearing in Eq.~(\ref{eq:gaugesym}). If we had not introduced $G_0$ in Eq.~(\ref{eq:rp}), we would have obtained an $\ell_1=1$ mode as well and a different amplitude for $A_4$. The above procedure explicitly shows that the mode with $\ell_1=1$ is pure gauge. The mode described in this section, i.e., the physical mode with $A_4\neq0$ and $V^{(0)}\neq0$, is the mode that we would like to EFT-correct. However, in order to do this, we first need to apply the EFT to the Kerr-Newman near-horizon geometry, which we do next.

\section[EFT-corrected near-horizon geometries of extremal Kerr-Newman \\black holes]{EFT-corrected near-horizon geometries of extremal Kerr-Newman\\ black holes} \label{sec:EFT_NHG}

We begin our investigation of EFT corrections to Kerr-Newman black holes by considering the near-horizon geometry.
All EFT-corrected near-horizon geometries take the same general form. In particular, they exhibit ${\rm O}(2, 1) \times {\rm U}(1) $ symmetry, and we can choose an angular coordinate $x\in[-1,1]$ so that
\begin{equation}
\label{eq:NH}
\begin{aligned}
{\rm d}s^2_{\rm NH}&=2\,M^2\,[{F_1(x)}]^2\bigg[-\rho^2 \mathrm{d}\tau^2+\frac{\mathrm{d}\rho^2}{\rho^2}+\frac{\Gamma_{\rm NH}^2\mathrm{d}x^2}{1-x^2}\\&\qquad\qquad\qquad\qquad +[{F_2(x)}]^2\left(1-x^2\right)\left(\mathrm{d}\varphi+\rho\,\omega_{\rm NH}\,\mathrm{d}\tau\right)^2\bigg]\\
A_{\rm NH}&=\frac{\sqrt{2}\,M}{\kappa}\left[Q_{\rm NH}\,\rho\,{\rm d}\tau+(1-x^2)F_2(x)F_3(x)\left(\mathrm{d}\varphi+\rho\,\omega_{\rm NH}\,\mathrm{d}\tau\right)\right]\,,
\end{aligned}
\end{equation}
where
\begin{equation}
\begin{aligned}
F_i(x)&=F^{(0)}_i(x)\left[1+\sum_{K=1}^{8}\frac{d_K}{M^2}\,\delta F^{(K)}_i(x)\right], \quad i=1,2,3\\
\Gamma_{\rm NH}&=1+\sum_{K=1}^{8}\frac{d_K}{M^2}\,\delta\Gamma^{(K)}_{\rm NH}\\
\omega_{\rm NH}&=\omega^{(0)}_{\rm NH}\left(1+\sum_{K=1}^{8}\frac{d_K}{M^2}\,\delta\omega^{(K)}_{\rm NH}\right)\\
Q_{\rm NH}&=Q^{(0)}_{\rm NH}\left(1+\sum_{K=1}^{8}\frac{d_K}{M^2}\,\delta Q^{(K)}_{\rm NH}\right)\,,
\end{aligned}
\end{equation}
and $d_K$ are the rescaled Wilsonian coefficients \eqref{eq:ddef}. The extremal horizon is located at the null hypersurface $\rho=0$, and it has spatial cross sections with $S^2$ topology. The coordinate $x$ parameterizes the deformed $S^2$, with $x=\pm1$ being the poles. Finally, the constant $\Gamma_{\rm NH}$ parameterizes the proper length from the north to the south pole of the squashed two-sphere of the EFT-corrected horizon along a constant-$\varphi$ slice. For a Kerr-Newman black hole, this is simply $2\,M\,{\rm E}\left(Z^2-1\right)$, where ${\rm E}$ is a complete elliptic integral. 

We then solve the equations of motion (\ref{eqs:EOM}) \emph{perturbatively} in the corresponding $d_K$. The explicit expressions for $\delta F^{(K)}_i(x)$ are not very illuminating, and we will refrain from presenting them in the main text. However, the resulting equations of motion can be explicitly integrated in full generality, and there are a few general results that we now outline.

The equations of motion yield three linear coupled ordinary differential equations for $\{\delta F^{(K)}_1,\delta F^{(K)}_2,\delta F^{(K)}_3\}$, which depend on up to first derivatives with respect to $\delta F^{(K)}_1$ and up to second derivatives with respect to $\{\delta F^{(K)}_2,\delta F^{(K)}_3\}$. Naturally, these equations also depend on $\delta \Gamma_{\rm NH}^{(K)}$ and $\delta \omega^{(K)}_{\rm NH}$ as well as on $\delta Q^{(K)}_{\rm NH}$. After integrating, for each $K$ the full solutions depend on \emph{eight arbitrary constants}: five integration constants, along with $\delta \omega^{(K)}_{\rm NH}$, $\delta \Gamma_{\rm NH}^{(K)}$, and $\delta Q_{\rm NH}^{(K)}$. At this stage, we impose regularity at the poles (i.e., at $x=\pm1$) and require that $\varphi$ have period $2\pi$, that is to say,
\begin{equation}
\left|F_2(\pm1)\right|=\left|\Gamma_{\rm NH}\right|\implies \delta F_2^{(K)}(\pm1)=\delta \Gamma^{(K)}_{\rm NH}
\end{equation}
to linear order in $d_{K}$. These conditions fix all of the integration constants as well as $\delta \Gamma_{\rm NH}^{(K)}$. However, both $\delta Q_{\rm NH}^{(K)}$ and $\delta \omega_{\rm NH}^{(K)}$ are left arbitrary. In principle, we could determine $\delta Q_{\rm NH}^{(K)}$ and $\delta \omega_{\rm NH}^{(K)}$ by gluing our near-horizon geometries to an asymptotically flat end and constructing the full extremal black hole solution. We will see in the next section that the shifts in scaling exponents $\delta \gamma$ are independent of these two constants.

We find that the $\delta \Gamma_{\rm NH}^{(K)}$ take the following simple form:
\begin{equation} 
\begin{aligned}
\label{eqs:deltas}
\delta \Gamma^{(1)}_{\rm NH}&=0\\
\delta \Gamma^{(2)}_{\rm NH}&=\frac{3 Z^2}{(2-Z^2)^{1/2}} \left(1-Z^2\right)\\
\delta \Gamma^{(3)}_{\rm NH}&=\frac{12 Z^2}{(2-Z^2)^{1/2}} \left(1-Z^2\right)\\
\delta \Gamma^{(4)}_{\rm NH}&=0\\
\delta \Gamma^{(5)}_{\rm NH}&=\frac{Z^2}{\left(2-Z^2\right)^{3/2}} \left(3 Z^4-8 Z^2+3\right)\\
\delta \Gamma^{(6)}_{\rm NH}&=\frac{Z^2}{(2-Z^2)^{7/2}} \left(6 Z^8-40 Z^6+94 Z^4-85 Z^2+21\right)\\
\delta \Gamma^{(7)}_{\rm NH}&=\frac{2 Z^4}{(2-Z^2)^{11/2}} \left(3 Z^{10}-31 Z^8+128 Z^6-268 Z^4+267 Z^2-107\right)\\
\delta \Gamma^{(8)}_{\rm NH}&=\frac{Z^4}{2 (2-Z^2)^{11/2}} \left(9 Z^{10}-93 Z^8+384 Z^6-796 Z^4+811 Z^2-331\right).
\end{aligned}
\end{equation}
We note the useful relation $2(F_{ab}F^{ab})^2 + (F_{ab} \widetilde F^{ab})^2 = 4F_{ab}F^{bc}F_{cd}F^{da}$~\cite{Cheung:2014ega}, so for geometries where $F_{ab}\widetilde F^{ab}$ vanishes---e.g., the static and purely electrically charged (or purely magnetically charged) black hole---one expects a simple factor-of-two relation between the higher-derivative corrections generated by $d_7$ and $d_8$~\cite{Kats:2006xp}.
However, in the stationary case with nonzero spin (with either electric and/or magnetic charge), this simplification does not hold. Indeed, no such simple relation exists between $\delta\Gamma^{(7)}_{\rm NH}$ and $\delta\Gamma^{(8)}_{\rm NH}$.

\section[Scaling dimensions for EFT-corrected extremal Kerr-Newman \\near-horizon geometries]{Scaling dimensions for EFT-corrected extremal Kerr-Newman near-horizon geometries} \label{sec:scaling_EFT}
After identifying the EFT-corrected near-horizon geometries, we aim to understand the rate at which deformations, preserving $\partial/\partial \tau$ and $\partial/\partial \varphi$, are permitted to decay as we approach these near-horizon configurations. 
Once again, modes with $\gamma^{(0)}\neq1$ and modes with $\gamma^{(0)}=1$ will receive EFT corrections computed through different methods. Our primary focus will be in EFT-correcting modes with $\gamma^{(0)}=1$, as these are present in the full asymptotically flat solution and may induce significant tidal deformations.

We first consider an ansatz similar to Eq.~(\ref{eq:fixed}), but with the $f_i^{(0)}(x,\rho)$ replaced by $f_i(x,\rho)$,
\begin{equation}
\begin{aligned}
{\rm d}s^2&=2\,M^2\,{f_1(x,\rho)}^2\bigg[-\rho^2 \mathrm{d}\tau^2+\frac{\mathrm{d}\rho^2}{\rho^2}+\frac{\mathrm{d}x^2}{f_6(x,\rho)\left(1-x^2\right)}
\\
&\qquad\qquad\qquad\qquad +{f_2(x,\rho)}^2\left(1-x^2\right)\left(\mathrm{d}\varphi+\rho\,f_4(x,\rho)\,\mathrm{d}\tau\right)^2\bigg],
\label{eq:fixed2}
\end{aligned}
\end{equation}
and take instead the following form for the deformations to the $f_i(x,\rho)$ around the background near-horizon solution,
\begin{equation}
\begin{aligned}
f_i(x,\rho)&=F_i(x)\left[1+\delta \hat{f}_i(x,\rho)\right]\,,\quad i = 1,2,3 \\ f_4(x,\rho)&=\omega_{\rm NH}\left[1+\delta \hat{f}_4(x,\rho)\right]\\
f_5(x,\rho)&=Q_{\rm NH}\left[1+\delta \hat{f}_5(x,\rho)\right]\\
f_6(x,\rho)&=\frac{1}{\Gamma_{\rm NH}^2}\left[1+\delta \hat{f}_6(x,\rho)\right]\,,
\end{aligned}
\end{equation}
with the $F_i(x)$ matching those appearing in Eq.~(\ref{eq:NH}). Once again, we consider the $\delta\hat{f}_i(x,\rho)$ to be perturbative, anticipating that the EFT corrections have not destabilized the near-horizon geometry of the extremal Kerr-Newman black hole. 

Since the EFT-corrected near-horizon geometries still enjoy ${\rm O}(2,1)$ symmetry, we can decompose the $\delta\hat{f}_i(x,\rho)$ into harmonics of ${\rm O}(2,1)$. As before, these harmonics will each be labeled by a real number $\gamma$, the scaling exponent, and again we must consider $\gamma\neq1$ and $\gamma=1$ differently.

\subsection{Modes with $\gamma\neq1$}
Using ${\rm O}(2,1)$ symmetry and the fact that the $\delta\hat{f}_i(x,\rho)$ are perturbatively small, we set
\begin{equation}
\delta\hat{f}_i(x,\rho)=\rho^{\gamma}\delta f_i(x)\,.
\end{equation}
We then expand the equations of motion~(\ref{eqs:EOM}) to linear order in $\delta\hat{f}_i(x,\rho)$ and use the decomposition above, finding a rather complicated system of equations for the $\delta f_i(x)$ that strongly depend on $\gamma$. Indeed, these equations depend on higher derivatives of the $\delta f_i(x)$, but nevertheless can be seen as a generalized eigenvalue problem for eigenvalues $\gamma$ and eigenfunctions $\delta f_i(x)$. So far we have not done any expansion in the Wilson coefficients $d_{K}$. To progress, one sets
\begin{equation}
\delta f_{i}(x)= \delta f^{(0)}_{i}(x)+\sum_{K=1}^{8}\frac{d_K}{M^2}\delta f^{(K)}_{i}(x)\,,
\end{equation}
together with
\begin{equation}
\gamma=\gamma^{(0)}+\sum_{K=1}^{8}\frac{d_K}{M^2}\delta \gamma^{(K)}\,,
\end{equation}
as well as Eq.~(\ref{eq:NH}). Since we are taking the $d_K$ to be perturbative, we are left with equations for the $\delta f^{(K)}_{i}(x)$ that do not explicitly depend on $d_K$. Indeed, these equations take the following schematic form,
\begin{equation}
{}^{(0)}\Delta_{ij} \delta f_{j}^{(K)}=T^{(K)}_i\,,\label{eq:withsource}
\end{equation}
with ${}^{(0)}\Delta_{ij}$ being the same operator as in Eq.~(\ref{eq:newzeroth}), $T^{(K)}_i$ are some complicated source terms that depend on the $\delta f_{i}^{(0)}$ and their first derivatives,\footnote{Higher-order derivatives can be eliminated by using the equations for $\delta f_{i}^{(0)}$.} along with $\delta \omega_{\rm NH}^{(K)}$, $\delta Q_{\rm NH}^{(K)}$, and $\delta \gamma^{(K)}$. 

One can manipulate these differential equations to find that
\begin{equation}
\label{eq:gaugelot}
\delta f^{(K)}_6(x)=2\left[2 \delta f^{(K)}_{1}(x)+\delta f^{(K)}_{2}(x)\right]+\frac{W_K(\delta f_i^{(0)},{\delta f_i^{(0)}}^\prime,x)\,d_K}{M^2}\,,
\end{equation}
where {$W_K(\delta f_i^{(0)},{\delta f_i^{(0)}}^\prime,x)$ is a complicated function of the $\delta f_i^{(0)}$, their first derivatives, and $x$, and are nonzero for $K=1,\ldots,6$ only}.  This relation is the EFT equivalent of Eq.~(\ref{eq:algebraic}). One then writes $\delta f_{i}^{(K)}$ in terms of new functions $v_i^{(K)}$, just as in Eq.~(\ref{eq:magic}), to determine the corresponding equations for the $v_i^{(K)}$. These take the same form as in Eq.~(\ref{eq:simplev}) with $v_i$ replaced by $v_i^{(K)}$ \emph{and} having on the right-hand side a complicated source term as in Eq.~\eqref{eq:withsource}. These equations can also be solved for $v_i^{(K)}$ once a particular value of $\gamma^{(0)}$ is given.

For a given value of $\gamma^{(0)}$, one can have more than one mode. For instance, if we fix $\gamma^{(0)}=2$, we can have the following modes: $\ell_1=2$, $\ell_2=1$, $\ell_3=2$, and $\ell_4=2$. Modes with different parity properties decouple from each other. However, within a given parity sector, one must start with a general linear combination of modes that share the same value of $\lambda^{(0)}$, in line with standard degenerate perturbation theory. The relative contribution of each mode is then found via solving the corresponding equations of motion for the $\delta f^{(K)}_{i}(x)$ after imposing regularity at the poles $x=\pm1$.

For instance, let us consider parity-even deformations with $\gamma^{(0)}=2$,  which remain invariant under sending $x\rightarrow -x$.  Recalling that modes $v_{1,2,4}$ are axial and $v_3$ is polar, there are two such modes with this property: $\ell_2=1$ and $\ell_3=2$. This means that for $\delta f^{(0)}_{i}(x)$ we should take
\begin{equation}
\begin{aligned}
\delta f^{(0)}_{1}(x)&=\frac{(1-x^2)^2}{1+(1-Z^2) x^2} A_2+\frac{2Z x^2 (1-x^2)}{1+(1-Z^2) x^2} A_3\\
\delta f^{(0)}_{2}(x)&=-\frac{2(1-x^2)^2}{1+(1-Z^2) x^2} A_2-\frac{4Zx^2 (1-x^2)}{1+(1-Z^2) x^2} A_3\\
\delta f^{(0)}_3(x)&=-\frac{Z^2 (1-x^2)}{1-Z^2} A_2+\frac{2(2-3 Z^2)x^2}{Z (1-Z^2)} A_3\\
\delta f^{(0)}_4(x)&=\frac{1}{3}\frac{(2-Z^2)(1+3 x^2)}{1-Z^2} A_2+\frac{2Z}{3}\frac{1-3 x^2}{1-Z^2}A_3\\
\delta f^{(0)}_5(x)&=-\frac{2(2+Z^2) (1-3 x^2)}{3Z^3}A_3\\
\delta f^{(0)}_6(x)&=0\,,
\end{aligned}
\end{equation}
with $A_2$ and $A_3$ being constant. Solving for the corresponding $\delta f^{(K)}_{i}(x)$ and imposing regularity at the poles determines both $\delta \gamma^{(K)}$ and $r^{(K)}\equiv A_2/A_3$. Since we have a twofold degeneracy within the same symmetry class, we are expecting to find two possible values for $\{\delta \gamma^{(K)},r^{(K)}\}$, which we label as $\{\delta \gamma_{\pm}^{(K)},r_{\pm}^{(K)}\}$. These corrections can be found by solving an eigenvalue problem that takes the following form,
\begin{equation}
\sum_{K=1}^{8}d_K\,\left(L^{(K)}-\delta \gamma^{(K)}I\right)\cdot \left[
\begin{array}{c}
A_2
\\
A_3
\end{array}\right]=0\,,
\label{eq:system}
\end{equation}
where $I$ is a $2\times2$ identity matrix and $L^{(K)}$ are \emph{symmetric} $2\times2$ matrices whose explicit expressions can be found in App.~\ref{app:crazynuts}. 

At this stage we note one of the most stringent tests of our calculation. Namely, it is a simple exercise to check, using the explicit expressions given in App.~\ref{app:crazynuts}, that Eq.~(\ref{eq:system}) is invariant under field redefinitions and that
\begin{equation}
\begin{aligned}
\delta \gamma^{(1)}&=\delta \gamma^{(4)}=0\\
\delta \gamma^{(2)}&=\delta \gamma^{(5)}= 4\,\delta\gamma^{(8)}-\delta\gamma^{(7)} =\frac{1}{4} \delta\gamma^{(3)}\,.
\end{aligned}
\end{equation}
In particular, using the above relations, we can write Eq.~(\ref{eq:system}) in terms of the field redefinition invariant basis $\{d_0,d_6,d_9\}$,
\begin{equation}
\left[d_0\left(\tilde{L}^{(0)}-\delta \tilde{\gamma}^{(0)}I\right)+d_6\left(\tilde{L}^{(6)}-\delta \tilde{\gamma}^{(6)}I\right)+d_9\left(\tilde{L}^{(9)}-\delta \tilde{\gamma}^{(9)}I\right)\right]\cdot \left[
\begin{array}{c}
A_2
\\
A_3
\end{array}\right]=0\,,\label{eq:systeminv}
\end{equation}
where $\tilde{L}^{(0)}$, $\tilde{L}^{(6)}$, and $\tilde{L}^{(9)}$ are given in App.~\ref{app:crazynuts} and
\begin{equation}
\begin{aligned}
\delta \tilde{\gamma}^{(0)}&=\frac{1}{4}\delta\gamma^{(7)}\\ \delta \tilde{\gamma}^{(6)}&=\delta\gamma^{(6)}+\frac{3}{4}\delta\gamma^{(7)}-2\,\delta\gamma^{(8)}\\ \delta \tilde{\gamma}^{(9)}&=\delta\gamma^{(8)}-\frac{1}{2}\delta\gamma^{(7)}\,.
\end{aligned}
\end{equation}
One might wonder why $d_3$ plays no role, but we remind the reader that the Gauss-Bonnet term is topological in four spacetime dimensions, a fact that we can use to show that the dependence on $d_3$ must drop out from any calculation that stems from the classical equations of motion. From the calculation above, it is clear that we can use $\delta \gamma_{\pm}^{(K)}$ for $K=6,7,8$ as a basis for the scaling exponents.

\begin{figure}[!h]
    \centering    \includegraphics[width=\textwidth]{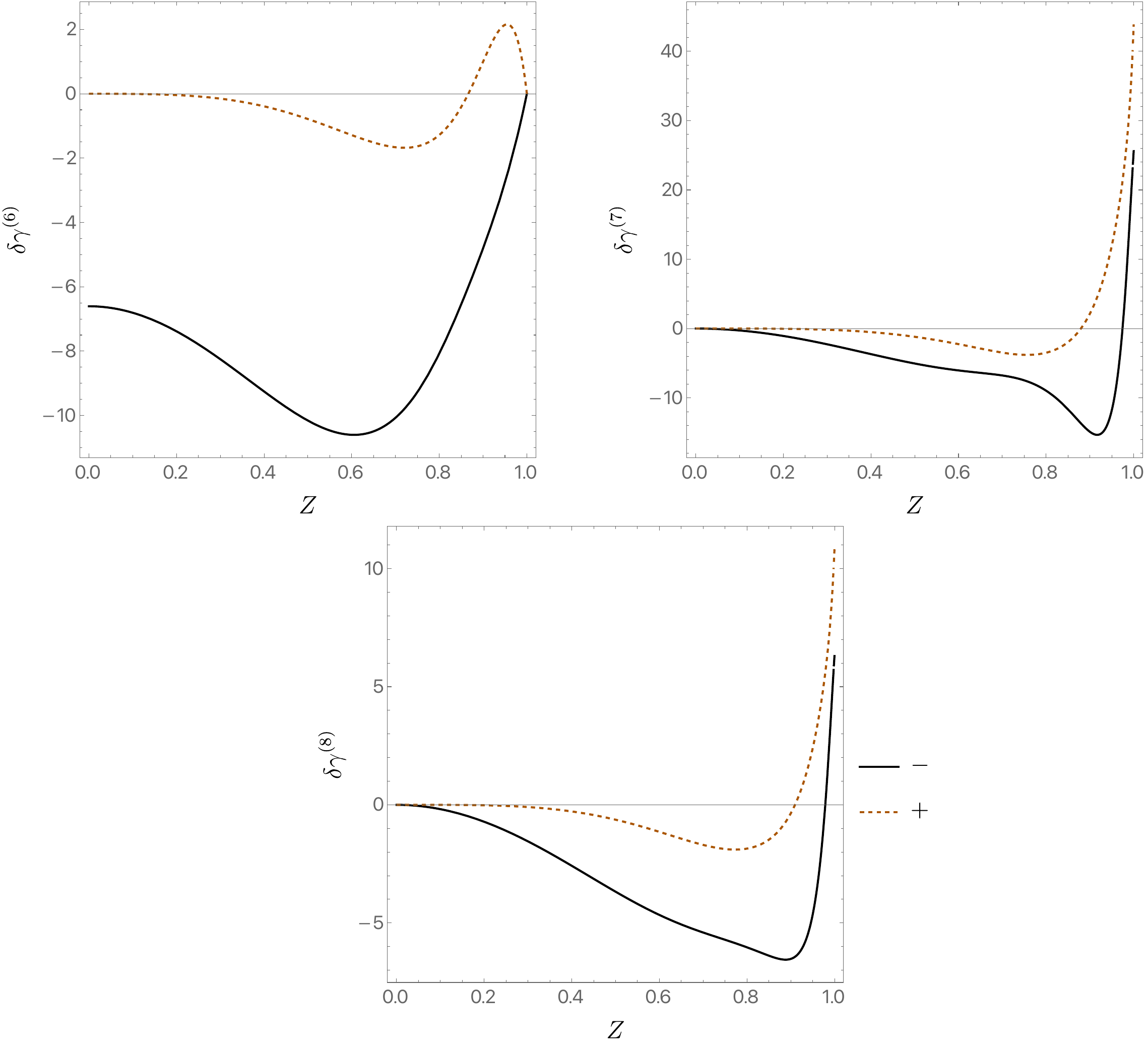}
    \caption{The change in the scaling dimensions $\delta\gamma^{(K)}_{\pm}$ for the two even-parity modes with $\gamma^{(0)} = 2$, as a function of $Z = Q/M$. The label $K=6,7,8$ corresponds to the three higher-derivative corrections shown in Eq.~\eqref{eq:xhd}.}
    \label{fig:all}
\end{figure}

In Fig.~\ref{fig:all}, we plot $\delta \gamma^{(K)}$ for $K=6,7,8$ as a function of $Z$. Note that since the $\tilde{L}^{(K)}$ are $2\times2$, we expect $\delta \gamma^{(K)}$ to take two values for fixed $K$. We label these by $\delta \gamma_{\pm}^{(K)}$ and chose $\delta \gamma_{-}^{(K)}\leq\delta \gamma_{+}^{(K)}$. There are several features to note about these corrections to the scaling exponents: (1) The $\delta \gamma_{\pm}^{(K)}$ remain real as we vary $Z$, as a consequence of the fact that the $L^{(K)}$ are real symmetric matrices. (2)
The $\delta \gamma_{\pm}^{(K)}$ are not monotonic in $Z$ and do not have a definite sign.
(3) While $\delta \gamma_{\pm}^{(6)}$ vanishes at $Z=1$, the same is not true for $\delta \gamma_{\pm}^{(7)}$ and $\delta \gamma_{\pm}^{(8)}$. 
This last point might seem to contradict the results in Ref.~\cite{Horowitz:2023xyl}, but it is important to note that Ref.~\cite{Horowitz:2023xyl} only analyzed the scalar sector of perturbations, whereas when $Z \to 1$ the modes we are investigating here approach two vector-type deformations in the decomposition used in Ref.~\cite{Horowitz:2023xyl}.
We will see that the effect of this shift in the scaling exponents on the tidal forces is suppressed in the Reissner-Nordstr\"om limit by a factor proportional to $1-Z^2$.

\subsection{Modes with $\gamma=1$\label{eq:gamma1}}
Modes with $\gamma=1$ are more subtle, but are also the most interesting ones since they have the potential to cause large tidal forces. We know that these modes, even for extremal Kerr-Newman black holes, are coupled to $\rho \log \rho$ modes. For this reason, we can envisage that such a $\rho \log \rho$ term will appear in the EFT-corrected modes. To EFT-correct modes with $\gamma^{(0)}=1$, we take 
\begin{equation}
\begin{aligned}
\label{eq:crazylogs}
\delta \hat{f}_1(x,\rho)=&\,\rho^{\gamma}\left[\delta f_1^{(0)}(x)+\sum_{K=1}^{8}\frac{d_K}{M^2}\delta f_1^{(K)}(x)\right]\\&+\rho\,\log \rho \left(V^{(0)}+\sum_{K=1}^{8}\frac{d_K}{M^2}V^{(K)}+ \sum_{K=1}^{8}b_1^{(K)}(x)\right)\\
\delta \hat{f}_2(x,\rho)=&\,\rho^{\gamma}\left[\delta f_2^{(0)}(x)-\frac{V^{(0)}}{2}+\sum_{K=1}^{8}\frac{d_K}{M^2}\delta f_2^{(K)}(x)\right]\\&-\rho\,\log \rho \left(V^{(0)}+\sum_{K=1}^{8}\frac{d_K}{M^2}V^{(K)}-\sum_{K=1}^{8}b_2^{(K)}(x)\right)\\
\delta \hat{f}_3(x,\rho)=&\,\rho^{\gamma}\left[\delta f_3^{(0)}(x)+\frac{V^{(0)}}{Z^2}+\sum_{K=1}^{8}\frac{d_K}{M^2}\delta f_2^{(K)}(x)\right] \\&+\rho\,\log \rho \left(V^{(0)}+\sum_{K=1}^{8}\frac{d_K}{M^2}V^{(K)} + \sum_{K=1}^{8}b_3^{(K)}(x)\right) \\
\delta \hat{f}_4(x,\rho)=&\,\rho^{\gamma}\left[\delta f_4^{(0)}(x)+\sum_{K=1}^{8}\frac{d_K}{M^2}\delta f_4^{(K)}(x)\right]\\&+\rho\,\log \rho \left(V^{(0)}+\sum_{K=1}^{8}\frac{d_K}{M^2}V^{(K)} +\sum_{K=1}^{8}b_4^{(K)}(x)\right)\\
\delta \hat{f}_5(x,\rho)=&\,\rho^{\gamma}\left[\delta f_5^{(0)}(x)-\frac{(2-Z^2)(1-Z^2)(1+x^2)}{4Z^4}V^{(0)}+\sum_{K=1}^{8}\frac{d_K}{M^2}\delta f_5^{(K)}(x)\right]
\\
&+\rho\,\log \rho \left(V^{(0)}+\sum_{K=1}^{8}\frac{d_K}{M^2}V^{(K)}+\sum_{K=1}^{8}b_5^{(K)}(x)\right)\\
\delta \hat{f}_6(x,\rho)=&\,\rho^{\gamma}\left[\delta f_2^{(0)}(x)+V^{(0)}+\sum_{K=1}^{8}\frac{d_K}{M^2}\delta f_6^{(K)}(x)\right]\\&+\rho\,\log \rho \left(2V^{(0)}+2\sum_{K=1}^{8}\frac{d_K}{M^2}V^{(K)}+\sum_{K=1}^{8}b_6^{(K)}(x)\right)\,,
\end{aligned}
\end{equation}
together with
\begin{equation}
\gamma=1+\sum_{K=1}^{8}\frac{d_K}{M^2}\delta \gamma^{(K)}\,.
\end{equation}
Let us unpack the above expression. First, we note that if we set $d_K=0$ and $b_i^{(K)}=0$, we recover the mode with $\gamma^{(0)}=1$ given in Eq.~(\ref{eq:expand}). This makes sense, as this is the mode whose EFT expansion we want to determine. Next, we comment on the terms proportional to $\rho \log \rho$. The new terms proportional to $d_K$ are natural, since we add a mode just like the one for Kerr-Newman, except perhaps with a different amplitude. The terms proportional to $b_{i}^{(K)}(x)$ are a little less obvious. They arise because corrections to $\gamma$ induce another set of $\rho \log \rho$ terms in the equations of motion. 
We would like these $\rho \log \rho$ terms to be absent altogether, since we want to reduce the calculation to finding the $\delta f^{(K)}_{i}(x)$, which depend on $x$ only.   Requiring the $\rho \log \rho$ terms to cancel determine the $b_{i}^{(K)}(x)$ to be
\begin{equation}
\begin{aligned}
b_1^{(K)}(x)&=b_4^{(K)}(x)=0\,,& b_2^{(K)}(x)&=\frac{V^{(0)}\delta \gamma^{(K)}}{2}\,,&\hspace{4mm} b_3^{(K)}(x)&=-\frac{V^{(0)}\delta \gamma^{(K)}}{Z^2}\,,
\\
b_5^{(K)}(x)&=\frac{(2-Z^2)(1-Z^2)}{4Z^4}(1+x^2)V^{(0)}\delta \gamma^{(K)}\,, \hspace{-35mm}& & &b_6^{(K)}(x)&=-V^{(0)}\delta \gamma^{(K)}\,. &\quad
\end{aligned}
\end{equation}

At this stage, we restrict to parity-even deformations since these are the ones generated by adding the asymptotically flat region. We are left with a single mode,
which is an $\ell_4=1$ mode with amplitude $A_4$. (Recall that the other mode in this symmetry class, $\ell_1=1$, is pure gauge.)  We are thus left with a coupled set of linear ordinary differential equations for $\delta f^{(K)}_{i}(x)$ that can be schematically written as
\begin{equation}
{}^{(0)}\tilde{\Delta}_{i j}\delta f^{(K)}_{j}(x)=J^{(K)}\,,
\end{equation}
where $J^{(K)}$ is a complicated, but known, source term that depends on $\delta \gamma^{(K)}$, $V^{(0)}$, $A_4$, $V^{(K)}$, $\omega^{(K)}_{\rm NH}$, $Q^{(K)}_{\rm NH}$, and $x$, and ${}^{(0)}\tilde{\Delta}_{i j}$ is the same operator that governs the $\gamma^{(0)}=1$ deformations for an extremal Kerr-Newman black hole. This operator has a nontrivial kernel due to the gauge freedom (\ref{eq:gaugesym}). Indeed, this gauge freedom can be generalized mutatis mutandis even for nonzero $d_K$, with the functions $F^{(0)}_i$ replaced by the corresponding $F_i(x)$. We fix this freedom by choosing $\delta f^{(K)}_6$ to satisfy the same relation as in Eq.~(\ref{eq:gaugelot}). Note, however, that for $\gamma\neq1$ this relation was a consequence of the equations of motion, whereas here it is a gauge choice. Just like for the extreme Kerr-Newman black hole, the gauge is fixed up to a constant term. This constant term will play no role in what follows but could have been used to eliminate the $\ell_1=1$ mode in $\delta f^{(K)}_{i}(x)$.

Once the dust settles, we find two first-order differential equations for $\delta f^{(K)}_1(x)$ and $\delta f^{(K)}_2(x)$ and three second-order equations for $\delta f^{(K)}_3(x)$, $\delta f^{(K)}_4(x)$, and $\delta f^{(K)}_5(x)$. These equations can be solved in terms of dilogarithmic functions ${\rm Li}_2$ and other elementary functions such as $\arcsin$ and $\arctan$ via the map in Eq.~(\ref{eq:magic}). The $v_i$ variables are now replaced by $v_i^{(K)}$, and the corresponding equations can again be found for each of these variables, with source terms that depend on $J^{(K)}$. Once  the equations are integrated in full generality, and regularity at the poles is imposed, one determines $V^{(K)}$ and $\delta \gamma^{(K)}$ in terms of $V^{(0)}$ and $A_4$. These expressions are rather lengthy at this stage. However, we note that we \emph{know} what $V^{(0)}$ and $A_4$ are for an extremal Kerr-Newman black hole (see Eq.~(\ref{eq:money})), and as such, we find an expression for $\delta \gamma^{(K)}$ that depends on $Z$ only.  It is convenient to write the result in terms of\footnote{We define $\mathfrak{a}$ in terms of $r_+$, since this will be useful in Sec. \ref{sec:finite_temp} when we discuss nonextremal solutions. In the extremal limit considered here,  $\mathfrak{a} =\sqrt{1-Z^2}$.}
\begin{equation}\label{eq:deffraka}
\mathfrak{a} \equiv \frac{a}{r_+}\,,
\end{equation}
with the result,
\begin{equation}\label{eq:deltasgamma}
\begin{aligned}
\delta\gamma^{(6)} = &\,\frac{3(\mathfrak{a}^2-1)}{10\mathfrak{a}^4(\mathfrak{a}^2+1)^4}\left(15+25\mathfrak{a}^2-201\mathfrak{a}^4+89\mathfrak{a}^6-187\mathfrak{a}^8+195\mathfrak{a}^{10}+245\mathfrak{a}^{12}+75\mathfrak{a}^{14}\right)
\\& + \frac{9(\mathfrak{a}^2-1)^2(\mathfrak{a}^2+1)(1-2\mathfrak{a}^2+5\mathfrak{a}^4)}{2\mathfrak{a}^5}\arctan\mathfrak{a}\\
\delta\gamma^{(7)}=&\,\delta\gamma^{(6)} +\frac{16(\mathfrak{a}^2-1)}{5(\mathfrak{a}^2+1)^6}(149-522\mathfrak{a}^2+436\mathfrak{a}^4-166\mathfrak{a}^6+7\mathfrak{a}^8)\\
\delta\gamma^{(8)}=&\,\frac{3}{4}\delta\gamma^{(6)} + \frac{4(\mathfrak{a}^2-1)}{5(\mathfrak{a}^2+1)^6}(167-558\mathfrak{a}^2+316\mathfrak{a}^4-226\mathfrak{a}^6+13\mathfrak{a}^8)\,,
\end{aligned}
\end{equation}
together with
\begin{equation}
\delta \gamma^{(1)}=\delta \gamma^{(4)}=0\,,\quad \delta \gamma^{(2)}=\delta \gamma^{(5)}=\frac{1}{4} \delta\gamma^{(3)}\quad\text{and}\quad \delta\gamma^{(5)}=4\delta\gamma^{(8)}-\delta\gamma^{(7)}\,.
\end{equation}

The algebraic relations ensure that $\sum_{K=1}^{8}d_K \delta \gamma^{(K)}$ remains field redefinition invariant. Indeed, it is straightforward to demonstrate that it can be expressed solely in terms of the basis that is invariant under field redefinitions, namely $\{d_0,d_6,d_9\}$, as
\begin{equation}
\sum_{K=1}^{8}d_K \delta \gamma^{(K)}=\frac{1}{4}\delta \gamma^{(7)} d_0+\left(\delta \gamma^{(6)}+\frac{3}{4}\delta \gamma^{(7)}-2\, \delta \gamma^{(8)}\right)d_6+\left(\delta \gamma^{(8)}-\frac{1}{2}\delta \gamma^{(7)}\right)d_9\,.\label{eq:deltagammafull}
\end{equation}
The above relation also shows that we can use $\delta\gamma^{(K)}$ with $K=6,7,8$ as a basis to understand the EFT corrections to the (parity-even) $\gamma^{(0)}=1$ mode.

\begin{figure}[ht]
    \centering
    \includegraphics[width=\textwidth]{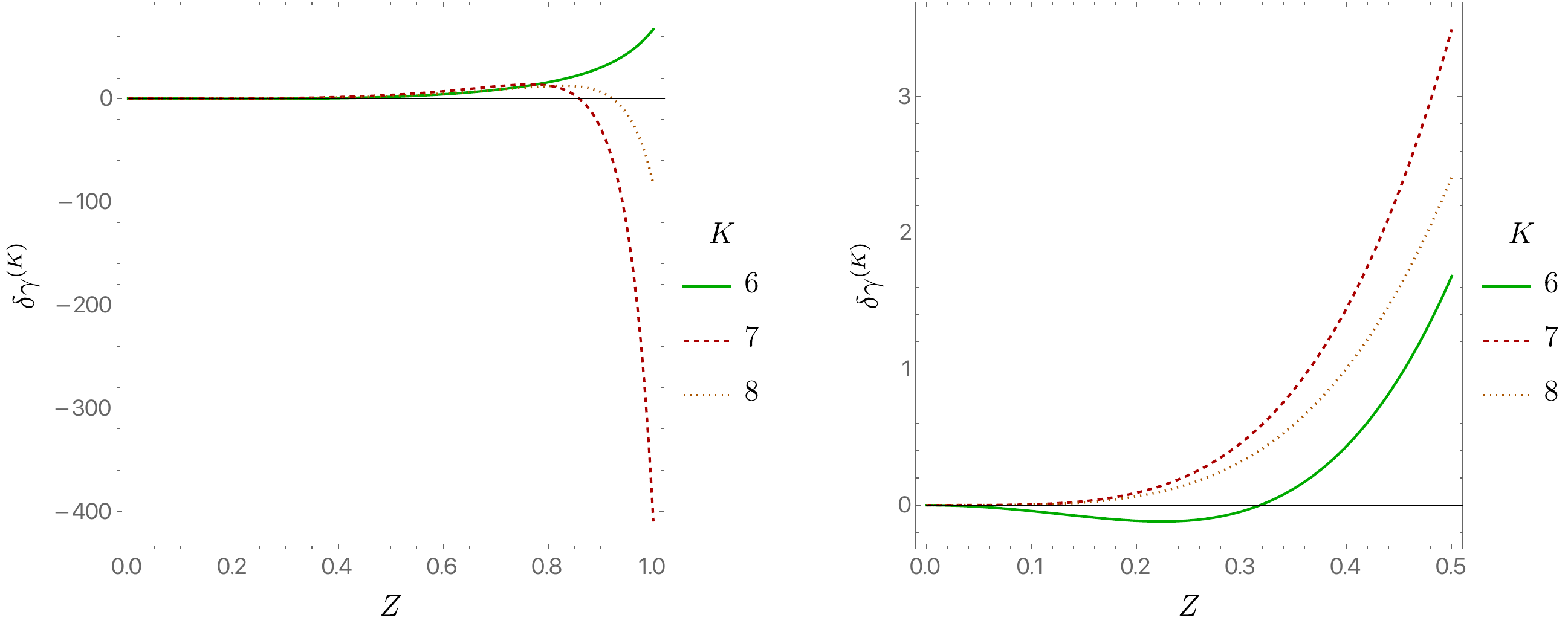}
    \caption{The change in the scaling dimension, $\delta\gamma^{(K)}$, for the  even-parity mode with $\gamma^{(0)} = 1$, as a function of $Z = Q/M$. The label $K=6,7,8$ corresponds to the three higher derivative corrections shown in Eq. \eqref{eq:xhd}. The right hand plot is a blow-up of a region on the left. }
    \label{fig:1}
\end{figure}

The most important outcome is that $\delta\gamma^{(K)}\neq 0$ for the (even parity) $\gamma^{(0)} =1$ mode, and as such we expect divergent tidal forces scaling as $1/\rho$ as we approach the extremal horizon. In Fig.~\ref{fig:1}, we plot $\delta \gamma^{(K)}$ for $K=6,7,8$ (with the right panel being a zoomed-in version of the left near $Z\sim0$), and we see that these are \emph{not} monotonic in $Z$ and additionally change sign as we vary $Z$ (but not all at the same value of $Z$).

Since we now have the EFT-corrected near-horizon geometry together with the leading deviations coming from the asymptotically flat region, we can compute the tidal forces and verify the above expectation. We first change to Bondi-Sachs coordinates, where the metric should take the simple form,
\begin{equation}
{\rm d}s_{\rm BS}^2=e^{2\beta}(-V{\rm d}v^2+2 {\rm d}v {\rm d}\rho)+e^{2\chi}h_{IJ}({\rm d}y^{I}+U^I \mathrm{d}v)({\rm d}y^{J}+U^J \mathrm{d}v)\quad\text{with}\quad I=\{x,\psi\}.
\end{equation}
We do this by setting
\begin{equation}
t=v+\frac{1}{\rho}\quad\text{and}\quad \varphi = \psi+\omega_{\rm NH} \log \rho-\lambda(\rho,x)
\end{equation}
and imposing
\begin{equation}
\frac{\partial \lambda}{\partial \rho}=\frac{\omega_{\rm NH}-f_4(\rho,x)}{\rho}\,.
\end{equation}
With the above choice, we find
\begin{equation}
\beta = \chi = \log (\sqrt{2} M f_1)\quad\text{and}\quad V=\rho^2\,,
\end{equation}
along with
\begin{equation}
\begin{aligned} 
h_{IJ} &=\left[
\begin{array}{cc}
\displaystyle \frac{\Gamma^2_{\rm NH}}{(1-x^2)f_6}+f_2^2 \left(\frac{\partial \lambda}{\partial x}\right)^2 & \displaystyle -f_2^2\left(\frac{\partial \lambda}{\partial x}\right)
\\
\\
 \displaystyle -f_2^2\left(\frac{\partial \lambda}{\partial x}\right) &\displaystyle f_2^2
\end{array}
\right] \\
U^{I} &=\left[\begin{array}{cc}
\displaystyle0 &\;\; \displaystyle\rho\,f_4
\end{array}\right]\,.
\end{aligned}
\end{equation}
Since $V=\rho^2$, in Bondi-Sachs coordinates, the future extremal event horizon is the null hypersurface $\rho=0$.

It is now a simple exercise to show that, to linear order in $d_K$, the Weyl tensor satisfies
\begin{equation}
C_{\rho I \rho J}= \frac{1-Z^2}{4 \rho }\left[
\begin{array}{cc}
\displaystyle -\frac{1}{2-Z^2} & \displaystyle -\frac{\sqrt{1-Z^2}\,x}{1+(1-Z^2) x^2}
\\
\\
\displaystyle -\frac{\sqrt{1-Z^2}\,x}{1+(1-Z^2) x^2} & \displaystyle \frac{(2-Z^2) (1-x^2)}{\left[1+(1-Z^2) x^2\right]^2}
\end{array}\right]\sum_{K=1}^{8}d_K \delta \gamma^{(K)},
\label{eq:tidalfinal}
\end{equation}
demonstrating that it indeed diverges like $1/\rho$ as expected. 
 Note that $C_{\rho I \rho J}$ is still proportional to $\delta \gamma^{(K)}$, despite the presence of the $\rho \log \rho$ terms.  However,  there is an overall factor of $(1-Z^2)$ in front, showing that there are no tidal force singularities for the extreme Reisner-Nordstr\"om black hole, in accordance with Ref.~\cite{Horowitz:2023xyl}, even though $\delta \gamma^{(K)}$ is  nonzero in that limit.\footnote{An analogous calculation of the 
Maxwell component $F_{\rho I}$ yields a $\log \rho$ divergence for either sign of $\delta \gamma$. But this is just a result of working to first order in the Wilson coefficients, since $A_I \sim\rho^{1 + \epsilon}= \rho(1+\epsilon \log \rho)$ to first order.} 

\section[Constructing EFT-corrected Kerr-Newman black holes at \\finite temperature]{Constructing EFT-corrected Kerr-Newman black holes at finite \\temperature} \label{sec:finite_temp}
Having analytically found the EFT corrections to the extremal Kerr-Newman black hole in the near-horizon limit, let us now turn to the challenging but more physically realistic question of constructing these solutions away from extremality, at finite black hole temperature. 
Here, we will find that a combination of numerical and analytical methods are necessary.

All of our results will automatically be field redefinition invariant, and therefore expressible in the manifestly invariant basis $d_{0,3,6,9}$ discussed previously.
We have seen this explicitly in Sec.~\ref{sec:scaling_EFT}, where we found algebraic relations among the scaling dimension contributions $\delta\gamma^{(K)}$ generated by all eight terms in the action~\eqref{eq:hd}, which allowed us to express the field redefinition invariant scaling dimension in Eq.~\eqref{eq:deltagammafull}.
While this exercise was a useful consistency check of our computations, as discussed previously we could instead have initially applied field redefinitions at the level of the action.
Doing so allows us to absorb any terms containing the Ricci tensor or scalar into gauge fields via Eq.~\eqref{eq:redef}.
The Riemann-squared operator, meanwhile, can be combined with other curvature-squared combinations to give the topological Gauss-Bonnet term, which as noted above cannot contribute to any physical bulk phenomena.
As a result, without loss of generality we can henceforth reduce to only the terms $d_{6,7,8}$ as in Eq.~\eqref{eq:xhd}---the Riemann-$F^2$ term and the two $F^4$ terms---with the understanding that these Wilson coefficients are now the new ones obtained after field redefining away the others.

The solutions we seek are stationary and axisymmetric, i.e., they possess two commuting Killing vector fields $k$ and $m$ generating time translations and rotations. We can introduce coordinates $(t,\phi,x_1,x_2)$ such that $k=\partial/\partial t$ and $m\equiv \partial/\partial \phi$, where $\phi \sim\phi+2\pi$. We assume that the solutions enjoy the so-called $t$-$\phi$ symmetry, i.e., they are invariant under the simultaneous reflection symmetry $(t,\phi)\to-(t,\phi)$. 
We choose the coordinates $(x_1,x_2)$ so that the metric on surfaces of constant $(t,\phi)$ is manifestly \emph{conformally flat}. All told, we are assuming that the metric and Maxwell potential can be written as
\begin{equation}
\begin{aligned}
\mathrm{d}s^2 &= G_{I J}(x_1,x_2)\mathrm{d}y^I\mathrm{d}y^J+\Phi(x_1,x_2)(\mathrm{d}x_1^2+\mathrm{d}x_2^2)\\
\label{eq:general}
A&=\tilde{A}_I(x_1,x_2)\,\mathrm{d}y^I
\end{aligned}
\end{equation}
with $y^I=(t,\phi)$\,.

Recall the metric for an electrically charged Kerr-Newman black hole \cite{Newman:1965my} written in Boyer-Lindquist coordinates  \cite{Boyer:1966qh} given in Eq.~\eqref{eqs:KNgA}. Since at finite temperature we will ultimately be using different coordinates than those of Secs.~\ref{sec:settingscene} through \ref{sec:scaling_EFT} once we introduce the EFT deformation of the metric, in order to disambiguate the notation let us replace the angular coordinate $x=\cos\theta$ with $X$.
The background is given by
\begin{equation}
\begin{aligned}
\mathrm{d}s^2_{\rm KN} =& -\frac{\Delta(r)}{\Sigma(r,X)}\left[\mathrm{d}t-\bar{a}(1-X^2)\mathrm{d}\phi\right]^2+\frac{1-X^2}{\Sigma(r,X)}\left[\bar{a}\,\mathrm{d}t-(r^2+\bar{a}^2)\mathrm{d}\phi\right]^2
\\
&+\Sigma(r,X)\left[\frac{\mathrm{d}r^2}{\Delta(r)}+\frac{\mathrm{d}X^2}{1-X^2}\right]
\label{eq:kerr}
\end{aligned}
\end{equation}
and
\begin{equation}
A_{\rm KN}=-\frac{\sqrt{2}\,\bar{Q}\,r}{\kappa\,\Sigma(r,X)}\left[{\rm d}t-\bar{a}\,(1-X^2)\, {\rm d}\phi\right],
\end{equation}
where
\begin{equation}
\Delta(r) = r^2+\bar{a}^2+\bar{Q}^2-2 \bar{M} r\,,\qquad \Sigma(r,X) = r^2+\bar{a}^2 X^2\,,\qquad\text{and}\qquad X\in[-1,1]\,.
\end{equation}
We use bars to distinguish the mass, angular momentum, and charge of the Kerr-Newman solution from those of the EFT-corrected solution. The event horizon is the null hypersurface $r=r_+$ with $r_{\pm} = \bar{M}\pm\sqrt{\bar{M}^2-\bar{a}^2-\bar{Q}^2}$. The event horizon becomes degenerate in the extremal limit, when $\bar{M}=\sqrt{\bar{a}^2+\bar{Q}^2}$.
One can bring Eq.~(\ref{eq:kerr}) to the form in Eq.~(\ref{eq:general}) by defining 
\begin{equation}
\label{x1x2}
x_1 = \int_{r_+}^r \frac{\mathrm{d}\widetilde{r}}{\sqrt{\Delta(\widetilde{r})}}\quad \text{and}\quad x_2 = \int_{X}^1 \frac{\mathrm{d}\widetilde{X}}{\sqrt{1-\widetilde{X}^2}}\,.
\end{equation}

Now consider the EFT-corrected black hole, which we assume is nonextremal throughout this section. We start with a metric in the form of Eq.~(\ref{eq:general}), apply the coordinate transformation in Eq.~\eqref{x1x2}, and parameterize $G_{IJ}$, $\Phi$,  and $\tilde{A}_I$ as follows,
\begin{equation}
\begin{aligned}\label{eqs:ansat}
\mathrm{d}s^2 =& -\frac{\Delta(r)}{\Sigma(r,X)}F_1(r,X)\left[\mathrm{d}t-(1-X^2)F_4(r,X)\mathrm{d}\phi\right]^2
\\
&+\frac{1-X^2}{\Sigma(r,X)}F_3(r,X)\left[F_4(r,X)\,\mathrm{d}t-(r^2+a^2)\mathrm{d}\phi\right]^2 \\&+\Sigma(r,X)F_2(r,X)\left[\frac{\mathrm{d}r^2}{\Delta(r)}+\frac{\mathrm{d}X^2}{1-X^2}\right]\\
A =&-\frac{\sqrt{2}\,r\,F_5(r,X)}{\kappa\,\Sigma(r,X)}\left[{\rm d}t-F_4(r,X)\,(1-X^2)\, {\rm d}\phi\right]
\\
&-\frac{\sqrt{2}\,(1-X^2)\,F_6(r,X)}{\kappa\,\Sigma(r,X)}\left[F_4(r,X)\,{\rm d}t-(r^2+a^2){\rm d}\phi\right],
\end{aligned}
\end{equation}
where $F_{i}$, $i=1,\ldots,6$, are functions of $r$ and $X$ to be determined below.

We are working to first order in the higher-derivative corrections, so we can write
\begin{equation}
\label{eqs:expan}
\begin{aligned}
F_{i}(r,X) &= 1+\sum_{K=6}^{8}\frac{d_K}{M^2}f_i^{(K)}(r,X),\quad i=1,2,3\\
F_{4}(r,X) &= \bar{a}+\left(1-\frac{r_+}{r}\right)\sum_{K=6}^{8}\frac{d_K}{M^2}f_4^{(K)}(r,X)\\
F_{5}(r,X) &= \bar{Q}+\left(1-\frac{r_+}{r}\right)\sum_{K=6}^{8}\frac{d_K}{M^2}f_5^{(K)}(r,X)\\
F_{6}(r,X) &= \sum_{K=6}^{8}\frac{d_K}{M^2}f_6^{(K)}(r,X)\,,
\end{aligned}
\end{equation}
where $M$ is the mass of the corrected solution; since we are working to first order, it does not matter whether we write $M$ or $\bar{M}$ above. The factors of $(1-r_+/r)$ were introduced to ensure that the EFT-corrected solutions have the same angular velocity and chemical potential as the background Kerr-Newman black hole.

The procedure is now clear. We input the ansatz~(\ref{eqs:ansat}) into the equations of motion derived from the action in Eq.~(\ref{eq:hd}) and linearize in $\{c_6,c_7,c_8\}$ to find the equations for $f_{i}^{(K)}(r,X)$. For each $K$, we obtain eight nontrivial components and thus eight partial differential equations in $f_{i}^{(K)}(r,X)$. However, we only have six $f_{i}^{(K)}(r,X)$. This is to be expected, since we have fixed the gauge, and so some of the equations must be redundant after a suitable choice of boundary conditions for $f_{i}^{(K)}(r,X)$. 

Let us denote the linearization of the Einstein and Maxwell equation in the absence of higher-derivative terms by $\delta G_{ab}$ and $\delta Z_a$, respectively. For each $K$, the Einstein-Maxwell equations reduce to an inhomogeneous linear system of partial differential equations of the form
\begin{equation}
\delta E^{ab}_{(K)}\equiv \delta G^{ab}-\delta T^{ab}_{(K)}=0\quad \text{and}\quad \delta P^{a}_{(K)}\equiv\delta Z^a-\delta J^{a}_{(K)}=0\,,
\end{equation}
where the source terms $\delta T^{ab}_{(K)}$ and $\delta J^{a}_{(K)}$ are obtained by varying the higher-derivative terms in the action and evaluating the result on the Kerr-Newman metric.

For the six equations that we will solve numerically, we choose
\begin{equation}
E_1^{(K)} \equiv \delta E^{tt}_{(K)}\,, \;\; E_2^{(K)} \equiv \delta E^{t\phi}_{(K)}\,, \;\; E_3^{(K)} \equiv \delta E^{\phi\phi}_{(K)}\,,\;\; E_4^{(K)} \equiv \delta E^{rr}_{(K)}\,g_{rr}+\delta E^{XX}_{(K)}\,g_{XX}\,,
\end{equation}
where in the last equation $g_{rr}$ and $g_{XX}$ are computed using the Kerr-Newman metric, together with the two nontrivial components of the Maxwell equation,
\begin{equation}
E_5^{(K)}\equiv \delta P^{t}_{(K)}\quad \text{and}\quad E_6^{(K)}\equiv \delta P^{\phi}_{(K)}\,.
\end{equation}
We write the remaining two equations as
\begin{equation}\label{eq:defCs}
C_1^{(K)} \equiv \frac{\sqrt{\Delta}\sqrt{1-X^2}}{2}\tilde{C}_1^{(K)}\quad\text{and}\quad C_2^{(K)} \equiv (1-X^2)\tilde{C}_2^{(K)}\,,
\end{equation}
where
\begin{equation}
\tilde{C}_1^{(K)}\equiv\Sigma(r,X)\left(\delta E^{rr}_{(K)}\,g_{rr}-\delta E^{XX}_{(K)}\,g_{XX}\right)\quad \text{and}\quad \tilde{C}_2^{(K)}=\Sigma(r,X)\,g_{XX}\,\delta E^{rx}_{(K)}.
\end{equation}
It is then a simple exercise to use the Bianchi identities to show that
\begin{equation}
\begin{aligned}\label{eq:simple}
\sqrt{1-X^2}\;\partial_X C_1^{(K)}-\sqrt{\Delta}\;\partial_r C_2^{(K)}&=0
\\
\sqrt{\Delta}\;\partial_r C_1^{(K)}+\sqrt{1-X^2}\;\partial_X C_2^{(K)}&=0\,.
\end{aligned}
\end{equation}
We can use the equations above, along with commutation of partial derivatives, to show that
\begin{equation}
{}^{(2)}\Delta C_1^{(K)}=0\qquad \text{and}\qquad
{}^{(2)}\Delta C_2^{(K)}=0\,,
\end{equation}{}
where ${}^{(2)}\Delta$ is the Laplacian computed with the two-dimensional metric, 
\begin{equation}
\mathrm{d}s^2_{(2)}\equiv{}^{(2)}g_{\check{i}\check{j}}\mathrm{d}z^{\check{i}}\mathrm{d}z^{\check{j}}=\frac{\mathrm{d}r^2}{\Delta}+\frac{\mathrm{d}X^2}{1-X^2}\,,
\end{equation}
where $\check{i}\in\{1,2\}$ and $z^{\check{i}}=\{r,X\}$. 

We will now show that $C_1^{(K)}$ and $C_2^{(K)}$ will vanish everywhere in our integration domain so long as they 
vanish at the boundary of the domain. In order to do this, we will look at the quantity,
\begin{equation}
\int_{\mathcal{I}}\mathrm{d}^2 z\; \sqrt{{}^{(2)}g}\;{}^{(2)}\nabla_{\check{i}}C_1^{(K)}\;{}^{(2)}\nabla^{\check{i}}C_1^{(K)}=\int_{\partial \mathcal{I}}\mathrm{d} \Sigma^{\check{i}}\,C_1^{(K)} \nabla_{\check{i}}C_1^{(K)}\,,
\label{eq:inte}
\end{equation}
where $\mathcal{I}=[-1,1]\cup [r_+,+\infty]$ is the domain of integration. So long as the integrand on the right vanishes on each segment of the boundary $\partial \mathcal{I}$, the right-hand side vanishes.\footnote{For the component of the boundary at $r=\infty$, by ``vanishing'' we mean that the integrand should decay sufficiently rapidly as $r \rightarrow \infty$.}  This means the left-hand side must also vanish, and since it is positive, we must have $C_1^{(K)}=0$ everywhere in $\mathcal{I}$. The same applies for $C_2^{(K)}$. As a check on our numerics, once we solve for the $f_{i}^{(K)}(r,X)$, we can a posteriori check that $C_1^{(K)}$ and $C_2^{(K)}$ do indeed approach zero in the continuum limit. Note that from Eq.~(\ref{eq:simple}) it is also easy to see that when \emph{one} of the $C_i^{(K)}$ vanishes on the integration domain, the other is forced to be a constant. Thus, we only need to impose that \emph{one} of the constraints 
vanishes on all boundaries while the other only needs to be imposed (at least) at a single point in the integration domain~\cite{Wiseman:2002zc}.

Before discussing the boundary conditions, we note that $r$ is not a useful coordinate to implement in a numerical grid, since $r$ is noncompact. We thus define
\begin{equation}
r \equiv \frac{r_+}{1-Y}\,,
\label{eq:num}
\end{equation}
with $Y\in[0,1]$. Spatial infinity is now located at $Y=1$ and the event horizon at $Y=0$.
This coordinate transformation also needs to be applied to the constraints $C^{(K)}_1$ and $C^{(K)}_2$. In particular, for the asymptotic boundary now located at $Y=1$, the spatial infinity component of the integrand on the right-hand side of Eq.~(\ref{eq:inte}) scales as
\begin{equation}
\frac{\tilde{C}_1^{(K)}\partial_Y \tilde{C}_1^{(K)}}{1-Y}+\frac{\tilde{C}_1^{(K)}{}^2}{(1-Y)^2},
\label{eq:expan}
\end{equation}
and we need both terms to vanish asymptotically. A similar expansion holds for $\tilde{C}_2^{(K)}$.

We now turn our attention to the choice of boundary conditions. At the axis, located at $X=1$, we demand
\begin{equation}
f_2^{(K)}(1,Y)=f^{(K)}_3(1,Y)\,,
\label{eq:BCXP}
\end{equation}
while for the remaining variables we get a set of rather complicated Robin boundary conditions. The latter conditions arise from evaluating the equations of motion at $X= 1$, using the fact that regularity on the axis implies that the $f_i^{(K)}$ are finite at $X=1$. The former condition, in Eq.~\eqref{eq:BCXP}, in turn ensures that $\phi$ has period $2\pi$ despite the inclusion of the higher-derivative corrections. This choice of boundary conditions is such that both $\tilde{C}^1_{(K)}$and $\tilde{C}^2_{(K)}$ vanish linearly at $X=1$, leading to a vanishing integrand of the left-hand side of Eq.~(\ref{eq:inte}) along this boundary. 

Since the solution we seek preserves parity with respect to the reflection symmetry $X\to-X$, we reduce our integration domain so that $X\in[0,1]$, with pure Neumann boundary conditions for all $f_i^{(K)}$ at $X=0$. These boundary conditions are enough to ensure that $\tilde{C}_1^{(K)}$ has a vanishing derivative at $X=0$, while $\tilde{C}_2^{(K)}$ vanishes linearly there.

Smoothness at the nonextremal horizon, together with the equations of motion, demands that all the $f_i^{(K)}$ obey Robin boundary conditions so that both $\tilde{C}^1_{(K)}$ and $\tilde{C}^2_{(K)}$ are finite at $Y=0$. Note, however, that this implies that the integrand on the right-hand side of Eq.~(\ref{eq:inte}) vanishes at the horizon (see the definition of $C_1^{(K)}$ in terms of $\tilde{C}_1^{(K)}$ in Eq.~(\ref{eq:defCs}) and take into account the measure ${\rm d}\Sigma^{\check{i}}$).

At spatial infinity, we expand the functions as
\begin{equation}
f_i^{(K)}(X,Y) = \sum_{p=0}^{+\infty}(1-Y)^p \tilde{f}^{(p,K)}_i(X)\,.
\end{equation}
We then choose
\begin{equation}
\tilde{f}_{1}^{(0,K)}(X)=\tilde{f}_{3}^{(0,K)}(X)=\tilde{f}_{6}^{(0,K)}(X)=0\,,
\label{eq:BCI}
\end{equation}
with the third condition imposing the absence of magnetic charges and the first two preserving asymptotic flatness. At each $p$-order in $(1-Y)$, one finds a system of sourced second-order ordinary differential equations for $\tilde{f}^{p,K}_i(X)$, which can be readily solved.

For instance, at first order in $(1-Y)$, one finds 
\begin{equation}
\begin{aligned}
\tilde{f}_{2}^{(0,K)}(X)&=\tau^K_0
\\
\tilde{f}_{4}^{(0,K)}(X)&=\omega^K_0
\\
\tilde{f}_{5}^{(0,K)}(X)&=\rho^K_0
\\
\tilde{f}_{1}^{(1,K)}(X)&=-\tilde{f}_{3}^{(1,K)}(X)=\alpha^K_0
\\
\tilde{f}_{2}^{(1,K)}(X)&=-\alpha^K_0+\beta^K_0\sqrt{1-X^2}+X\,\gamma^K_0
\\
\tilde{f}_{4}^{(1,K)}(X)&=\frac{1}{1+\mathfrak{a}^2+\mathfrak{q}^2}\left[\omega^K_0(1+\mathfrak{q}^2)+3 \omega^K_0\mathfrak{a}^2-\mathfrak{q}\mu^K_0-\mathfrak{a} \mu^K_0\right]
\\
\tilde{f}_{5}^{(1,K)}(X)&=\rho^K_0+\frac{\mathfrak{q}}{2}\alpha^K_0
\\
\tilde{f}_{6}^{(1,K)}(X)&=\mu^K_0
\\
\tilde{f}_{3}^{(2,K)}(X)&=-2\omega^K_0 \mathfrak{a}\,X^2+\chi_0^K\,,
\end{aligned}
\label{eq:asym}
\end{equation}
where we have defined the parameters 
\begin{equation}\label{eq:defaq}
    \mathfrak{a}=\frac{\bar{a}}{r_+},  \qquad  \mathfrak{q}=\frac{\bar{Q}}{r_+},
\end{equation}
and $\alpha^K_0$, $\beta^K_0$, $\gamma^K_0$, $\lambda^K_0$, $\omega^K_0$, and $\tau^K_0$ are constants of integration. In deriving Eq.~(\ref{eq:asym}) we assumed finiteness of the $f_{i}^{(1,K)}(X)$ at $X=\pm1$. Using the boundary conditions at $X=1$, one then finds
\begin{equation}
\tau^K_0=\beta^K_0=\gamma^K_0=0.
\end{equation}
 The last two conditions can be readily summarized in the boundary condition for $f_{2}^{(K)}(X,Y)$ at $Y=1$,
\begin{equation}
\left.\frac{\partial f_{2}^{(K)}}{\partial Y}+\frac{\partial f_{1}^{(K)}}{\partial Y}\right|_{Y=1}=0\,,
\label{eq:nonc}
\end{equation}
which we directly impose on the grid. With this choice of boundary conditions, $\tilde{C}_1^{(K)}$ vanishes quadratically at $Y=1$, while $\tilde{C}_2^{(K)}$ vanishes linearly there. This means that the  right-hand side of the integrand in Eq.~(\ref{eq:inte}) vanishes on all four boundaries for $C^{(K)}_1$. For $C^{(K)}_2$ this need not be the case, since the second term in Eq.~(\ref{eq:expan}) does not necessarily vanish at $Y=1$. 
At spatial infinity, we thus require
\begin{equation}
\begin{aligned}
f_1^{(K)}(X,1)=f_3^{(K)}(X,1)=f_6^{(K)}(X,1)&=0 \\ 
\frac{\partial f_{4}^{(K)}(X,Y)}{\partial Y}+\frac{1+\mathfrak{q}^2+3\mathfrak{a}^2-2X^2\mathfrak{a}^2}{1+\mathfrak{a}^2+\mathfrak{q}^2}f_{4}^{(K)}(X,Y)\hspace{24mm}
\\
-\frac{\mathfrak{a}}{2}\frac{1}{1+\mathfrak{a}^2+\mathfrak{q}^2}\frac{\partial^2 f_3^{(K)}(X,Y)}{\partial Y^2}+\frac{\mathfrak{q}}{1+\mathfrak{a}^2+\mathfrak{q}^2}\frac{\partial f_6^{(K)}(X,Y)}{\partial Y}\Bigg|_{Y=1}&=0 \vphantom{\Bigg]}\\ 
\left.\frac{\partial f_{5}^{(K)}(X,Y)}{\partial Y}-\frac{\mathfrak{q}}{2}\frac{\partial f_{1}^{(K)}(X,Y)}{\partial Y}+f_{5}^{(K)}(X,Y)\right|_{Y=1}&=0 \,,
\end{aligned}
\end{equation}
together with Eq.~(\ref{eq:nonc}).

Our boundary conditions are such that $C_1^{(K)}$ vanishes or has zero derivative on all boundary segments of our integration domain, and as such, remains zero in the whole integration domain. Furthermore, $C_2^{(K)}$ is vanishing on several of the boundaries. We thus conclude that our boundary conditions enforce $C_1^{(K)}$ and $C_2^{(K)}$ to vanish everywhere. We have explicitly checked that this is the case for all of the solutions reported in this paper.

There is one final technical complication. We want to investigate solutions that are very close to extremality. These solutions develop large gradients close to the horizon. To deal with this final hurdle, we use \emph{multiple} Chebyshev-Gauss-Lobatto grids, which connect at interfaces (see Ref.~\cite{Dias:2015nua} for more details on how to deal with patching procedures).
\subsection{Thermodynamics}
Once we have found the solutions, we proceed to determine their thermodynamic properties. This is instructive, since we can compare our results with the analytic expressions reported in Refs.~\cite{Cheung:2018cwt,Cheung:2019cwi}. To compute the energy, angular momentum, and electric charge of our solutions, we use Komar integrals evaluated at spatial infinity. These charges are defined just as in Eq.~(\ref{eqs:komar}), despite the presence of the higher-derivative terms.

We note that, from our ansatz, the chemical potential and angular velocity of the corrected solutions read
\begin{equation}
\mu = \frac{\bar{Q}}{r_+^2+\bar{a}^2}\quad\text{and}\quad \Omega=\frac{\bar{a}}{r_+^2+\bar{a}^2}\,,
\end{equation}
showing that, as promised, our boundary conditions ensure that the EFT-corrected solution has the same chemical potential and angular velocity as the background Kerr-Newman black hole.  The constants $\mathfrak{a}\equiv \bar{a}/r_+$ and $\mathfrak{q}\equiv \bar{Q}/r_+$ defined previously serve as the parameters that move along the moduli space of solutions. Note that the electric charge of the corrected Kerr-Newman black hole will no longer be simply given by $\bar{Q}$. For these solutions, $\bar{Q}$ is just a dial to change the chemical potential.

Expanding our equations of motion for the finite-temperature solutions near spatial infinity and using our boundary conditions in Eq.~(\ref{eq:BCI}) yields
\begin{equation}
\left.\frac{\partial f_{2}^{(K)}}{\partial Y}\right|_{Y=1}\!\!\!\!{=}\left.\frac{\partial f_{3}^{(K)}}{\partial Y}\right|_{Y=1}\!\!\!\!{=}-\left.\frac{\partial f_{1}^{(K)}}{\partial Y}\right|_{Y=1}\!\!\!\!{=}\,\alpha^{K}_0\,,\;f_4^{(K)}(X,1)=\omega^{K}_0\,,\,\;\text{and}\,\;f_5^{(K)}(X,1)=\rho^{K}_0\,.
\end{equation}
Inputting these asymptotic expansions into Eq.~(\ref{eqs:komar}) yields the energy, angular momentum, and electric charge of the nonextremal solutions,
\begin{equation}
\begin{aligned}
E&= \bar{E}-\frac{4\pi}{\kappa^2\,r_+}\sum_{K=1}^{8}\alpha_0^K d_K\\
J&=\bar{J}+\frac{4\pi}{\kappa^2}\sum_{K=1}^{8}\left[\omega_0^K(1+\mathfrak{a}^2+\mathfrak{q}^2)-2 \mathfrak{a} \alpha_0^K\right]d_K\\
Q_e&=\frac{4\pi \sqrt{2}}{\kappa^2}\left[\bar{Q}+\frac{1}{r_+}\sum_{K=1}^{8}\rho_0^K d_K\right]\,.
\end{aligned}
\end{equation}

To compute the temperature of the nonextremal solutions, we simply evaluate the surface gravity at $r=r_+$ and find
\begin{equation}
T = \frac{1-\mathfrak{a}^2-\mathfrak{q}^2}{4\pi r_+(1+\mathfrak{a}^2)}\left(1+\frac{1}{2}\sum_{K=6}^{8}\lambda^{K}\frac{d_K}{r_+^2}\right)\,,
\label{eq:hawkingT}
\end{equation}
with
\begin{equation}
\lambda^{K}\equiv f_1^{(K)}(X,0)-f_2^{(K)}(X,0)\,,
\end{equation}
We note that $\lambda^{K}$ is independent of $X$ by virtue of one of the constraint equations, and thus that $T$ is independent of $X$, as demanded by the rigidity theorems \cite{Hawking:1971vc,Hawking:1973uf,Hollands:2006rj}.

Finally, we need to determine the entropy. For this task, we use the Wald entropy~\cite{Wald:1993nt}, which accounts for the effect of the higher-derivative corrections. As noted earlier, terms generated by $d_7$ and $d_8$ will not explicitly modify the entropy functional, since there is no explicit dependence on the Riemann tensor, though they will modify the value of the entropy by changing the horizon size, via the modifications to the equations of motion. For terms proportional to $d_6$, the Wald formula generates explicit corrections to the entropy functional itself. Indeed, we find 
\begin{multline}
S = \frac{8\pi^2}{\kappa^2}(1+\mathfrak{a}^2)r_+^2\Bigg\{1+\frac{1}{4 r_+^2}\sum_{K=6}^{8}\int_{-1}^1\mathrm{d}X\left[f^{(K)}_2(X,0)+f^{(K)}_3(X,0)\right]d_K
\\
-\frac{2 \mathfrak{q}^2}{3r_+^2}\left[\frac{9+4 \mathfrak{a} ^2+3 \mathfrak{a} ^4}{\left(1+\mathfrak{a} ^2\right)^3}+\frac{3\, {\rm arctan}\,\mathfrak{a}}{\mathfrak{a} }\right]d_6\Bigg\}\,,
\end{multline}
where the last term comes from the Wald entropy shift induced by $d_6$. 

\begin{figure}[h]
    \centering
    \includegraphics[width=\textwidth]{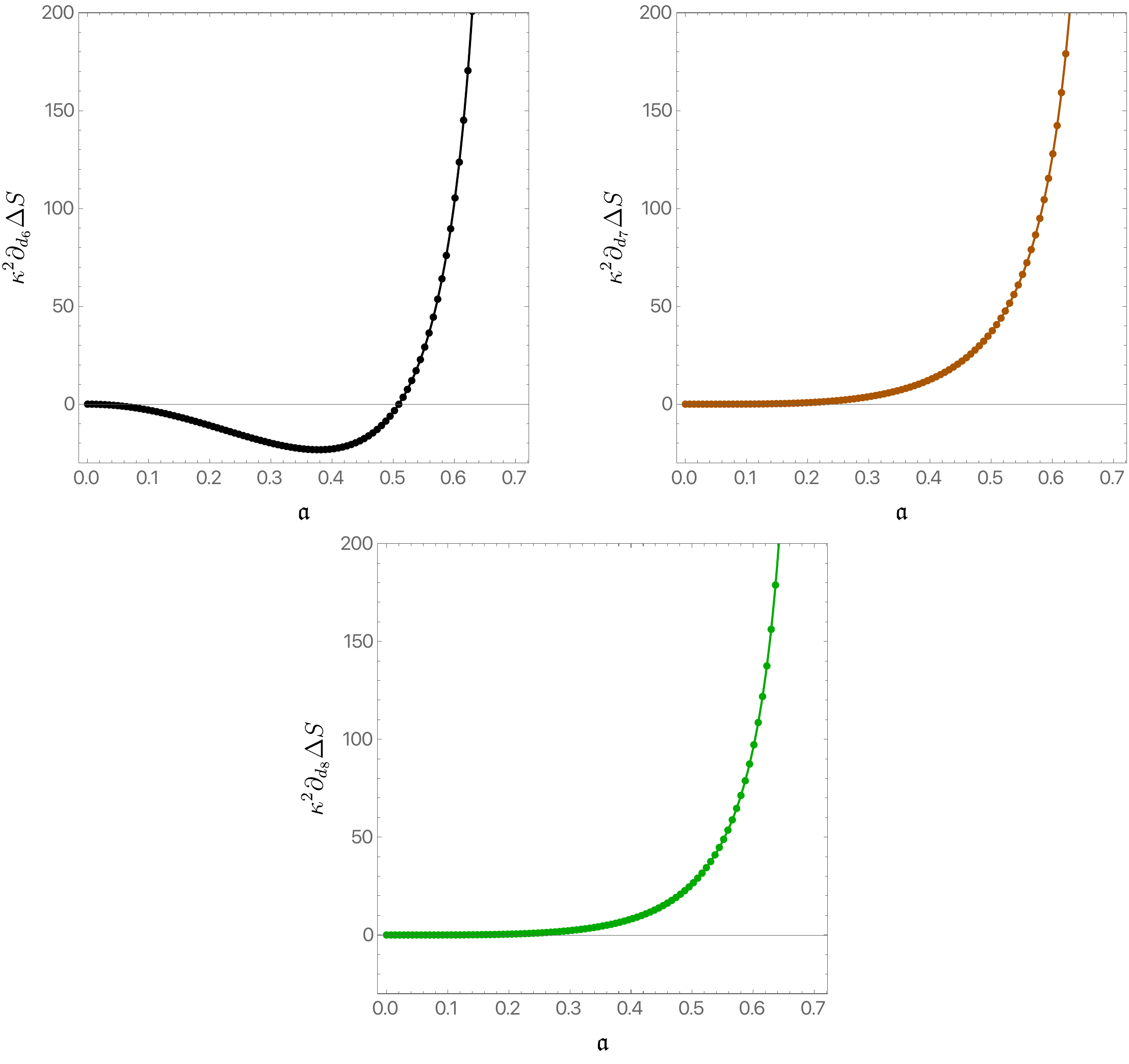}
    \caption{The thermodynamic predictions of Ref.~\cite{Cheung:2019cwi}, represented as the solid lines, against the numerical data, represented as the colored disks. These figures were all generated for the parameter choice $\mathfrak{a}=\mathfrak{q}$, where $\mathfrak{a}$ and $\mathfrak{q}$ are defined in Eq.~\eqref{eq:defaq}. }
    \label{fig:thermodynamics}
\end{figure}

In Ref.~\cite{Cheung:2019cwi}, the entropy difference between the EFT-corrected Kerr-Newman black hole and a Kerr-Newman black hole with the same electric charge, mass, and angular momentum was computed using the on-shell action methods of Ref.~\cite{Reall:2019sah}. This turns out to be given by\footnote{It turns out that Eq.~(11) in Ref.~\cite{Cheung:2019cwi} has a small typo, namely, their general expression for $\Delta S$ should be multiplied by an overall factor of $(1+\mathfrak{a}^2)$.}
\begin{equation}
\hspace{-2mm}\begin{aligned}
&\Delta S(\mathfrak{a},\mathfrak{q})=\frac{16 \pi^2 (\mathfrak{a} ^2-3) (3 \mathfrak{a} ^2-1) [1-\xi -\mathfrak{a}^2 (1+\xi )]^2 }{15 \kappa ^2 \xi  (1+\xi ) (1+\mathfrak{a} ^2)^4}(d_0+d_6-d_9)
\\
&+\frac{\pi^2\left[1{-}\xi {-}\mathfrak{a} ^2 (1{+}\xi )\right]^2 [\mathfrak{a}  (3{+}2 \mathfrak{a} ^2{+}3 \mathfrak{a} ^4){+}3 (\mathfrak{a} ^2{-}1) \left(1{+}\mathfrak{a} ^2\right)^2{\rm arctan}\,\mathfrak{a}]}{2\kappa ^2 \xi  (1+\xi ) \mathfrak{a} ^5}  (d_0+d_6+d_9)
   \\
& +\frac{64 \pi^2}{\kappa^2}d_3  +\frac{32 \pi ^2 [1-\xi -\mathfrak{a} ^2 (1+\xi )] [\mathfrak{a} ^2 (3+4 \xi )-1-4 \xi ]}{5 \kappa ^2 \xi  (1+\xi ) (1+\mathfrak{a} ^2)^2} d_6\,,
\end{aligned}\hspace{-1mm}
\end{equation}
with $\xi$ an extremality parameter, 
\begin{equation}
\xi=\frac{\sqrt{\bar M^2 - \bar{a}^2 - \bar Q^2}}{\bar M} =\frac{1-\mathfrak{a}^2-\mathfrak{q}^2}{1+\mathfrak{a}^2+\mathfrak{q}^2} = \frac{\kappa^2}{4\pi \bar M}\bar T \bar S\,,
\end{equation}
where $\bar T$ and $\bar S$ are the temperature and entropy of the uncorrected Kerr-Newman black hole.\footnote{As shown in Ref.~\cite{Cheung:2019cwi}, there is also an explicit contribution to the entropy going like $d_3$, i.e., sensitive to the Gauss-Bonnet term.
This is not inconsistent, since the black hole horizon is a boundary of the exterior spacetime, on which topological terms can have support. However, this term cannot affect any physical observable. In any case, its contribution to $\Delta S$ is subdominant in the extremal limit, and furthermore $d_3$ is small compared to $d_{6,7,8}$ in realistic completions, so we drop it as before.} Note that $\Delta S$ diverges in the extremal limit $\xi \to 0$, but this is simply a consequence of comparing black holes at the same mass, and the fact that the extremal mass changes when we include EFT corrections \cite{Reall:2019sah}. If we compared black holes at the same temperature, $\Delta S$ does not diverge.
In Fig.~\ref{fig:thermodynamics}, we plot $\Delta S$ for several different EFT terms for fixed $\mathfrak{a}=\mathfrak{q}$. The solid curves represent the analytic predictions of Ref.~\cite{Cheung:2019cwi}, while the colored disks represent our numerical solutions for EFT corrections proportional to $d_6$, $d_7$, and $d_8$. The agreement between the numerical data and the analytic curves provides one of the most stringent tests of our numerical solutions. 
\subsection{Tidal forces\label{sec:tidal}}
Let us now turn to the physical question of the tidal forces on ingoing light rays in the EFT-corrected spacetime.
There are several approaches to computing tidal forces associated with ingoing null geodesics. One could  (numerically) switch to Bondi-Sachs-type coordinates and compute the associated tidal forces there, or compute ingoing null geodesics in our numerically determined spacetimes. Our approach will be based on the latter technique, but we will restrict ourselves to certain totally geodesic submanifolds to make our calculation more tractable. In particular, it is a simple exercise to show that any geodesic that starts at $X=0$ (i.e., $\theta = \pi/2$) with a tangent vector in the equatorial plane  will remain within this plane. In what follows, we will therefore consider ingoing null geodesics restricted to the equatorial plane. 

We start with the Lagrangian for null geodesics,
\begin{equation}
L=\dot{x}^a\dot{x}^b g_{ab}\,,\label{eq:nullL}
\end{equation}
with $\dot{x}^a=\{\dot{t}(\lambda),\dot{r}(\lambda),\dot{X}(\lambda),\dot{\phi}(\lambda)\}^a$ and $\lambda$ an affine parameter. Next, we again note that $\partial/\partial t$ and $\partial/\partial \phi$ are commuting Killing vector fields in our spacetime. As such, we have two conserved quantities,
\begin{equation}
E_n\equiv -\frac{1}{2}\frac{\partial L}{\partial \dot{t}}\quad\text{and}\quad L_n \equiv \frac{1}{2}\frac{\partial L}{\partial \dot{\phi}}\,.
\end{equation}
Note that $\lambda$ is only defined up to affine reparameterizations, so that $E_n$ and $L_n$ have no physical meaning, but their ratio $L_n/E_n$ is invariant under affine transformations and can be interpreted as the impact parameter of the ingoing geodesic.

Next, we focus on  geodesics with $L_n=0$ and adjust the affine parameter so that, without loss of generality, $E_n=1$. Furthermore, we take $X(\lambda)=0$, which we can show is consistent with the geodesic equation, as expected. From the definition of $E_n$ we can read off $\dot{t}$, and from $L_n=0$ (for a null geodesic) we determine $\dot{r}$ up to an overall sign. To decide which sign to take, we introduce coordinates in which our line element and gauge potential (\ref{eqs:ansat}) are regular on the horizon, namely,
\begin{equation}
\begin{aligned}
&{\rm d}t={\rm d}v-\frac{{\rm d}r}{\Delta(r)}(r_+^2+\bar{a}^2)\left(1-\frac{1}{2 r_+^2}\sum_{K=6}^{8} \lambda^K d_K\right)\,,
\\
&{\rm d}\phi={\rm d}\varphi-\bar{a}\frac{{\rm d}r}{\Delta(r)}\left(1-\frac{1}{2 r_+^2}\sum_{K=6}^{8} \lambda^K d_K\right)\,.
\end{aligned}
\end{equation}
Since constant-$v$ hypersurfaces require ${\rm d}r/{\rm d}t<0$, the above are ingoing coordinates. It is in these coordinates that we want the ingoing geodesics to be regular. This fixes the overall sign of $\dot{r}$, and one finds
\begin{equation}
\label{eq:dotxa}
\dot{x}^a=\Theta(r)\left[
{\renewcommand{\arraystretch}{1.8}
\begin{array}{c}
\frac{\left(\bar{a}^2+r^2\right)^2 f_3(r,0)-\Delta (r) f_1(r,0) f_4(r,0){}^2}{\bar{a}^2+r^2-f_4(r,0){}^2}
\\
-\frac{\Delta (r) \sqrt{f_1(r,0) f_3(r,0)} \sqrt{\left(\bar{a}^2+r^2\right)^2 f_3(r,0)-\Delta (r) f_1(r,0) f_4(r,0){}^2}}{r^2 \sqrt{f_2(r,0)}}
\\
0
\\
\frac{\left(\bar{a}^2+r^2\right) f_3(r,0)-\Delta (r) f_1(r,0)}{\bar{a}^2+r^2-f_4(r,0){}^2} f_4(r,0)
\end{array}}
\right]^a\,,
\end{equation}
where
\begin{equation}
\Theta(r)\equiv\frac{r^2}{\Delta (r) f_1(r,0) f_3(r,0) \left[r^2+\bar{a}^2-f_4(r,0){}^2\right]}\,.
\end{equation}
Note also that our choice of $E_n=1$ makes $\dot{x}^a$ future-directed. 

With the above expressions, we can compute the quantity
\begin{equation}
C_{\varphi \varphi}\equiv \dot{x}^a \dot{x}^b C_{\varphi a \varphi b}\label{eq:Cphiphi}
\end{equation}
on the EFT-corrected black hole horizon, as a measure of the tidal forces experienced by an ingoing congruence of null geodesics as it crosses the future event horizon. From Eq.~(\ref{eq:tidalfinal}), we expect the tidal forces to be largest at $X=0$ (the black hole equatorial plane). As such, we will monitor $C_{\varphi \varphi}$ evaluated on the black hole event horizon and at $X=0$. We will be interested in comparing the tidal forces experienced by an ingoing congruence of null geodesics crossing an EFT-corrected Kerr-Newman black hole with those crossing a standard Kerr-Newman black hole with the same temperature, electric charge, and angular momentum. We thus define  
\begin{equation}
\delta C^{(K)}\equiv \frac{Q^2}{d_K}\frac{C^{\mathcal{H}}_{\varphi \varphi}-\bar{C}^{\mathcal{H}}_{\varphi \varphi}}{\bar{C}^{\mathcal{H}}_{\varphi \varphi }}\,,
\label{eq:deltaCK}
\end{equation}
where the superscript $\mathcal{H}$ denotes evaluation on the horizon and at $X=0$, and $\bar{C}^{\mathcal{H}}_{\varphi \varphi}$ is computed for a standard Kerr-Newman black hole (with the same temperature, electric charge, and angular momentum  as the EFT-corrected one). The superscript ${}^{(K)}$ reminds us that we should compute $C^{\mathcal{H}}_{\varphi \varphi}$ for each of the EFT terms under consideration. 

\begin{figure}[ht]
    \centering
    \includegraphics[width=\textwidth]{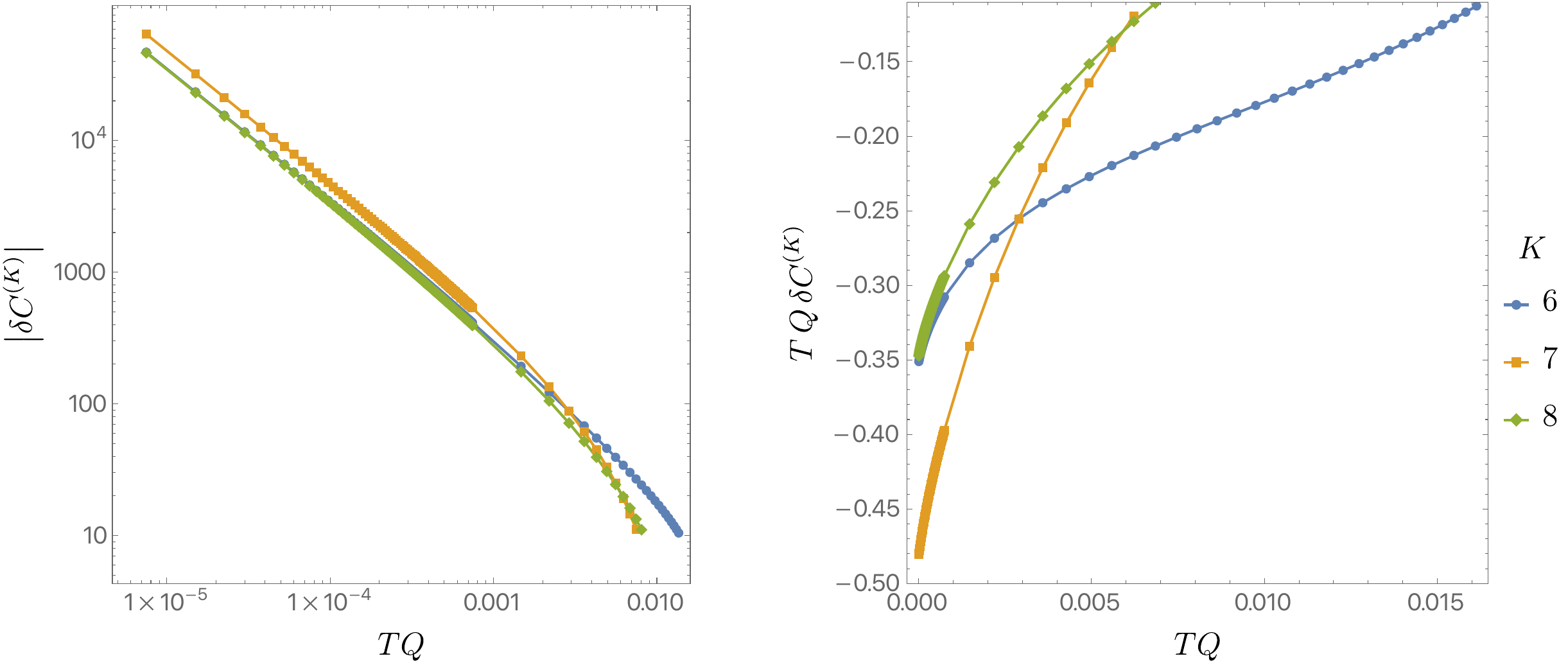}
    \caption{The (rescaled) change in the tidal force  $\delta C^{(K)}$ defined in Eq.~(\ref{eq:deltaCK}), as a function of $TQ$ and plotted for the parameter choice $\mathfrak{a}=\mathfrak{q}$. For a given value of $TQ$, there are two Kerr-Newman black holes, and we only show the family for which $TQ\to 0$ because $T \to 0$ (not $Q\to 0$). Other ways to approach extremality exhibit qualitatively similar behavior. Our numerical results are given by the filled points, with the thin curves present to guide the eye. The right-hand panel shows $TQ\,\delta C^{(K)}$, which approaches a constant as $T\rightarrow 0$, confirming the predicted $1/T$ scaling.}
    \label{fig:tidal}
\end{figure}

 To present our results, we choose a one parameter family of black holes that approach extremality. In Fig.~\ref{fig:tidal}, we plot $\delta C^{(K)}$ against $T Q$ for  $\mathfrak{a}=\mathfrak{q}$.  (Other choices produce similar results.) Note that for fixed $T Q$ there are two Kerr-Newman black holes (for instance, both Schwarzchild and extreme Kerr-Newmman black holes have $T Q=0$). We are interested in the family of solutions for which $T Q\to 0$ because $T\to 0$, and in Fig.~\ref{fig:tidal} we only plot this family. The fact that $TQ \,\delta C^{(K)}$ approaches a constant value at low temperatures shows that the tidal forces are  diverging as $1/T$. This behavior is consistent with the near-horizon analysis in Eq.~(\ref{eq:tidalfinal}), since the scaling symmetry of the near-horizon geometry allows one to relate scaling with $\rho$ in the extremal solution to scaling with $T$ in the near-extremal solution~\cite{Horowitz:2022mly}. We can also confirm the
 prediction made in Eq.~(\ref{eq:tidalfinal}) directly, by  constructing the EFT-corrected black hole at extremality. We discuss this calculation in the next section. 

\vspace{5mm}

\section{Extremal solutions}\label{sec:extremal}
Much of the numerical construction of Sec.~\ref{sec:finite_temp} goes through to the extremal case, but there are a few important differences that we now outline.

We generalize our metric ansatz to
\begin{equation}
\begin{aligned}
\mathrm{d}s^2 =& -\frac{\Delta(r)}{\Sigma(r,X)}F_1(r,X)\left[\mathrm{d}t-(1-X^2)F_4(r,X)\mathrm{d}\phi\right]^2
\\
&+\frac{1-X^2}{\Sigma(r,X)}F_3(r,X)\Xi(r)\left[F_4(r,X)\,\mathrm{d}t-(r^2+\bar{a}^2)\mathrm{d}\phi\right]^2\\&+\Sigma(r,X)F_2(r,X)\left[\frac{\mathrm{d}r^2}{\Delta(r)}+\Xi(r)\frac{\mathrm{d}X^2}{1-X^2}\right]\,,
\label{eq:kerrmodextre}
\end{aligned}
\end{equation}
with $\bar{a}=\sqrt{\bar{M}^2-\bar{Q}^2}$. Note that we have added a function $\Xi(r)$  with respect to Eq.~(\ref{eqs:ansat}). We choose
\begin{equation}
\Xi(r)=1+\sum_{K=6}^{8}\frac{r_+^4}{r^4}\frac{d_K}{M^2}a^{(K)}\,,
\end{equation}
with $a^{(K)}$ being real numbers to be determined in what follows. The function $\Xi(r)$ is chosen so that at asymptotic infinity $r\rightarrow\infty$, we have $\Xi\rightarrow 1$. As in the nonextremal case, we write
\begin{equation}
\begin{aligned}
\label{eqs:expan}
F_{i}(r,X) &= 1+\sum_{K=6}^{8}\frac{d_K}{M^2}f_i^{(K)}(r,X)\,,\quad i=1,2,3\\
F_{4}(r,X) &= \bar{a}+\left(1-\frac{r_+}{r}\right)\sum_{K=6}^{8}\frac{d_K}{M^2}f_4^{(K)}(r,X)\\
F_{5}(r,X) &= \bar{Q}+\left(1-\frac{r_+}{r}\right)\sum_{K=6}^{8}\frac{d_K}{M^2}f_5^{(K)}(r,X)\\
F_{6}(r,X) &= \sum_{K=6}^{8}\frac{d_K}{M^2}f_6^{(K)}(r,X)\,.
\end{aligned}
\end{equation}
As before, the factors of $(1-r_+/r)$ in the definitions of $F_4(r,X)$ and $F_5(r,X)$ ensure that the solutions we find have the same angular velocity and chemical potential as the extremal Kerr-Newman black hole. Again, we work with a compact coordinate $Y$, defined as
\begin{equation}
r=\frac{r_+}{1-Y^2}\,,
\label{eq:rextreme}
\end{equation}
with $Y=0$ being the extremal horizon and $Y=1$ asymptotic infinity. Note that this is slightly different from the coordinate choice defined in Eq.~\eqref{eq:num}.

The boundary conditions at the axis of symmetry, located at $X=\pm1$, and asymptotic infinity are unchanged. In particular, at asymptotic infinity the angular momentum and total charge are not fixed.
We now come to the issue of boundary conditions at the extremal horizon. A careful Frobenius analysis reveals that
\begin{equation}
\left.\frac{\partial f_{i}^{(K)}}{\partial Y}\right|_{Y=0}=0\,.
\end{equation}
None of the boundary conditions so far has enforced $C_2^{(K)}=0$. It turns out for the extremal case it is paramount to impose this condition on the event horizon (since we cannot fix the energy or angular momentum at infinity when we are trying to fix the temperature and the horizon location). This in turn imposes
\begin{equation}
f_{4}^{(K)}(X,0)=\Omega^{(K)}_0\,,\quad f_{5}^{(K)}(X,0)=\frac{2 \mathfrak{q} (2-\mathfrak{q}^2)}{\sqrt{1-\mathfrak{q}^2} \left[1+(1-\mathfrak{q}^2) X^2\right]}\Omega^{(K)}_0+\mu^{(K)}_0\,,
\end{equation}
and
\begin{equation}
f_{1}^{(K)}(X,0)=A^{(K)}+f_{2}^{(K)}(X,0)\,,
\end{equation}
with $A^{(K)}$, $\Omega^{(K)}_0$, and $\mu^{(K)}_0$ being constants. In particular, we need to determine $A^{(K)}$ to be able to give a Dirichlet-type boundary condition for $f_{1}^{(K)}(X,0)$, as well as all the $a_{(K)}$.

On the horizon, we find three ordinary differential equations in $f_{2}^{(K)}(X,0)$, $f_{3}^{(K)}(X,0)$, and $f_{6}^{(K)}(X,0)$. These take a fairly complicated form, which we refrain from presenting here. However, if we further change variables via $f_{2}^{(K)}(X,0)=\alpha_1^{(K)}+\alpha_2^{(K)}$ and $f_{3}^{(K)}(X,0)=\alpha_1^{(K)}-\alpha_2^{(K)}$, the equation for $\alpha_1^{(K)}$ takes a rather beautiful form,
\begin{equation}
\frac{\mathrm{d}}{\mathrm{d}X}\left[(1-X^2)^{3/2}\frac{\mathrm{d}\alpha_1^{(K)}(X)}{\mathrm{d}X}\right]=S_{K}(X;a^{(K)})\,,\label{eq:exta}
\end{equation}
where the source $S_{K}(X;a^{(K)})$ depends explicitly on which higher-derivative correction we are considering. Smooth solutions of Eq.~(\ref{eq:exta}) will only exist if
\begin{equation}
\int_{-1}^1\mathrm{d}X\,S(X;a^{(K)})=0\,,
\end{equation}
since we can integrate either side of Eq.~(\ref{eq:exta}), and the left-hand side is zero so long as the $\alpha^{(K)}(X)$ are finite at $X=\pm1$. This in turn determines all of the $a^{(K)}$ to be
\begin{equation}   
\begin{aligned}
\label{eqs:as}
a^{(6)}&=\frac{2\mathfrak{q}^2}{(2-\mathfrak{q}^2)^{7/2}} \left(6 \mathfrak{q}^8-40 \mathfrak{q}^6+94 \mathfrak{q}^4-85 \mathfrak{q}^2+21\right)\\
a^{(7)}&=\frac{4 \mathfrak{q}^4}{(2-\mathfrak{q}^2)^{11/2}} \left(3 \mathfrak{q}^{10}-31 \mathfrak{q}^8+128 \mathfrak{q}^6-268 \mathfrak{q}^4+267 \mathfrak{q}^2-107\right)\\
a^{(8)}&=\frac{\mathfrak{q}^4}{(2-\mathfrak{q}^2)^{11/2}}\left(9 \mathfrak{q}^{10}-93 \mathfrak{q}^8+384 \mathfrak{q}^6-796 \mathfrak{q}^4+811 \mathfrak{q}^2-331\right)\,.
\end{aligned}
\end{equation}

To determine $A^{(K)}$, we need to solve the equations on the extremal horizon. Remarkably, we were able to analytically solve for $\alpha_1^{(K)}(X)$, $\alpha_2^{(K)}(X)$, and $f_6^{(K)}(X,0)$ for all of the higher-derivative corrections. The expressions for $A^{(K)}$ are not particularly illuminating, so we will not present them here. Once the dust settles, the new near-horizon geometries become functions of $\Omega^{(K)}_0$ and $\mu^{(K)}_0$, which one can show control the perturbed angular momentum and charge of the corresponding solution.

\begin{figure}[ht]
    \centering
    \includegraphics[width=0.55\textwidth]{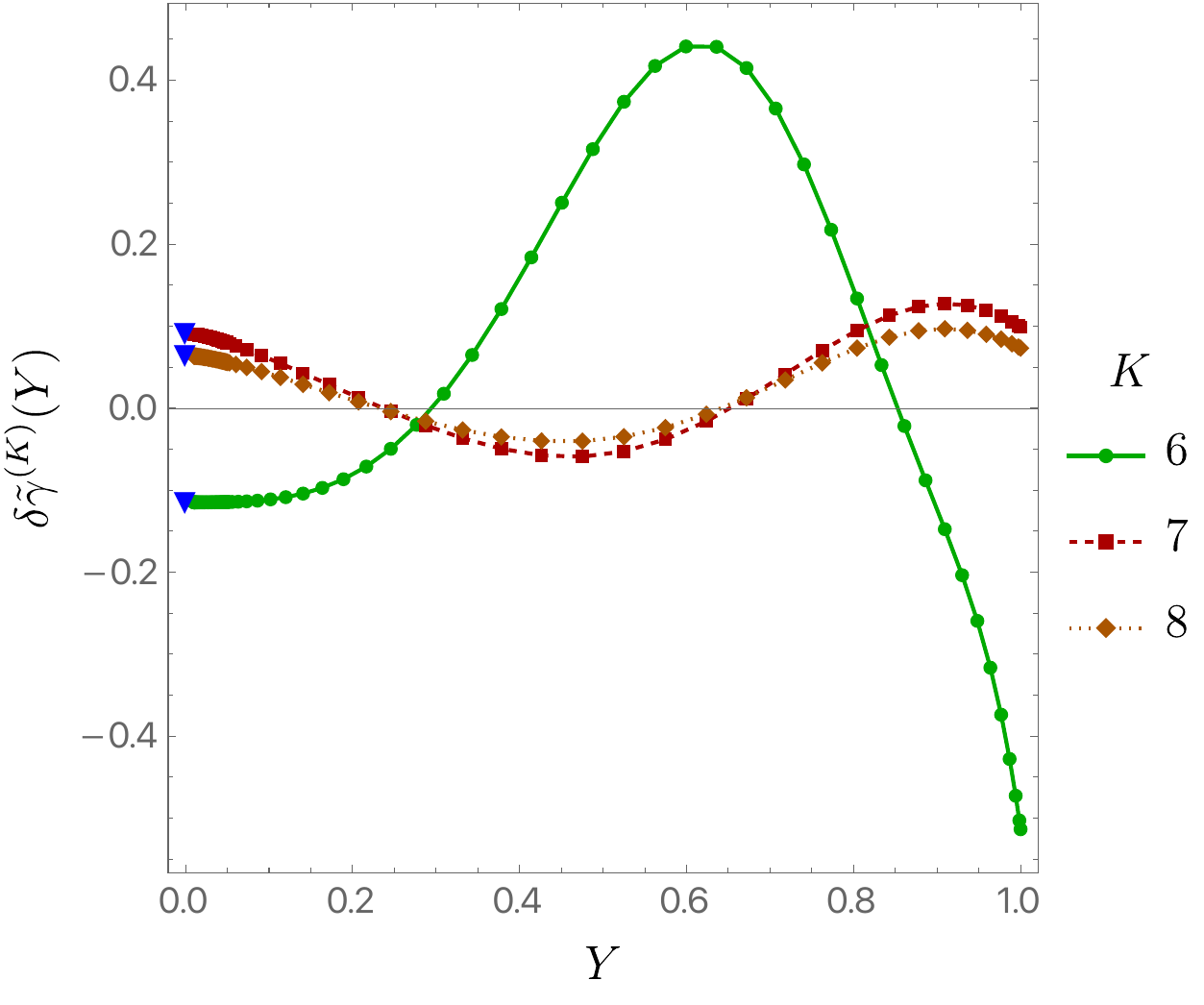}
    \caption{The quantity $\delta\tilde{\gamma}^{(K)}(Y)$ defined in Eq.~(\ref{eq:crazydeltay}) as a function of the radial coordinate $Y$, plotted for $\mathfrak{q}=0.2$.
    The blue inverted triangles are the predictions based on our near-horizon analysis in Eq.~(\ref{eq:deltasgamma}). The agreement at $Y=0$ shows that the near-horizon analysis correctly captures the change in the scaling dimensions of the full extremal solution.}
    \label{fig:scalings}
\end{figure}

The main advantage of constructing the extremal solutions numerically is that we can try to test the predictions of Sec.~\ref{eq:gamma1} in great detail. However, in order to accomplish this task, we have to overcome one last hurdle. In particular, we would like to see how the change in the scaling $\delta \gamma^{(K)}$ enters the near-horizon behavior of our metric functions $f^{(K)}_i$. In order to do this, we have to change from the coordinates used in Sec.~\ref{eq:gamma1} to those used here. It turns out to be a lot easier to transform from $(\rho,x)$ to $(r,X)$ rather than the other way around. One sets 
\begin{equation}
\label{eq:expansionsrhox}
\begin{aligned}
&\rho = \kappa_1(r-r_+)+\kappa_2(X)(r-r_+)^2+\tilde{\kappa}_2(X) (r-r_+)^2\log(r-r_+)+o[(r-r_+)^2]
\\
&x=X+\lambda_1(X)(r-r_+)+\lambda_2(X)(r-r_+)^2+o[(r-r_+)^2]
\end{aligned}
\end{equation}
and demands that the metric ansatz~(\ref{eq:fixed2}) with $\delta f_i(\rho,x)$ given in Eq.~(\ref{eq:crazylogs}) match the line element~(\ref{eq:kerrmodextre}) to linear order in $d_K$ and to order $r-r_+$ away from the extremal horizon. This procedure is rather tedious, but determines $\kappa_2(X)$, $\tilde{\kappa}_2(X)$, $\lambda_1(X)$, and  $\lambda_2(X)$ as functions of $\kappa_1$, $V^{(0)}$, and $Z$. It also reveals that $a^{(K)}=2 \delta \Gamma^{(K)}_{\rm NH}$, which one can check by equating Eq.~(\ref{eqs:deltas}) with Eq.~(\ref{eqs:as}) (note that, at extremality, $\mathfrak{q}=Z$). It also shows that $f_3^{(K)}(r,X)$ admits the following expansion near the extremal horizon,
\begin{equation}
f_3^{(K))}(r,X)=z_0^{(K)}(X)+z_1^{(K)}(X)(r-r_+)+\mathfrak{q}^{(K)}_1(X)(r-r_+)\log(r-r_+)+o[(r-r_+)]\,,
\end{equation}
with
\begin{equation}
\mathfrak{q}^{(K)}_1(X)=-\frac{2 (1-\mathfrak{q}^2) (1-X^2)}{(2-\mathfrak{q}^2) r_+^3 \left[1+(1-\mathfrak{q}^2) X^2\right]}\delta \gamma^{(K)}\,.
\end{equation}
In terms of our $Y$ variables, we can use the above expansion to read off $\delta \gamma^{(K)}$,
\begin{equation}
\delta \gamma^{(K)}=\lim_{Y\to0}\delta \tilde{\gamma}^{(K)}(Y)\,,
\end{equation}
where we have defined
\begin{equation}
\delta\tilde{\gamma}^{(K)}(Y)\equiv\left.-\frac{(2-\mathfrak{q}^2)}{8 (1-\mathfrak{q}^2)}Y \frac{\partial^3 f_3}{\partial Y^3}\right|_{X=0}.
\label{eq:crazydeltay}
\end{equation}
The above quantity can be computed from our numerically determined solutions. If our near-horizon analysis is correct, $\delta\tilde{\gamma}^{(K)}(Y)$ should approach the values given in Eq.~(\ref{eq:deltasgamma}). In Fig.~\ref{fig:scalings}, we plot $\delta\tilde{\gamma}^{(K)}(Y)$ for $\mathfrak{q}=0.2$ for all three Wilson coefficients under consideration. The blue inverted triangles are the predictions based on our near-horizon analysis given in Eq.~(\ref{eq:deltasgamma}). The green disks, red squares, and orange diamonds show $\delta\tilde{\gamma}^{(K)}(Y)$ computed for $K=6,7,8$, respectively. The agreement between $\delta\tilde{\gamma}^{(K)}(0)$ and $\delta \gamma^{(K)}$ given in Eq.~(\ref{eq:deltasgamma}) shows that our near-horizon analysis indeed captures the shift in the scaling dimensions of the full asymptotically flat solution.

{We were also able to compute the tidal forces directly at extremality and observe the expected divergence when $\rho\sim 0$. We first note that if we use Eq.~(\ref{eq:expansionsrhox}) together with Eqs.~(\ref{eq:rextreme}) and (\ref{eq:tidalfinal}), we expect the tidal forces to diverge as $Y^{-2}$ near $Y=0$. One can repeat the same steps as in Sec.~\ref{sec:tidal} to compute the relevant future-directed ingoing null geodesics. In fact, the expression for $\dot{x}^a$ is almost the same as the one given in Eq.~(\ref{eq:dotxa}), but with $\Xi(r)$ also playing a role, namely,
\begin{equation}
\label{eq:dotxaex}
\dot{x}^a=\Theta(r)\left[
{\renewcommand{\arraystretch}{1.8}
\begin{array}{c}
\frac{\left(\bar{a}^2+r^2\right)^2 \Xi(r)f_3(r,0)-\Delta (r) f_1(r,0) f_4(r,0){}^2}{\bar{a}^2+r^2-f_4(r,0){}^2}
\\
-\frac{\Delta (r) \Xi(r) \sqrt{f_1(r,0) f_3(r,0)} \sqrt{\left(\bar{a}^2+r^2\right)^2 f_3(r,0)-\frac{\Delta (r) f_1(r,0) f_4(r,0){}^2}{\Xi(r)}}}{r^2 \sqrt{f_2(r,0)}}
\\
0
\\
\frac{\left(\bar{a}^2+r^2\right)\Xi(r) f_3(r,0)-\Delta (r) f_1(r,0)}{\bar{a}^2+r^2-f_4(r,0){}^2} f_4(r,0)
\end{array}
}
\right]^a\,,
\end{equation}
where
\begin{equation}
\Theta(r)\equiv\frac{r^2}{\Xi(r)\Delta (r) f_1(r,0) f_3(r,0) \left[r^2+\bar{a}^2-f_4(r,0){}^2\right]}\,.
\end{equation}
Given the null geodesics, we can define $C_{\varphi\varphi} \equiv \dot{x}^a \dot{x}^b C_{\varphi a \varphi b} $ as before.

To measure the expected divergent behavior in the tidal forces, we define 
\begin{equation}
\delta \widetilde{C}^{(K)}\equiv \frac{\kappa^2J}{8\pi d_K}\frac{C^{X=0}_{\varphi \varphi}-\bar{C}^{X=0}_{\varphi \varphi}}{\bar{C}^{X=0 }_{\varphi \varphi }}\,,
\label{eq:deltaCtildeK}
\end{equation}
where $\bar{C}^{X=0}_{\varphi \varphi}$ is computed for a standard Kerr-Newman black hole with the same temperature (i.e., both the corrected and uncorrected black holes are at zero temperature), electric charge, and angular momentum as the EFT-corrected black hole.  Note that $\delta \widetilde{C}^{(K)}$ is defined almost identically to Eq.~(\ref{eq:deltaCK}), except that now we do not evaluate the tidal forces on the extremal horizon (but we still restrict to the equatorial plane, $X=0$) and use a different overall normalization.

In Fig.~\ref{fig:tidalextremal},  the left panel shows $\delta \widetilde{C}^{(K)}$ as a function of $Y$ in a log-log plot. The inverse power law behavior in $Y$ is clearly evident from the straight-line trend observed near $Y=0$. To confirm this expectation, in the right panel we plot $Y^2\delta \widetilde{C}^{(K)}$ as a function of $Y$. This quantity clearly approaches a constant value as $Y\to0^+$, showing that $\delta \widetilde{C}^{(K)}$ diverges as $Y^{-2}\propto \rho^{-1}$ as we approach the horizon of the EFT-corrected extremal black hole. Both plots were generated with $\mathfrak{q}=0.2$, but we found qualitatively similar behavior for other values of $\mathfrak{q}$.
\begin{figure}[ht]
    \centering
    \includegraphics[width=\textwidth]{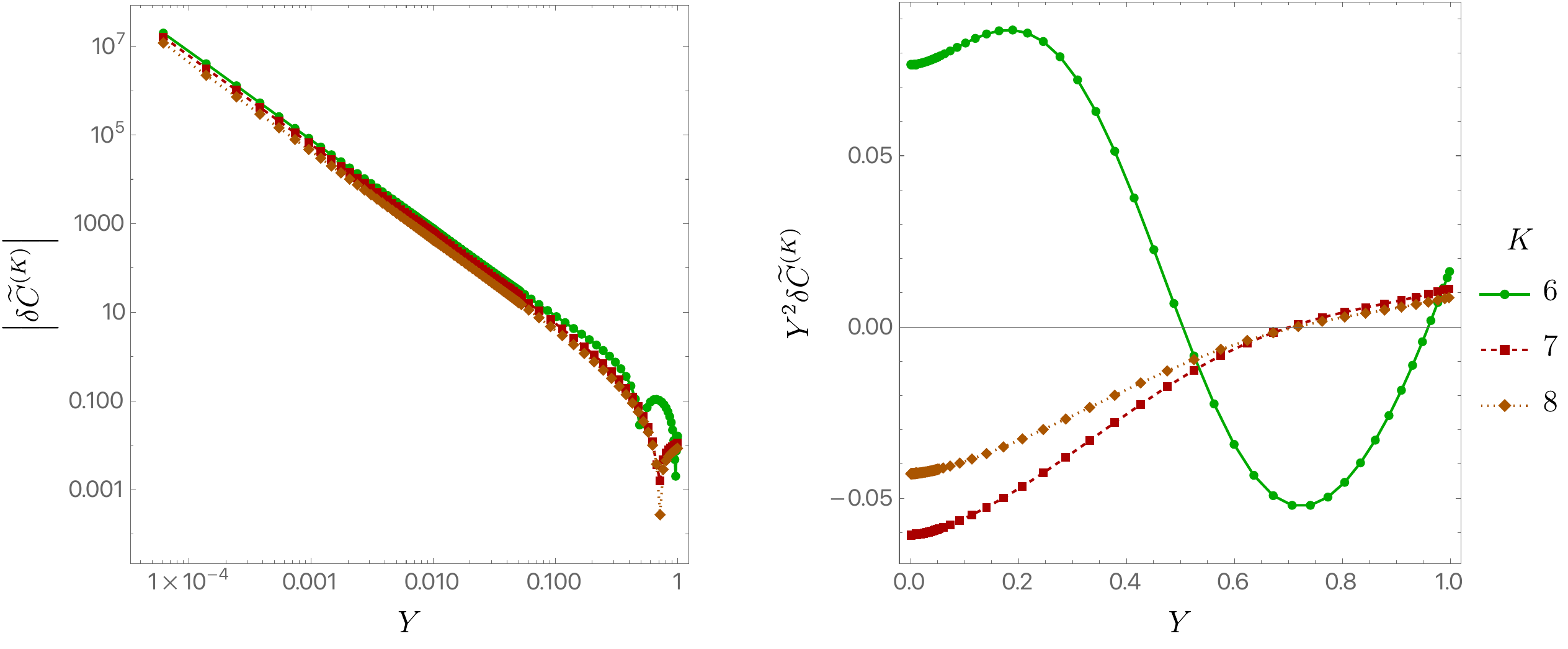}
    \caption{Left panel:  The change in the tidal force $\delta \widetilde{C}^{(K)}$ defined in Eq.~(\ref{eq:deltaCtildeK}), as a function of the radial coordinate $Y$. The linear behavior in this log-log plot shows that it diverges at the extremal horizon $Y=0$. Right panel: The quantity $Y^2\delta \widetilde{C}^{(K)}$ as a function of $Y$. The fact that it approaches a constant confirms the near-horizon result~\eqref{eq:tidalfinal}. Both panels were generated with  $\mathfrak{q}=0.2$.}
    \label{fig:tidalextremal}
\end{figure}

Though the tidal forces diverge, we expect \emph{all} curvature scalar invariants to remain small at the extremal horizon. This is essentially a consequence of the ${\rm O}(2,1)$ symmetry of the near-horizon geometry and the fact that the equations of motion are given in terms of a second-rank tensor~\cite{Hadar:2017ven}. This expectation is validated by our numerical data. To see this, we define the following auxiliary quantity,
\begin{equation}
\delta \mathcal{R}^{(K)}\equiv \frac{\kappa^2J}{8\pi d_K}\frac{\mathcal{R}^{X=0}-\bar{\mathcal{R}}^{X=0}}{\bar{\mathcal{R}}^{X=0}}\quad \text{with}\quad \mathcal{R}=R_{abcd}R^{abcd}\quad\text{and}\quad \bar{\mathcal{R}}=\bar{R}_{abcd}\bar{R}^{abcd},
\label{eq:deltaRK}
\end{equation}
where $\bar{R}_{abcd}$ is computed as before for a standard Kerr-Newman black hole with the same (zero) temperature, electric charge, and angular momentum as the black hole with EFT corrections.  In Fig.~\ref{fig:curvature}, we plot $\delta \mathcal{R}^{(K)}$ for $K=6,7,8$ as a function of the radial coordinate $Y$ and find that it remains finite at the extremal horizon, as expected. This figure was generated for $\mathfrak{q}=0.2$, but we have checked that similar behavior occurs for other values of $\mathfrak{q}$.
\begin{figure}[ht]
    \centering
    \includegraphics[width=0.55\textwidth]{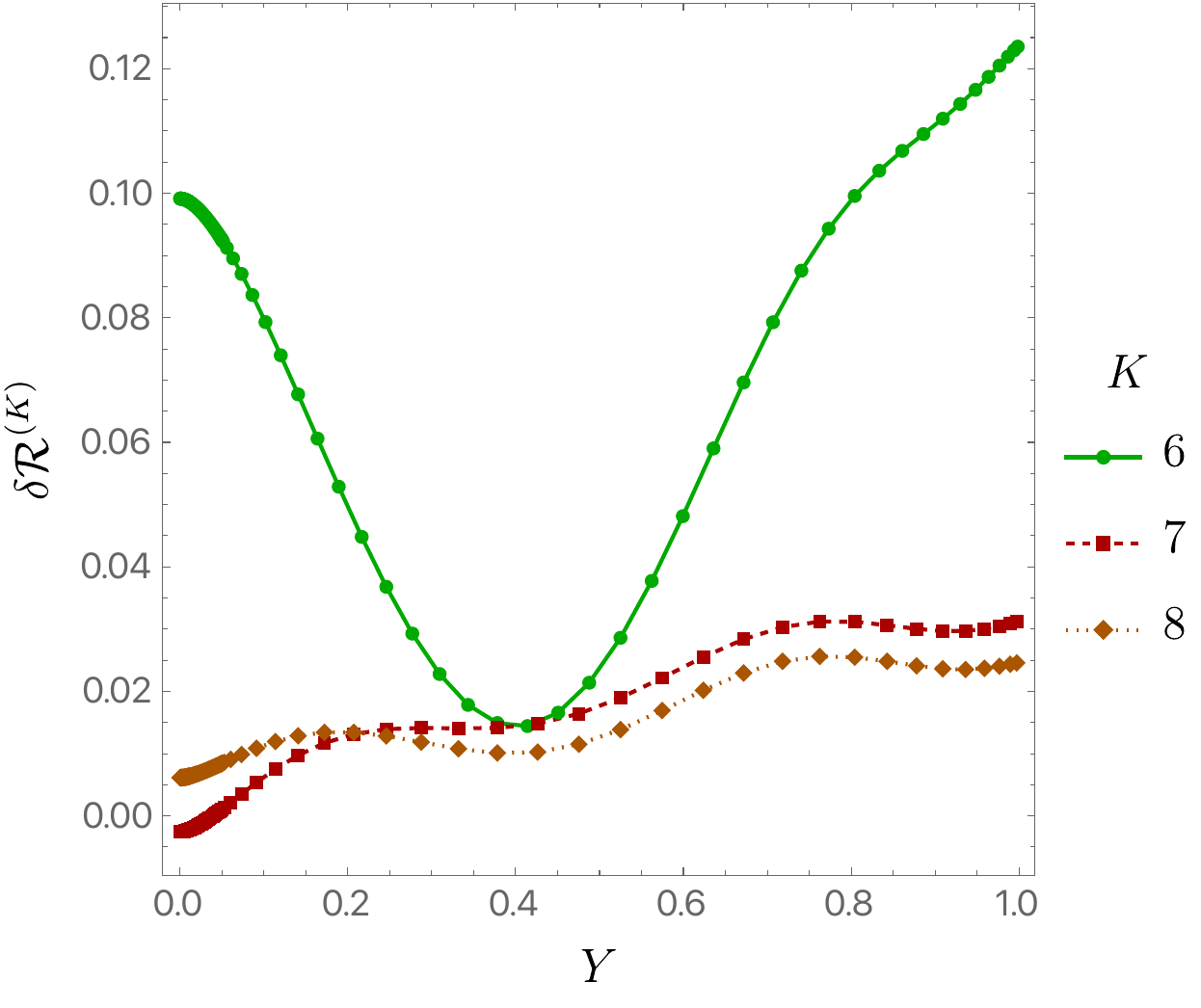}
    \caption{The change in the square of the Riemann tensor $\delta \mathcal{R}^{(K)}$ defined in Eq.~(\ref{eq:deltaRK}), as a function of the radial coordinate $Y$, for $\mathfrak{q}=0.2$. Note that this quantity and all other scalar invariants remain finite at the horizon.}
    \label{fig:curvature}
\end{figure}}

\section{Numerical estimates for astrophysical black holes}\label{sec:numerics}
With the numerical solutions from Sec. \ref{sec:finite_temp} in hand for nonextremal black holes, we can ask whether there is an astrophysical scenario where these quantum corrections---that is, higher-derivative terms in the Einstein-Maxwell action---become important for realistic black holes.
The rough estimates in Sec.~\ref{sec:rough} indicate that such a scenario may be rare in nature.
While numerical analysis bears this conclusion out, we will find hints that, under optimistic assumptions, detection of the enhanced near-horizon effects near extremality may be conceivable.

\begin{figure}[t]
    \centering \hspace{-3mm}   \includegraphics[height=7cm]{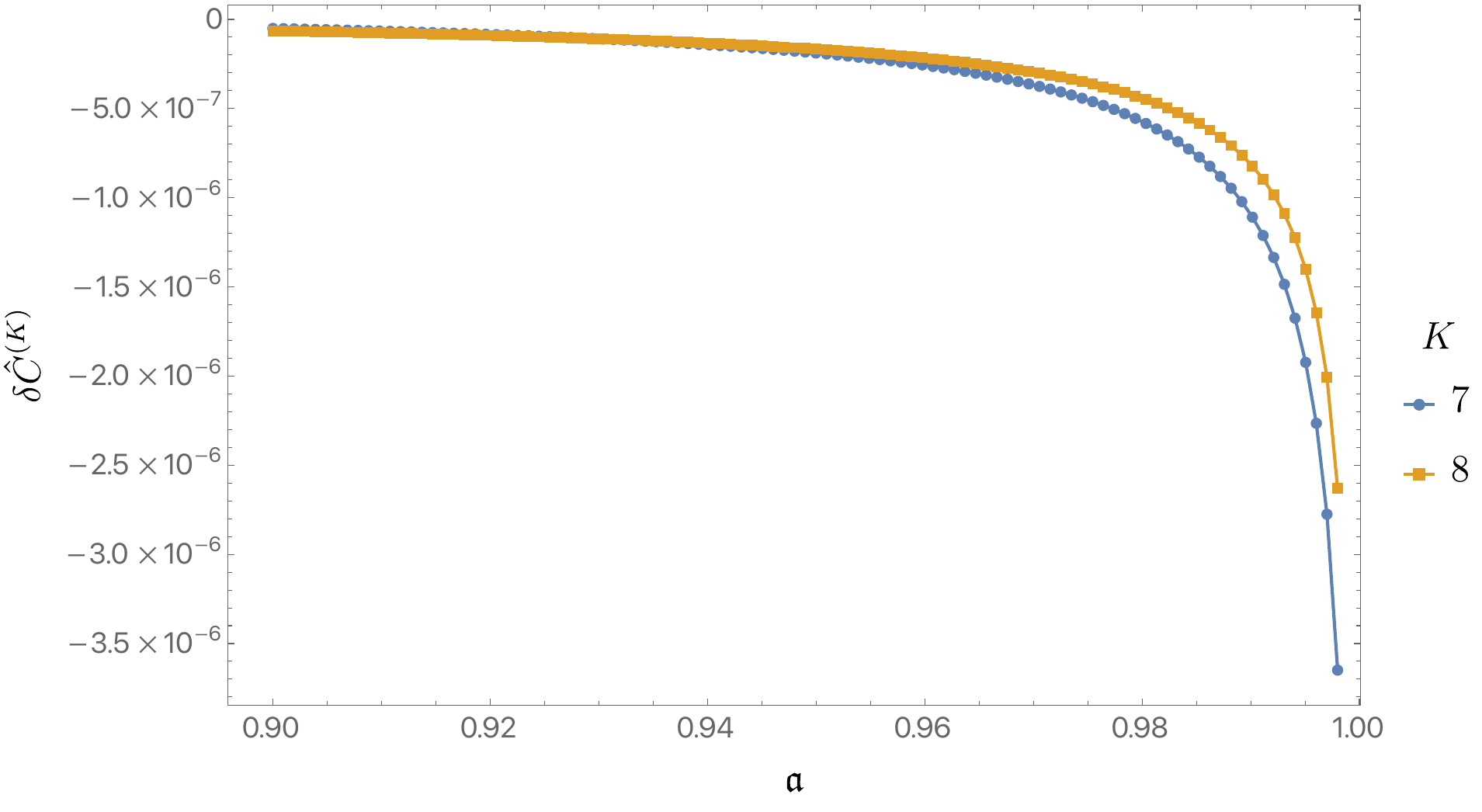}\\  \hspace{3.9mm} \includegraphics[height=7.09cm]{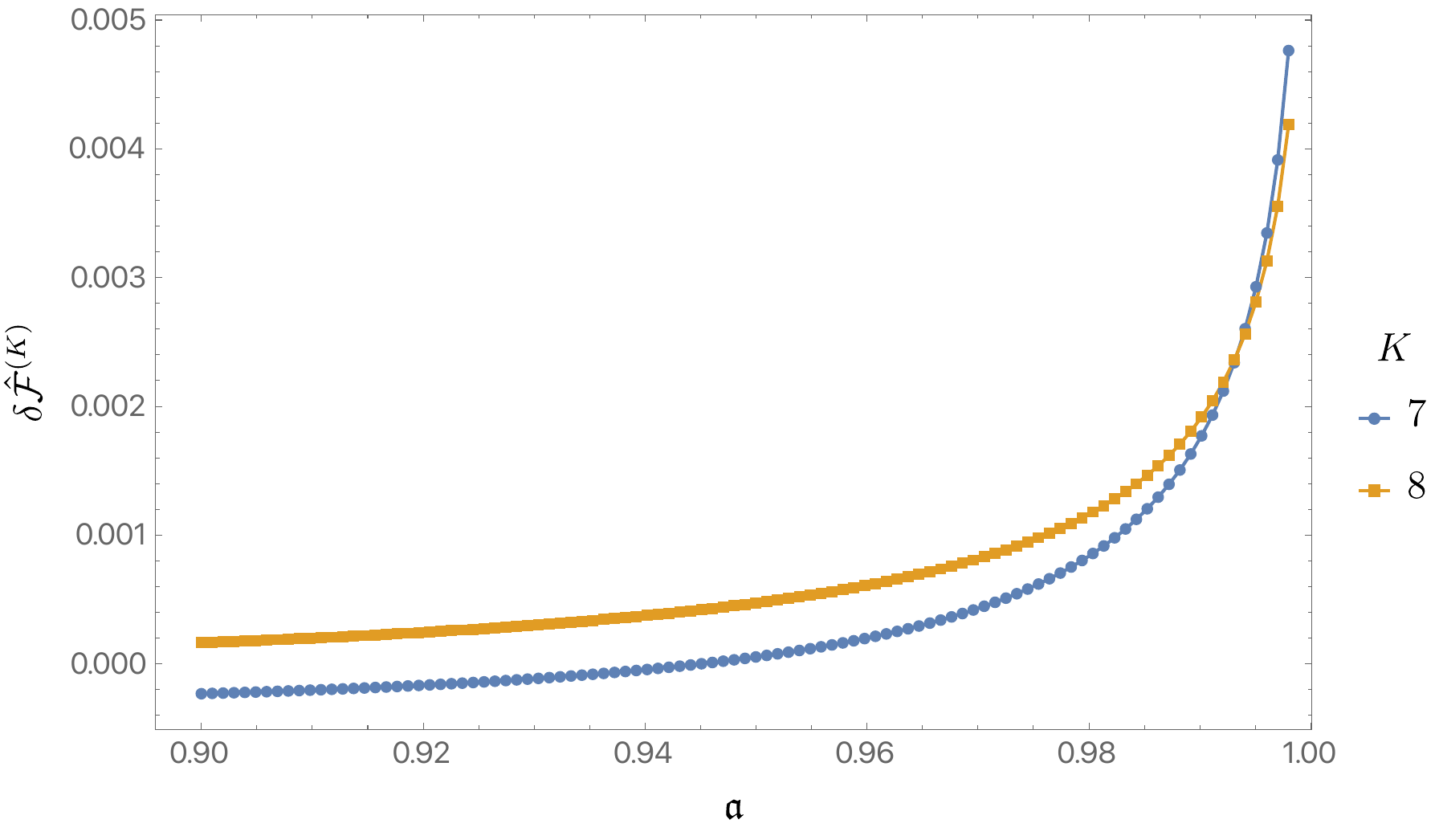}
    \caption{Parameters $\delta \hat C^{(K)}$ and $\delta \hat {\cal F}^{(K)}$ defined in Eqs.~\eqref{eq:deltaChat} and \eqref{eq:deltaFhat} measuring the change in tidal force and gauge field strength between the EFT-corrected black hole and a Kerr-Newman black hole of the same mass, charge, and angular momentum. The EFT corrections clearly increase near extremality. The numerical results shown here were computed for a charge parameter $\mathfrak{q}=0.01$.}
    \label{fig:hats}
\end{figure}

Taking the $\mathfrak{a}$ parameter to range from $0.9$ (the typical scale of observed high-spin black holes) to $0.998$ (the limit predicted by Thorne~\cite{Thorne}, where spin-up from a hot accretion disk is balanced by torque from thermal radiation), our numerical results give 
\begin{equation}
\delta \hat{C}^{(K)}\equiv \frac{\kappa^2J}{8\pi d_K}\frac{C^{\mathcal {H}}_{\varphi \varphi}-\bar{C}^{\mathcal {H}}_{\varphi \varphi}}{\bar{C}^{\mathcal {H}}_{\varphi \varphi }}  \sim (5 \;\;{\rm to}\;\; 300)\times \mathfrak{q}^4\,,\label{eq:deltaChat}
\end{equation}
for $K=7,8$, which we checked for $\mathfrak{q}=10^{-2}$ and $10^{-1}$; see Fig.~\ref{fig:hats}  for an illustration.
Here, $\delta \hat C^{(K)}$ is defined as the difference in tidal force---with and without the EFT corrections---for black holes with the angular momentum, charge, and {\it mass} held fixed, not the temperature, distinguishing it from $\delta C^{(K)}$ in Eq.~\eqref{eq:deltaCK} and $\delta\widetilde{C}^{(K)}$ in Eq.~\eqref{eq:deltaCtildeK}.
(We do not include $K =6$, since the effect of the $d_6$ term is suppressed in the standard model by additional factors of the electron charge-to-mass ratio per Eq.~\eqref{eq:cvalue}.)
That is, one-loop effects from the standard model in the $d_{7,8}$ terms yield a fractional deviation from the naive tidal force at the horizon going like
\begin{equation}
\frac{C_{\varphi\varphi}^{{\cal H}} - \bar C_{\varphi\varphi}^{{\cal H}}}{\bar C_{\varphi\varphi}^{{\cal H}}}\sim  (10^5\;\;{\rm to}\;\;10^{7}) \times \left(\frac{10\,M_\odot}{M}\right)^2\times \mathfrak{q}^4.
\end{equation}
In order to find the best-case scenario for an observable effect, we must now ask what the largest realistic astrophysical charge can be.

The estimate in Eq.~\eqref{eq:bestcase} above from Ref.~\cite{Levin:2018mzg} for the charge induced on a black hole in a magnetic field from the Wald effect---which in fact represents an upper limit on the accreted charge~\cite{Komissarov:2021vks}---used typical pulsar parameters: a magnetic field of $B \sim 10^{12}\,{\rm G}$ and a neutron star about to merge with the horizon a $10\,M_\odot$ black hole.
Assuming a $10\,{\rm km}$ radius for our neutron star and taking a slightly lighter black hole of the same size $M \approx 7\;M_{\odot}$,  we find that the deviation from general relativity that goes like
\begin{equation}
\frac{C_{\varphi\varphi}^{{\cal H}} - \bar C_{\varphi\varphi}^{{\cal H}}}{\bar C_{\varphi\varphi}^{{\cal H}}}\sim  (10^{-8}\;\;{\rm to}\;\;10^{-6}) \times (B/10^{16}\;{\rm G})^4.
\end{equation}
Magnetars are known to sustain magnetic fields in the range of $10^{15}\,{\rm G}$~\cite{Kaspi:2017fwg}, and field strengths as high as $10^{16}\,{\rm G}$ are believed possible in certain extreme scenarios~\cite{Raynaud:2020ist}.
In this case, the induced charge-to-mass ratio of the black hole is on the order of $10^{-4}$, and the electric field at the horizon is $\sim 10^{19}\,{\rm V/m}$.
Of course, in this case the electric field is above the Schwinger limit at which the electrovacuum breaks down, and the magnetic field dramatically exceeds it (i.e., both are larger than $m_e^2$), though this may be somewhat ameliorated by loop factors.\footnote{Strictly speaking, only the electric field ionizes the vacuum. In purely magnetic backgrounds, as long as $\nabla B/B \ll m_e$ so that the magnetic field is effectively constant, the full one-loop Euler-Heisenberg Lagrangian~\cite{Huet:2011kd} can be resummed at all orders in $F$; see, e.g., App. C of Ref.~\cite{Arkani-Hamed:2021ajd}.}
In any case, we see that even in physically realizable situations in which the vacuum is ionized, the relative deviation of the tidal force compared to general relativity remains small, unless the spin of the black hole is taken unphysically large, above the Thorne limit.
It would be interesting to see if such small effects could be observable with LIGO or future gravitational wave detectors.

However, we can also investigate electromagnetic effects of the singular extremal limit, though the observable consequences in astrophysical scenarios may be less immediately clear.
In particular, our numerical analysis yields 
\begin{equation}
\delta\hat{\cal F}^{(K)} \equiv \frac{\kappa^2 J}{8\pi d_K}\frac{\mathcal{F}-\bar{\mathcal{F}}}{\bar{\mathcal{F}}} \sim (2\;\;{\rm to}\;\; 50)\times \mathfrak{q}^2\,, \label{eq:deltaFhat}
\end{equation}
as shown in Fig.~\ref{fig:hats}, for spin parameters $\mathfrak{a}\in(0.9,0.998)$ for $K=7,8$, where $\mathcal{F}\equiv \dot{x}^a F_{a \varphi}$, and as in Eq.~\eqref{eq:deltaChat} we are comparing two black holes with all of the same asymptotic charges (mass, angular momentum, and electric charge).
That is, for the same parameters as above, we find percent-level deviations from the expected Kerr-Newman background,
\begin{equation}
\frac{\mathcal{F}-\bar{\mathcal{F}}}{\bar{\mathcal{F}}} \gtrsim \text{few percent.}
\end{equation}
More moderate choices of the external magnetic field will still yield appreciable results, due to the weaker $\mathfrak{q}$-scaling of $\delta \hat {\cal F}^{(K)}$ compared to $\delta \hat C^{(K)}$.
Importantly, the observed effect as $\mathfrak{a}\rightarrow 1$ is larger than what one would expect from power counting alone. 
The equations of motion~\eqref{eqs:EOM} go like $\nabla {\cal F} \sim 8 c_K \nabla {\cal \bar F}^3$, so we expect differences from the Coulomb-like solution going like ${\cal F}-{\cal \bar F} \sim 8c_K \bar F^3$, which for $O(1)$ values of $\mathfrak{a}$ and a background solution going like $Q_e/4\pi r^2$ gives us $\delta{\cal F}^{(K)} \sim 16 \mathfrak{q}^2$ at the horizon.
While this is the same order of magnitude as seen in Eq.~\eqref{eq:deltaFhat}, the upward trajectory as $\mathfrak{a}$ increases---apparent in Fig.~\ref{fig:hats}---is indicative of the singularity in the $\mathfrak{a}\rightarrow 1$ limit (and even at the Thorne-limited value of $\mathfrak{a}=0.998$, $\delta F^{(K)}$ is already a factor of a few above EFT expectations).

\section{Discussion} \label{sec:Discussion}

We have seen that the leading effective field theory corrections to the Einstein-Maxwell equations produce tidal force singularities on the horizon of extreme Kerr-Newman black holes. These singularities are stronger than the ones that were found earlier for extreme Kerr. 

At first sight, the fact that the leading EFT corrections produce singularities suggests that the derivative expansion in our classical calculation must be breaking down, and that terms of yet higher order in derivatives will be important to the black hole solution. However, this is not the case. Our singularities have the property that all scalar curvature invariants remain small. The curvature only becomes large along the null horizon.
Intuitively, the correction $\delta g$ to the Kerr-Newman metric scales like $\rho^\gamma$ where $\gamma$ is slightly different from one. Two $\rho$ derivatives produce a curvature that diverges like $\rho^{\gamma -2}$, and one might think that more $\rho$ derivatives from higher-derivative terms will produce stronger singularities. But the equation of motion is a second-rank tensor, so all but two $\rho$ derivatives will have to be contracted with $g^{\rho\rho} \sim\rho^2$ and not produce a faster divergence. Since higher-derivative terms in the EFT appear with smaller coefficients, they will only produce small corrections to the singular solution we have found.
In other words, both the spacetime Lagrangian density and classical equation of motion have derivative expansions that remain under control, the singularity notwithstanding.

However, despite this classical control in the equation of motion and bulk Lagrangian, the singularity does indeed signal that the EFT is breaking down. We can see this, for example, in the worldline effective action of an infalling observer, the timelike analogue of Eq.~\eqref{eq:nullL}.
In the point-particle limit, the worldline action is just the proper time, but finite-size effects associated with tidal forces (the so-called {\it Love numbers}) are parameterized by higher-derivative corrections to this action in the form of Lorentz scalars involving the Weyl tensor~\cite{Cheung:2020sdj}, where components can be dotted into the observer's four-velocity (e.g., the square of Eq.~\eqref{eq:Cphiphi} or \eqref{eq:NP}), which can diverge. 
Terms of yet higher order in derivatives in the worldline EFT, e.g., higher powers of these divergent operators, will diverge more strongly, and the derivative expansion in the worldline EFT will break down, even though that of the bulk Lagrangian does not.
That is, any time an observable diverges, there is some operator in the EFT of some observer that also diverges, and hence some observer whose worldline EFT is breaking, even if the bulk action is healthy. 
This accords with the definition of a Wilsonian EFT within quantum field theory, in which the only modes within the Hilbert space of the EFT are those field configurations with significant support in Fourier space only below the cutoff.
While our classical calculation itself is robust under adding classical terms of yet higher order in derivatives than we have included, it remains the case that, in a quantum field theoretic sense, and for the classical predictions of the worldline effective action for an infalling observer, the EFT is indeed breaking down.

For near-extremal black holes with tidal forces large compared to the uncorrected solution but small compared to the Planck (or string) scale, the EFT should be replaced with an ultraviolet complete theory of the matter. When the tidal forces become larger than the string or Planck scale, a full theory of quantum gravity might be needed.  We can already see that strings might see a big effect.  The large tidal forces could cause infalling strings  to become highly excited. The following rough estimate indicates that this might be the case. Starting in an unexcited state, the string will follow an ingoing  null geodesic. Suppose we can approximate the spacetime seen by the string by taking a Penrose limit~\cite{Penrose1976}. The result will be a plane wave, and near the horizon it will take the form,\footnote{This form is required to match the curvature since for a plane wave in these Brinkman coordinates, $R_{\rho i\rho j} \sim \partial_i\partial_j g_{\rho\rho}$.}
\begin{equation}\label{eq:planewave}
    {\rm d}s^2 = -{\rm d}\rho\,{\rm d}v +{\rm d}x_i\,{\rm d}x^i + {\rho^{\gamma-2}} h_{ij} x^i x^j {\rm d}\rho^2
\end{equation}
for some constant $h_{ij}$. There is no particle (or string) production in plane waves, but strings can get excited.
Strings in singular plane waves have been extensively studied, with the result that if $\gamma \le 1$, $\langle M^2\rangle$ diverges, while for $\gamma > 1$ it remains finite \cite{deVega:1990ke}. We saw in Sec.~\ref{sec:scaling_EFT} that both of these options arise for certain Kerr-Newman black holes. 
We leave further exploration of this effect for future investigation.

In another direction, we know that black holes are sometimes immersed in magnetic fields. Wald~\cite{Wald:1974np} used a test magnetic field in a Kerr background to estimate the amount of charge a black hole might carry. However exact solutions are known for a Kerr-Newman black hole in a magnetic field \cite{Booth:2015nwa}. One might wonder how the magnetic field affects the singularities on the extremal horizon. With no charge, the magnetic field does not change the near-horizon geometry~\cite{Kunduri:2013gce}, so EFT corrections will produce singularities analogous to Kerr. When charge is added, we expect the singularities will be similar to those described here. In other words, treating the magnetic field exactly is unlikely to qualitatively change the   nature of the singularity.

We have seen that the tidal forces become large at black hole temperatures of order $1/M^3$ (see Eq. \eqref{eq:TEFT}). This corresponds to a timescale of order $M^3$, which is the black hole evaporation time. When a near-extremal black hole evaporates, quantum superradiance causes it to evolve away from extremality. One might thus wonder whether superradiance (which we have not included) might affect our results. Fortunately, the answer is no, since there is a large prefactor multiplying $1/M^3$, so the corresponding timescale is many orders of magnitude shorter than the evaporation time.

\begin{acknowledgments}
We thank Harvey Reall for suggesting a number of comments and participating in discussions related to Secs.~\ref{sec:finite_temp} and \ref{sec:extremal}.
G.H. and M.K. were supported in part by NSF Grant PHY-2107939. G.H. was also supported in part by grant NSF PHY-2309135 to the Kavli Institute for Theoretical Physics (KITP).
G.N.R. is supported by the James Arthur Postdoctoral Fellowship at New York University. 
J.E.S. has been partially supported by STFC consolidated grants ST/T000694/1 and ST/X000664/1.
\end{acknowledgments}

\appendix
\section{The $L^{(K)}$ matrices\label{app:crazynuts}}
In this appendix, we give explicit expressions for the $L^{(K)}$ matrices that appear in Eq.~\eqref{eq:system}.
We first note the following relations,
\begin{equation}
\begin{aligned}
L^{(1)}&=0\\ L^{(3)}&=4 L^{(2)}\\
L^{(4)} &= 0 \\
L^{(5)}&=L^{(2)}\,.
\end{aligned}
\end{equation}
For the remaining $L^{(K)}$, $K=2,6,7,8$, we will give the independent components of the $2\times 2$ matrices in terms of  $\mathfrak{a}$, which at extremality is equal to $\sqrt{1-Z^2}$ for charge-to-mass ratio $Z$.
We first note that the $\tilde L$ matrices appearing in Eq.~\eqref{eq:systeminv} are given by
\begin{equation}
\tilde{L}^{(0)}=\frac{1}{4}L^{(7)}\,,\quad \tilde{L}^{(6)}=L^{(6)}+\frac{3}{4}L^{(7)}-2 L^{(8)}\,,\quad \text{and}\quad \tilde{L}^{(9)}=L^{(8)}-\frac{1}{2}L^{(7)}\,.
\end{equation}

\noindent The components of $L^{(2)}$ are
\begin{equation}
\begin{aligned}
&L^{(2)}_{11}=\frac{3 Z^4 \left[\mathfrak{a} \left(105{-}35 \mathfrak{a}^2{+}39 \mathfrak{a}^4{-}45 \mathfrak{a}^6\right)-3 \left(35{+}6 \mathfrak{a}^4{-}8 \mathfrak{a}^6{+}15 \mathfrak{a}^8\right) \arctan \mathfrak{a}\right]}{16 \mathfrak{a}^7}
\\  
&L^{(2)}_{12}=-\frac{3 Z^3 \left[\mathfrak{a} \left(105{+}25 \mathfrak{a}^2{-}17 \mathfrak{a}^4{+}15 \mathfrak{a}^6\right)-3 \left(35{+}20 \mathfrak{a}^2{-}6 \mathfrak{a}^4{+}4 \mathfrak{a}^6{-}5 \mathfrak{a}^8\right)\arctan \mathfrak{a}\right]}{8 \mathfrak{a}^7}
\\ 
&L^{(2)}_{22}=\frac{3 Z^2 \left[\mathfrak{a} \left(105{+}85 \mathfrak{a}^2{+}17 \mathfrak{a}^4{-}15 \mathfrak{a}^6\right)-3 \left(35{+}40 \mathfrak{a}^2{+}12 \mathfrak{a}^4{-}4 \mathfrak{a}^6{+}5 \mathfrak{a}^8\right) \arctan \mathfrak{a}\right]}{4 \mathfrak{a}^7}\,.
\end{aligned}
\end{equation}
\medskip

\noindent The components of $L^{(6)}$ are
\begin{equation}
\begin{aligned}
&L^{(6)}_{11}=-\frac{3 Z^2}{40 \mathfrak{a}^7 \left(1+\mathfrak{a}^2\right)} \bigg[\mathfrak{a} \left(2835+3855 \mathfrak{a}^2+842 \mathfrak{a}^4+50 \mathfrak{a}^6+195 \mathfrak{a}^8-225 \mathfrak{a}^{10}\right)
\\
&\qquad\qquad\qquad-15 \left(1+\mathfrak{a}^2\right)^2 \left(189-58
   \mathfrak{a}^2+52 \mathfrak{a}^4-38 \mathfrak{a}^6+15 \mathfrak{a}^8\right) \arctan \mathfrak{a}\bigg]
\\
& L^{(6)}_{12}=\frac{3 Z}{20 \mathfrak{a}^7 \left(1+\mathfrak{a}^2\right)^2} \bigg[\mathfrak{a} \left(2835+3750 \mathfrak{a}^2+37 \mathfrak{a}^4-636 \mathfrak{a}^6-35 \mathfrak{a}^8-10 \mathfrak{a}^{10}+75 \mathfrak{a}^{12}\right)
\\
&\qquad\qquad\qquad-15 \left(1+\mathfrak{a}^2\right)^2
   \left(189-65 \mathfrak{a}^2+10 \mathfrak{a}^4-10 \mathfrak{a}^6+9 \mathfrak{a}^8-5 \mathfrak{a}^{10}\right)\arctan \mathfrak{a}\bigg]
\\
&
L^{(6)}_{22}=-\frac{3}{10 \mathfrak{a}^7 \left(1+\mathfrak{a}^2\right)^4} \bigg[\mathfrak{a} \big(2835+6480 \mathfrak{a}^2+1302 \mathfrak{a}^4-4460 \mathfrak{a}^6+368 \mathfrak{a}^8+5096 \mathfrak{a}^{10}
\\
&\qquad\qquad\qquad+1010 \mathfrak{a}^{12}-140 \mathfrak{a}^{14}-75
   \mathfrak{a}^{16}\big)-15 \big(189-261 \mathfrak{a}^2+124 \mathfrak{a}^4
   \\
&\qquad\qquad\qquad+16 \mathfrak{a}^6-9 \mathfrak{a}^8+5 \mathfrak{a}^{10}\big) \left(1+\mathfrak{a}^2\right)^4 \arctan \mathfrak{a}\bigg]\,.
\end{aligned}
\end{equation}
\medskip

\noindent The components of $L^{(7)}$ are
\begin{equation}
\begin{aligned}
&L^{(7)}_{11}=\frac{Z^4}{40 \mathfrak{a}^7 \left(1+\mathfrak{a}^2\right)^4} \bigg[\mathfrak{a} \big(30375+111375 \mathfrak{a}^2+148095 \mathfrak{a}^4+82511 \mathfrak{a}^6+7453 \mathfrak{a}^8
\\
&\qquad\qquad\qquad-2235 \mathfrak{a}^{10}-2115 \mathfrak{a}^{12}-675 \mathfrak{a}^{14}\big)
\\
&\qquad\qquad\qquad-45\left(1+\mathfrak{a}^2\right)^4 \left(675+6 \mathfrak{a}^4-8 \mathfrak{a}^6+15 \mathfrak{a}^8\right)\arctan \mathfrak{a}\bigg]
\\
&L^{(7)}_{12}=-\frac{Z^3}{20 \mathfrak{a}^7
   \left(1+\mathfrak{a}^2\right)^5} \bigg[\mathfrak{a} \big(30375+139770 \mathfrak{a}^2+249690 \mathfrak{a}^4+210706 \mathfrak{a}^6+78808 \mathfrak{a}^8
   \\
&\qquad\qquad\qquad-3186 \mathfrak{a}^{10}+1350 \mathfrak{a}^{12}+870 \mathfrak{a}^{14}+225\mathfrak{a}^{16}\big)
\\
&\qquad\qquad\qquad-45 \left(1+\mathfrak{a}^2\right)^5 \left(675-44 \mathfrak{a}^2-6 \mathfrak{a}^4+4 \mathfrak{a}^6-5 \mathfrak{a}^8\right) \arctan \mathfrak{a}\bigg]
\\
&L^{(7)}_{22}=\frac{Z^2}{10 \mathfrak{a}^7 \left(1+\mathfrak{a}^2\right)^6} \bigg[\mathfrak{a} \big(30375+168165 \mathfrak{a}^2+379050 \mathfrak{a}^4+438746 \mathfrak{a}^6+256656 \mathfrak{a}^8
\\
&\qquad\qquad\qquad+92380 \mathfrak{a}^{10}-26010 \mathfrak{a}^{12}+1350 \mathfrak{a}^{14}-1095\mathfrak{a}^{16}-225 \mathfrak{a}^{18}\big)
\\
&\qquad\qquad\qquad-45 \left(1+\mathfrak{a}^2\right)^6 \left(675-88 \mathfrak{a}^2+12 \mathfrak{a}^4-4 \mathfrak{a}^6+5 \mathfrak{a}^8\right) \arctan \mathfrak{a}\bigg]\,.
\end{aligned}
\end{equation}
Finally, the components of $L^{(8)}$ are
\begin{equation}
\begin{aligned}
&L^{(8)}_{11}=\frac{Z^4}{160 \mathfrak{a}^7 \left(1+\mathfrak{a}^2\right)^4} \bigg[\mathfrak{a} \big(33525+122925 \mathfrak{a}^2+163965 \mathfrak{a}^4+92141 \mathfrak{a}^6+8023 \mathfrak{a}^8
\\
&\qquad\qquad\qquad-6705 \mathfrak{a}^{10}-6345 \mathfrak{a}^{12}-2025 \mathfrak{a}^{14}\big)
\\
&\qquad\qquad\qquad-45\left(1+\mathfrak{a}^2\right)^4 \left(745+18 \mathfrak{a}^4-24 \mathfrak{a}^6+45 \mathfrak{a}^8\right) \arctan \mathfrak{a}\bigg]
\\
&L^{(8)}_{12}=-\frac{Z^3}{80 \mathfrak{a}^7
   \left(1+\mathfrak{a}^2\right)^5} \bigg[\mathfrak{a} \big(33525+156270 \mathfrak{a}^2+284430 \mathfrak{a}^4+247606 \mathfrak{a}^6+99208 \mathfrak{a}^8
   \\
&\qquad\qquad\qquad+3114 \mathfrak{a}^{10}+4050 \mathfrak{a}^{12}+2610 \mathfrak{a}^{14}+675\mathfrak{a}^{16}\big)
\\
&\qquad\qquad\qquad-45 \left(1+\mathfrak{a}^2\right)^5 \left(745-4 \mathfrak{a}^2-18 \mathfrak{a}^4+12 \mathfrak{a}^6-15 \mathfrak{a}^8\right) \arctan \mathfrak{a}\bigg]
\\
&L^{(8)}_{22}=\frac{Z^2}{40 \mathfrak{a}^7 \left(1+\mathfrak{a}^2\right)^6} \bigg[\mathfrak{a} \big(33525+189615 \mathfrak{a}^2+442110 \mathfrak{a}^4+542606 \mathfrak{a}^6+359856 \mathfrak{a}^8
\\
&\qquad\qquad\qquad+152980 \mathfrak{a}^{10}-8910 \mathfrak{a}^{12}+210 \mathfrak{a}^{14}-3285
   \mathfrak{a}^{16}-675 \mathfrak{a}^{18}\big)
   \\
&\qquad\qquad\qquad-45 \left(1+\mathfrak{a}^2\right)^6 \left(745-8 \mathfrak{a}^2+36 \mathfrak{a}^4-12 \mathfrak{a}^6+15 \mathfrak{a}^8\right) \arctan \mathfrak{a}\bigg]\,.
\end{aligned}
\end{equation}

\bibliography{extremal}{}
\bibliographystyle{utphys-modified}

\end{document}